\documentclass[3p]{elsarticle}
\usepackage{hyperref}
\usepackage{amsmath}
\usepackage{amssymb}
\usepackage{color}
\usepackage{mathdesign}

\begin{document}
%\newproof{proof}{Proof}
%\jshort
\def\rvtEPJWC{Eur. Phys. J. Web Conf.}
\newcommand{\araa}{Ann. Rev. Astron. Astrophys.}
\newcommand{\apj}{Astrophys. J.}
\newcommand{\apjl}{Astrophys. J. Lett.}
\newcommand{\apjs}{Astrophys. J. Suppl.}
\newcommand{\aap}{Astron. Astrophys.}
\newcommand{\aj}{Astron. J.}
\newcommand{\gca}{Geo. Ch. Act.}
\newcommand{\mnras}{Mon. Not. R. Astron. Soc.}
\newcommand{\na}{New Astron.}
\newcommand{\prc}{Phys.~Rev.~C }
\newcommand{\prd}{Phys.~Rev.~D }
\newcommand{\prl}{Phys.~Rev.~Lett.}
\newcommand{\pasa}{Publ. Astron. Soc. Aust.}
\newcommand{\pasp}{Publ. Astron. Soc. Pac.}
\newcommand{\pasj}{Publ. Astron. Soc. Japan}
\newcommand{\nat}{Nature}
\newcommand{\nphysa}{Nuclear Phys. A }
\newcommand{\physrep}{Phys. Rep.}
\newcommand{\rvtRModP}{Rev. Mod. Phys.}
\newcommand{\rvtPPNP}{Prog. Part. Nucl. Phys.}
\newcommand{\rmmu}{\mu}
\newcommand{\tplus}{$^+$}
\let\citep\cite

%\begin{macros}
%\end{macros}

\title{Current Status of r-Process Nucleosynthesis}
\author[NAOJ,UT,BU]{T. Kajino}
\ead{kajino@nao.ac.jp}
\author[NAOJ]{W. Aoki}
\ead{aoki.wako@nao.ac.jp}
\author[UW]{A. B. Balantekin}
\ead{baha@physics.wisc.edu}
\author[MPI,TUM]{R. Diehl}
\ead{rod@mpe.mpg.de}
\author[WMU]{M.A. Famiano}
\ead{michael.famiano@wmich.edu}
\author[UND]{G.J. Mathews}
\ead{gmathews@nd.edu}

\address[NAOJ]{National Astronomical Observatory of Japan, Mitaka, Tokyo, Japan}
\address[UT]{University of Tokyo, Tokyo, Japan}
\address[BU]{Beihang University, Beijing, China}
\address[UW]{University of Wisconsin, Madison 53706 U.S.A.}
\address[MPI]{Max Planck Institut f\"ur Extraterrestrische Physik, Garching, Germany}
\address[TUM]{Technical University of Munich, Germany}
\address[WMU]{Western Michigan University, U.S.A.}
\address[UND]{University of Notre Dame, U.S.A.}

\begin{abstract}
The rapid neutron capture process ($r$-process) is believed to be responsible for about half of the production of the elements heavier than iron  and contributes to abundances of some lighter nuclides as well. 
    A universal  pattern of $r$-process element abundances is observed in some metal-poor stars of the Galactic  halo. 
    This suggests that a well-regulated combination of astrophysical conditions and nuclear physics conspires  to produce such a universal abundance pattern. 
    The search for the astrophysical site for $r$-process nucleosynthesis has stimulated interdisciplinary research for more than six decades.  There is currently much enthusiasm surrounding evidence for $r$-process nucleosynthesis in binary neutron star mergers in the multi-wavelength follow-up observations of  kilonova/gravitational-wave GRB170807A/GW170817. Nevertheless, there remain questions as to the contribution over the history of the Galaxy to the current solar-system $r$-process abundances from other sites such as neutrino-driven winds or magnetohydrodynamical ejection of material from core-collapse supernovae.  In this review we highlight some current issues surrounding the nuclear physics input, astronomical observations, galactic chemical evolution, and theoretical simulations of  $r$~process astrophysical environments with goal of outlining a path toward resolving the remaining mysteries of the $r$-process. 
\end{abstract}
%\QUERY[4]
%% \keywords{kwd1 \sep kwd2 \sep ...}
\maketitle
\newpage
\tableofcontents
 \section{Introduction}

      The origin of most atomic nuclei with masses above that of iron group elements (as well as some lighter elements) is attributed to neutron capture nucleosynthesis. 
Two different categories for neutron capture have been
identified, depending upon the competition between neutron
capture and $\beta$~decay: If neutron capture rates are slow
compared to $\beta$~decay rates, only isotopes near stability
are synthesized.  This has been termed the slow-neutron capture
process,
$s$-\emph{process}~\citep{Burbidge:1957,Cameron57}. 

On the other hand, when neutron captures occur much more rapidly
than  $\beta$-decay rates, isotopes far from stability are
synthesized. This is called the rapid-neutron capture  process,
$r$-\emph{process}~\citep{Burbidge:1957}, or
\emph{fast}~\citep{Cameron57} process. Successive and rapid
neutron capture on a particular element can build up neutron-rich
isotopes to the point at which photo-disintegration reaction
rates equal those of neutron capture. At this point the
$(n,\gamma) \leftrightarrows (\gamma,n)$ equilibrium is maintained until the waiting-point
isotope can $\beta$-decay to the next element.  This process
of neutron capture and $\beta$ decay then repeats along a
reaction path far from stability.  This is called  the
\emph{r-process path}.  It is described in more detail in
\ref{models}. 

From the effect of nuclear closed shells on the final $r$-process abundances, one can deduce that the $r$-process path runs far from the line of nuclear stability, so that very high neutron fluxes must occur.  
Moreover, because of the short $\beta$-decay times far from stability, the process only requires a short duration, typically tens of seconds or less to synthesize nuclei all the way from the iron group to actinide elements. 

One way of regulating the $r$ process is via fission recycling.  As neutron captures and $\beta$~decays proceed to the heaviest nuclei, eventually the $r$ process is terminated by neutron-induced or $\beta$-induced fission.  The two fission fragments with intermediate-masses $120\lesssim A \lesssim 160$ can then continue to undergo neutron captures and $\beta$ decays.  
This re-cycles the $r$~process, and the 
repeated flow through the entire $r$-process path results in a reduced dependence on the astrophysical environment, thus enhances the importance  of nuclear properties for determining the isotopic abundances.

The $s$-process, on the other hand, can occur with
comparably low neutron fluxes inside stars. This process is
generally associated~\cite{Busso99} with episodic neutron
irradiation of pre-existing nuclear matter in stars during the
asymptotic giant branch stage of stellar evolution.

The nuclear physics and astrophysics for $s$-process nucleosynthesis are more accessible  both experimentally and theoretically than that for the isotopes relevant to the $r$ process. Therefore, one typically evaluates the $s$-process nucleosynthesis in the solar system abundances and subtracts them from the total of the heavy-element abundances, to deduce  the solar $r$-process abundance distribution; this is described below in \ref{observations}.

 An important stimulus for studies of the $r$-process has been the elemental abundance pattern seen in spectroscopic studies of emission from the stellar-atmospheres of metal-poor stars in the Galactic halo.  Such stars occasionally exhibit the same elemental $r$-process abundance pattern as known from the solar system abundances, This pattern appears to be rather universal even for stars at much younger ages and  metallicities than the Sun.
Even though the iron content, as a proxy for metallicity, varies
over orders of magnitude, this elemental abundance pattern is
rather robust for elements in the range of
$Z = 50\mbox{--}75$~\citep[e.g.~][]{Sneden95,sneden02}.
Thus, the enrichment from the $r$~process appears to have occurred very early in the cosmic chemical evolution in the Galaxy.  Moreover, it appears to to have operated the same way as it did to produce the solar abundances attributed to the $r$~process. 
These observational facts suggest a rather well-regulated origin of heavy elements beyond iron.
However, we do not know if this may be only an artifact of nuclear  properties such as binding energies and $\beta$-decay rates, or it may point to a single cosmic site with astrophysical conditions that are generated uniformly throughout cosmic time.

There appears to be a relatively large scatter of abundances at low metallicities compared to the light $\alpha$-elements produced in supernovae. This indicates that such $r$-process nucleosynthesis events may be attributable to only one or very few events, and that these events are rather rare compared to occurrence rate of supernovae. 

The absolute abundances of $r$-process elements are about seven orders of magnitude below those of abundant elements such as C and O.
Therefore, with a galactic total interstellar gas mass estimate of 10$^9$~M$_\odot$, the total amount of $r$-process material in the Galaxy is of order a few hundreds of M$_\odot$. 
Therefore, nucleosynthesis events at occurrence rates of 10$^{-4}$~y$^{-1}$ or below, ejecting amounts of 10$^{-4}$~M$_\odot$ or more are sufficient sources for the $r$~process.

 From a theoretical astrophysics point of view, a variety of sources have been proposed that can provide the required conditions of high neutron flux during a short time. 
 Among these are core collapse supernovae which form a proto-neutron star after collapse.  This nascent neutron star provides a plausible environment for the generation of the requisite high neutron fluxes. 
In such a final stage of stellar evolution, the stellar core that has transformed its composition during different nuclear burning stages in the previous massive-star evolution collapses when further nuclear energy release is impossible. This may occur as the composition has evolved to iron for more-massive stars ($>$11--25~M$_\odot$), or from electron captures de-stabilizing a degenerate core of O-Ne-Mg composition for stars with $m \sim 8\mbox{--}10$ M$_\odot$. 
The details of matter flows near a newly-formed neutron star,  however, are complex, and  neutrino interactions, hydrodynamic convection, and magnetic fields are among the challenging issues to affect the outcomes. Rare scenarios such as rapidly rotating cores with high magnetic fields may be required to generate significant amounts of $r$-processed ejecta as described in \ref{models}.

Another plausible scenario is the disruption of neutron stars, either in a collision with between two neutron stars or a neutron star plus a black hole, or from matter accretion onto a black hole, around which an extremely-compact accretion disk forms. 
Here again extreme nucleosynthesis and neutron rich conditions are expected.

As a third alternative that has been proposed is the  efficient  
production of neutrons in explosive  helium burning.  Although
this scenario seems no longer viable for the main $r$
process, it may be a source for the light-element primary process
\emph{LEPP}~\citep{Travaglio:2004}.

 During the last decades, astronomical observations have become sufficiently detailed to show that there is more complexity in the above-described picture:
The $s$~process itself was found to show a cosmic evolution that includes puzzles, as, for example, is evident in the C enhancements of very metal poor stars or in globular cluster metallicities. Hence, nuclei which are not due to the  $s$-process are not necessarily due  to a single $r$~process. 
Variability studies for $r$-process evolution have indicated that there are differences between the observed abundances of less massive and the most-massive elements. This suggests a distinction of a \emph{main}, and \emph{weak}, and possibly an \emph{intermediate} $r$-process, which are independently supported by theoretical modeling of the sources.
It is tempting to associate each of these to a particular and specific type of source.

 Chemical evolution studies have shown that each of the above candidate $r$-process sources has specific issues to consistently explain the observed abundance evolution. 
Enrichments with the typical $r$-process abundance pattern have been found at lowest metallicities. This suggests that the $r$-process events should be linked rather promptly to star formation, and not have a significant delay in ejecting their nucleosynthesis yields (as expected for all kinds of binaries involving a compact star remaining after stellar evolution has ended, as, for example, in type Ia supernovae, or neutron star mergers).
Constraining the early evolution of interstellar gas composition,
and matching such delay constraints after the onset of star
formation, could help to identify the type of
sources~\citep[e.g.~][]{Mathews90,Argast04,Wehmeyer15}. This has
been a major challenge, in particular not only for the neutron star merger
source, but also even for core collapse supernovae.
Suggestions have been made (e.g.~\cite{Shibagaki16}) for a
multi-component origin of the $r$-process, based on
neutron star mergers as a dominating source, but supplemented by
variants of  core collapse environments that added just what was
difficult to obtain from the neutron star merger model.
It remains to be seen if such a hybrid  model, with its larger number of degrees of freedom, can be sufficiently constrained by observation and theory to provide a firm and satisfactory explanation of $r$-process nucleosynthesis in the Galaxy.

 After decades of investigating different aspects of core collapse supernovae as potential sites for $r$-process, it is now thought that to find the right conditions in this environment is not easy.
In particular, the conditions of neutron richness in the
neutrino-driven wind~\cite{woosley94} may revert into proton
richness once neutrino (and antineutrino) interactions are
accounted for in detail as described in \ref{models}. 
Hence, only a weak $r$-process might occur in this environment.

 The collision of two neutron stars, also a called \emph{merger}
event, was recognized 40~years ago  as a promising scenario, with
ejection of enough matter from nucleosynthesis to consider those
events as significant contributors to the cosmic abundance of
$r$ elements~\citep{Symbalisty:1982}. 
 With the discovery of a binary neutron star merger through
gravitational waves in 2017~\citep{abbott17}, the hypothesized
occurrence  of neutron star mergers now also is an observational
fact. 
From the astronomical signature of this event in electromagnetic
emission, the \emph{macronova} or \emph{kilonova} associated with
the neutron star merger~\citep[see][and references therein, for a
review of kilonovae]{Metzger17} adds  a significant advance in
the quest for the origins of the $r$-process abundances. 
From the duration of the afterglow emission, it was clear that some injection of energy was powering the kilonova emission, radioactivity from nucleosynthesis being the most plausible.
Then the optical/IR spectra of the emission peak in the IR also suggested that atoms heavier than iron are responsible for the absorption and re-radiation of the emission from this interior energy source. Clearly, this single event matches all expectations for neutron star collisions contributing to cosmic $r$-process nucleosynthesis.
The current consensus  is that the \emph{main/heavy}
$r$~process may be linked to neutron star merging, while
the \emph{light/weak} $r$-process may be realized in
neutrino driven winds from core collapse
supernovae~\cite{Shibagaki16}. 

Studies of the $r$~process have brought together nuclear physicists, astronomers and astrophysicists under a common theme.
Even though it addresses only the origin of elements heavier than iron, whose abundances altogether are about 5 to 8 orders of magnitude below those of lighter cosmic nuclei, the attempts to understand the sites and physics of the $r$~process are still among the main drivers of the entire field of nuclear astrophysics.
 In this review, we begin by laying out the constraints on nuclear reactions from nuclear properties and theory. 
We show how nuclear experiments can continue to place the nuclear physics input for the $r$-process on a firm basis.
This puts us in a position to present the candidate sources that may implement conditions for such nuclear reactions.
Then we review the astronomical observations as they now constrain the $r$-process origins that we wish to model and explain.
After this, we review and discuss the variety of models for the candidate sources in detail. This includes the nucleosynthesis within a source, the astrophysics including processes and the rates of occurrence, and their observational support in terms of direct source observations and less direct abundance histories.
We conclude with a critical discussion of the achievements, status and prospects to solve the quest for the cosmic origins of the heaviest elements and isotopes in the $r$-process. 

 \section{Nuclear Theory}

Theoretical calculations of the r-process nucleosynthesis yields
provide crucial information not only to identify its sites, but
also to interpret observations. Nuclear network calculations are
not yet at a stage of precision for a comprehensive comparison
with the observations. In principle, these network calculations
involve a very large number  of nuclei ranging from the line of
stability to the neutron drip line. It is presently impossible to
measure reaction rates and nuclear properties for all these
nuclei.  Hence,  one needs reliable theoretical estimates of
their properties. Excellent reviews of the nuclear theory issues
in r-process nucleosynthesis are
available~\cite{Mumpower:2015ova,Kajino17}, so that here we only
summarize some salient points. 
 
Nuclear physics input into the network calculations include nuclear masses, $\beta$-decay properties, as well as neutron capture and fission rates. 
Theoretical approaches to each of these are summarized briefly here.

\subsection{Theoretical Neutron Capture Rates}

The nuclear reaction flow in the $r$ process occurs in the vicinity of the neutron drip line. 
There are two main theoretical approaches for neutron capture
reactions.  These are:  (1) via a compound nucleus (including
resonances) as in the Hauser--Feshbach estimates; (2)  direct
capture and  semi-direct (DSD) processes. For most applications 
of $r$-process nucleosynthesis,  nuclear cross sections
have been based upon  a simple estimate of the direct and
semi-direct cross sections in terms of pre-equilibrium
$\gamma$  emission~\citep{Akkermans85}, or in the context of
Hauser--Feshbach  theory,
[e.g.,~\cite{Koning04,Young92,Rauscher00,cyburt10}]. Such an
approach can be justified when it is applied to  nuclei in the
vicinity of the stability line. However, in the neutron-rich
region relevant to the $r$ process, the neutron
separation energies are diminished, so the compound nuclei may
not have enough level density to compete with the compound
elastic process. In this case, the compound capture cross section
may be suppressed, and direct capture becomes dominant even at
low energies.  

Normally, the direct process is not  very important because its
cross section is much smaller than the compound capture cross
sections.  However, in~\cite{Mathews83}  it was noted that far
from stability  where the level density is low, the direct 
capture process could be the dominant  mode of  neutron capture
reactions for  the $r$-process.

In~\cite{chiba08} the DSD components of the neutron capture cross
sections were calculated for a number  of tin isotopes by
employing a single-particle potential (SPP) that gives a good
reproduction of the known single-particle energies (SPEs) over a
wide mass region. The results were compared with the
Hauser--Feshbach  contribution in the energy region of
astrophysical interest. Their calculations showed that the
Hauser--Feshbach component drops off rapidly for the isotope
$^{132}$Sn and toward more neutron-rich nuclei, whereas the DSD
component decreases gradually and eventually becomes the dominant
reaction mechanism.  In~\citet{chiba08} the reason for the
difference in the isotopic dependence between the
Hauser--Feshbach and DSD components was discussed, and its
implication for $r$-process nucleosynthesis was given.

This result is consistent with those of previous studies, but the dependence of the DSD cross section on the target mass number is a feature of their  SPP that gave a smooth variation of SPEs. As a consequence, the direct portion of the DSD components gave the largest contribution to the total ($n, \gamma$) cross section for neutron-rich isotopes below a few MeV, and the direct capture process modifies significantly the astrophysical ($n, \gamma$) reaction rates. The semi-direct component, however, gives a negligible contribution to the astrophysical reaction rates, but its impact is significant above several MeV.

 Valuable studies of the impact of varying theoretical neutron
capture rates in $r$-process models can be found
in~\cite{Mumpower:2015ova}.  In those studies, a  Monte Carlo
variation of Hauser--Feshbach neutron capture rates within the
context of several mass models was explored.    Crucial isotopes
in the vicinity of the $r$-process peaks at $A=130$
and 195 were identified and also in the vicinity of the
rare-earth peak whose measurement would be most effective in
reducing the uncertainties in $r$-process abundance
calculations.

\subsection{Theoretical Nuclear Masses}

Experimentally determined masses~\citep{Audi95,ame03b,Wang12} should be  adopted if available. Otherwise, the theoretical predictions for nuclear masses are necessary. At present there are many available theoretical mass estimates far from stability.  A good resource for nuclear masses can be found at \texttt{http://nuclearmasses.org}.

For nuclear masses there are two basic theoretical approaches: either models which combine liquid-drop model with the nuclear Shell Model and pairing corrections, or models which are entirely microscopic. A review of these different approaches is given in Refs. 
\cite{Lunney:2003zz,Kajino17}. Different nuclear mass models vary
in their success in describing experimentally known nuclear
masses~\cite{Sobiczewski:2018fyd}.  
Hence, a comprehensive study of the sensitivity of
$r$-process nucleosynthesis to individual nuclear masses
was carried out in Refs.~\cite{Mumpower:2015hva}. 
It was found that mass variations of $\pm 0.5~{\rm MeV}$ can result in up
to an order of magnitude change in the final abundance pattern. A
more recent calculation using 10 different mass models found a
similar sensitivity~\cite{Cote:2017evr}. 
It is worth emphasizing that nuclear masses enter into the calculations of all other quantities relevant to the $r$~process.

Theoretical mass tables for  $r$-process nuclei are based upon macroscopic/microscopic methods depending upon a liquid droplet formula
plus  shell corrections.  The most popular adaptation of this is
the finite range droplet model (FRDM)~\citep{moller95,Moller12}.
Another variant of the macroscopic/microscopic approach is the
phenomenological hybrid KTUY model~\citep{Koura00,Koura05}.  A
third is the DZ model~\citep{Duflo99} based upon a
parametrization of multipole moments of the nuclear Hamiltonian.
In a sense  the DZ  is more fundamental than the
macroscopic/microscopic models.  However,  it is not strictly a
microscopic theory, since no explicit nuclear interaction appears
in the formulation.   

At the next level would be masses based upon the extended Thomas--Fermi
random phase approximation (ETSFI)  plus Strutinsky
integral semi-classical approximation to a Hartree--Fock (HF)
approach~\citep{aboussir95,Pearson96}.  The most microscopic
extrapolations generally available of masses for neutron rich
nuclei are those based upon the Skyrme Hartree--Fock--Bogoliubov
(HFB) method.   This is a fully variational, approach with
single-particle energies and pairing treated simultaneously and
on the same footing.  Some recent formulations include: HFB-19,
HFB-21~\citep{Goriely10}; Gogny HFB~\citep{Goriely09a}; the
Skyrme-HFB~\citep{Goriely09b,Chamel08};
HFB-15~\citep{Goriely08}; HFB-14~\citep{Goriely07}.  

A good comparison of the relative merits of each approach can be found in 
\cite{Pearson06}.  All approaches give a reasonable fit to known
nuclear masses.  However, there can be large deviations as one
extends the mass tables to unknown neutron rich nuclei.  Hence,
there is a need for experimental mass determinations for neutron
rich nuclei.  

A recent study has been made~\citep{martin16} of the impact of
nuclear mass uncertainties based upon six Skyrme energy density
functionals based on different optimization protocols.
Uncertainty bands related to mass modeling for
$r$-process abundances were determined for  realistic
astrophysical scenarios.  This work highlights the critical role  
of experimental nuclear mass determinations for understanding the site for $r$-process nucleosynthesis.

\subsection{Theoretical Nuclear Structure}

The level structure of nuclei along the $r$-process path
is important both as a means to determine the partition functions
and as a means to test the strength of shell closures.  Recently
a number of studies have been
completed~\citep{watanabe13,simpson14,Taprogge14} in the
neighborhood of the $N=82, A=130$ $r$-process peak.  In
particular,  the first ever studies~\citep{watanabe13} of the
level structure of the waiting-point nucleus $^{128}$Pd   and
$^{126}$Pd have been completed.  That study  indicated that the
shell closure at the neutron number $N=82$ is fairly robust.
Hence, there is conflicting  evidence between the nuclear masses
and nuclear spectroscopy as to the degree of shell quenching near
the $N=82$ closed shell.  It will be  important to clarify
this point as it has important implications for the site of
$r$-process nucleosynthesis as discussed below.

\subsection{Theoretical $\beta$-Decay Rates}

Treating $\beta$-decay properties for a wide range of nuclei
using a single model is not an easy task. The so-called gross
model was originally introduced to take both allowed and
first-forbidden transitions into account by introducing certain
approximations such as replacing sums by integrals and assuming
average values of matrix
elements~\cite{grosstheory,Takahashi:1973qiv}. This theory is
further refined by introducing shell effects of the parent
nuclei~\cite{Nakata:1997qvt}.
At present, a popular alternative is the quasi particle
random-phase approximation for the Gamow--Teller matrix
elements~\cite{Moller:2003fn}. Neutron capture cross sections are
calculated using the Hauser--Feshbach statistical model, again
necessitating theoretical input for the level densities and
gamma-strength functions for the experimentally inaccessible
neutron-rich nuclei.

The  $\beta$-decay rates, particularly at  waiting point nuclei, constitute  one of the most important nuclear physics inputs to  nucleosynthesis calculations in  the $r$-process. 
Theoretical investigations of the beta decay of isotones with
neutron magic number of $N = 82$ have been done by various
methods including the shell model~\citep{Zhi13}, quasiparticle
random-phase approximation (QRPA)/finite-range droplet model
(FRDM)~\citep{Moller03}, QRPA/extended Thomas--Fermi plus
Strutinsky integral (ETFSI)~\citep{Borzov00}, and
Hartree--Fock--Bogoliubov (HFB) ~+~ QRPA~\citep{Engel99}
calculations as well as in the continuum quasiparticle
random-phase approximation (CQRPA)~\citep{Borzov03}. The
half-lives of nuclei obtained by these calculations are rather
consistent with one another, and especially in shell-model
calculations experimental half-lives at proton numbers
$Z = 47, 48$, and $49$ are well
reproduced~\citep{Martinez99}.

For the $\beta$ decays at $N=126$ isotones, however,
half-lives obtained by various calculations differ from one
another~\citep{Langanke03,Grawe07}. First-forbidden (FF)
transitions become important for these nuclei in addition to the
Gamow--Teller (GT) transitions in 
contrast to the case of $N = 82$. 

A strong suppression of the half-lives has been predicted
in~\cite{Borzov03} for  $N = 126$ isotones due to the FF
transitions. Most shell-model calculations of the
$\beta$-decay rates of $N = 126$ isotones have been done
with only  the contributions from the GT transitions
included~\citep{Langanke03,Martinez01}. Moreover, as noted below
experimental data for the $\beta$ decays in this region of
nuclei are not yet available. The region near the waiting point
nuclei at $N = 126$ is therefore called the ``blank spot''
region.

In~\cite{Suzuki12}, $\beta$ decays of $N = 126$ isotones
were studied by taking into account both the GT and FF
transitions to evaluate their half-lives. 
Shell-model calculations were done with the use of shell-model interactions based upon modified G-matrix elements that reproduce well the observed energy levels 
of the isotones with a few (two to five) proton holes outside
$^{208}$Pb~\citep{steer08,Rydstrom90}. 

In~\cite{Marketin15} the impact of first-forbidden transitions on
decay rates was 
studied using a fully self-consistent covariant density functional theory (CDFT) framework to provide a table of $\beta$-decay half-lives and $\beta$-delayed neutron emission probabilities, including first-forbidden transitions.   This work demonstrated that there is a significant contribution of the first-forbidden transitions to the total decay rate of nuclei far from the valley of stability.  This also leads to  better agreement with experimentally determined half lives as discussed below.

In addition to ground state $\beta$ decay, at the high
temperatures of the $r$-process environment decay can
proceed through thermally excited states.  In~\cite{Famiano08} a
calculation was made  to evaluate the possible effects of the
$\beta$-decay of nuclei in excited-states on the
astrophysical $r$-process. Single-particle levels were
calculated in the FRDM model with quantum numbers determined
based upon their proximity to Nilsson model levels. The resulting
rates were  used in an $r$-process network calculation.
Even though  the decay rate model was simplistic,  this work did
provide a measure  of the possible effects of excited-state
$\beta$-decays on $r$-process freeze-out
abundances.  The main result of that work was that in the more
massive nuclei, the speed up of the decay rates  in the approach
to closed shells tended to exaggerate the underproduction of
nuclei below the nuclear closed shells as discussed below.

\subsection{$\beta$-\lowercase{Delayed Neutron Emission}}

Beta-delayed neutron emission is particularly important for the
freezeout of the $r$-process.   Recently,  beta-delayed
neutron emission probabilities of neutron rich Hg and Tl nuclei
have been measured~\citep{caballero16} together with beta-decay
half-lives for 20 isotopes of Au, Hg, Tl, Pb, and Bi in the
region of neutron number $N \ge 126$. These are the heaviest
nuclear species for which  neutron emission has been observed. 

Although not directly on the $r$-process path, these measurements have provided information with which to evaluate the viability of nuclear microscopic and phenomenological models for  the high-energy part of the $\beta$-decay strength distribution. Indeed, this study indicated that there is no global $\beta$-decay model that provides satisfactory $\beta$-decay half-lives and
neutron branchings on both sides of the $N = 126$ shell
closure. There was, however,  a slight preference for the
Hartree--Bogoliubov model  of~\cite{Marketin15}.

\subsection{Fission Barriers and Fission Fragment Distribution}

Fission is another important component of the $r$~process. During the evolution of the system along the $r$-process path there can be multiple fission cycles. Once this path reaches fissionable nuclei the resulting fission products continue capturing neutrons, potentially repeating the cycle a number of times 
\cite{Beun:2007wf,Mendoza-Temis:2014mja}. The relevant fission
cross sections are not yet experimentally accessible, requiring
theoretical input. Fission may also be $\beta$-delayed: a
parent nucleus first $\beta$ decays into a daughter nucleus
which then undergoes fission. Beta-delayed fission may be a
dominant fission channel in the termination of the
$r$~process~\cite{Mumpower:2018wjv,Shibagaki16}.

In $r$-process models with a very high neutron-to-seed
ratio (such in the ejecta from neutron star mergers) the
$r$-process path can proceed until neutron-induced or
beta-induced fission terminates the beta flow at $A \sim 300$.
Determining where this occurs can significantly impact the yields
from $r$-process models~\citep{Eichler15,Shibagaki16}.
Unfortunately there are no measurements of fission barriers or
fission fragment distributions (FFDs) for nuclei heavier than
$^{258}$Fm~\citep{schmidt12}.

This is a major uncertainty in all calculations of fission
recycling in the $r$-process~\cite{Shibagaki16}.
considered a FFD model based upon the KTUY model plus a
%\QUERY[5]
two-center shell model to predict both symmetric and asymmetric
FFDs with up to three components.  As such,  fissile nuclei could
span a wide mass range (A $=$ 100--180) of fission fragments as
demonstrated below.

On the other hand, the $r$-process models
of~\cite{Korobkin12} were  mostly based upon a simple two
fragment distribution as in~\cite{Panov01} (or alternatively the
prescription of~\cite{Kodama75}).  The assumption of only two
fission daughter nuclei  tends to place a large yield near  the
second $r$-process peak leading to a distribution that
looks rather more like the solar $r$-process abundances.
In contrast, the FFDs of~\cite{Goriely13} are based upon a rather
sophisticated SPY revision~\citep{SPY} of the  Wilkinson fission
model~\citep{Wilkins76}.  The main ingredient of this model is
that the individual potential of each fission fragment is
obtained as a function of its axial deformation from tabulated
values. Then a Fermi gas state density is used  to determine the
main fission distribution.   This leads to FFDs with  up to
four humps.   

An even more important  aspect is the termination of the $r$-process path and the number of fissioning nuclei that contribute to   fission recycling and the  freezeout of the $r$-process abundances.  
The $r$-process path in~\cite{Shibagaki16} proceeded
rather below the fissile region until nuclei with $A \sim 320$,
whereas the $r$-process path in~\citet{Goriely13}
terminates at $A \approx 278$ [or for a maximum $\langle Z \rangle$
for~\cite{Korobkin12}]. Moreover,~\cite{Shibagaki16} found that
only ${\sim}10$\% of the final yield comes from the termination
of the $r$-process path at N  $=$  212 and Z  $=$  111,  while
almost 90\% of the $A=160$ came  from the fission of  more
than 200  different parent nuclei mostly via beta-delayed
fission. 
On the other hand, the yields of~\cite{Goriely13} that are almost
entirely due to a few  $A \approx 278$ fissioning nuclei with a
characteristic four hump FFD.  As noted below, this has a
dramatic impact on the final   $r$-process abundance
distribution. 

Recently, there has been significant progress by the FIRE collaboration.
In reference~\cite{Vassh18} results were compared using the
Finite Range
Droplet Model, Thomas--Fermi (TF), Hartree--Fock--Bogoliubov (HFB-17), and Extended
Thomas--Fermi with Strutinsky Integral (ETFSI)
model masses and corresponding barriers along
with the phenomenological fission  yields described in that work. In particular, they  explored
how the different termination points for the r process
predicted by these models influence the final abundance
pattern, and they identified which  fissioning nuclei were most accessed in neutron-star merger conditions.

Similarly, in~\cite{Mumpower:2018wjv} an exploration of  the
impact of beta-delayed fission in the r-process in the tidal
ejecta of neutron-star
mergers  was made.  This study showed that beta-delayed fission is a key fission channel that shapes the final abundances near the second r-process peak.

\subsection{Neutrino Physics}

One common element for all possible $r$-process sites is the presence of an abundant number of neutrinos, making neutrino interactions a crucial part of the nuclear physics input. Since neutrinos only interact via weak interactions they can transfer energy and entropy over long distances. In addition, charged-current interactions affect the neutron-to-proton ratio in the r-process. This implies that fundamental neutrino parameters such as the neutrino mass hierarchy, mixing angles and the possible existence of sterile neutral fermions which mix with the neutrinos but do not take part in the weak interactions can influence $r$-process nucleosynthesis. To illustrate how neutrinos impact element synthesis consider the mass fraction of the species $i$ in the nucleosynthesis environment:
\begin{equation}
X_i = \frac{N_iA_i}{B}
\end{equation}
where $N_i$ is the number of species $i$ per unit volume, $A_i$ the atomic weight of these species, and
\begin{equation}
B = \sum_j N_jA_j
\end{equation}
is the total baryon number. If the medium is locally charge neutral the electron fraction is given by
\begin{equation}
\label{Yevalue}
Y_e = X_p + \frac{1}{2} X_{\alpha} + \sum_{h\neq p, \alpha} \left( \frac{Z_h}{A_h} \right) X_h 
\end{equation}
where $Z_h$ is the charge of the nucleus $h$. Note that $X_h$ for the heavier nuclei is exceedingly small. The rate of change of the number of free protons is given by
\begin{equation}
\label{NPchange}
\frac{dN_p}{dt} = - \lambda_- N_p + \lambda_+ N_n
\end{equation}
where $\lambda_-$ is the total rate for the reactions $\overline{\nu}_e+ p \rightarrow n+e^+$ and $e^- + p \rightarrow n+ \nu_e$ which destroys protons and $\lambda_+$ is the total rate for the reactions $\nu_e+ n \rightarrow p+e^-$ and $e^+ + n \rightarrow p+ \overline{\nu}_e$ which creates protons. In writing Eq.~\ref{NPchange} one assumes that neutrinos do not dissociate alpha particles and ignores spallation reactions on heavier nuclei which could knock out nucleons. Both are reasonable approximations given the peak of the neutrino energy spectra during the r-process  occurs at energies well below the alpha particle binding energy and spallation thresholds for most heavier nuclei. 

From Eqs. \ref{Yevalue} and \ref{NPchange} we can write the equilibrium value of the electron fraction as
\begin{equation}
Y_e = \frac{\lambda_+}{\lambda_- +\lambda_+} +\frac{1}{2} 
\frac{\lambda_- - \lambda_+}{\lambda_-+\lambda_+} X_{\alpha} + \sum_{h\neq p, \alpha} 
\left[ \left( \frac{Z_h}{A_h} \right)  - \frac{\lambda_+}{\lambda_- +\lambda_+} \right] X_h .
\end{equation}
Clearly the electron fraction in the nucleosynthesis environment is determined by the neutrino reaction rates. These rates in turn are determined by the neutrino cross sections and neutrino fluxes for the appropriate flavors:
\begin{equation}
\lambda = \int dE_{\nu}\> \sigma (E_{\nu}) \> \frac{d\Phi_{\nu}}{dE_{\nu}} .
\end{equation}
The fundamental neutrino properties such as the masses and mixing
angles determine the flavor content of the neutrino energy
spectrum which then determines the value of the electron
fraction, the controlling parameter of the
$r$-process~\cite{Balantekin:2013gqa}. 
In many astrophysical environments the sheer number of neutrinos
produced necessitate including neutrino--neutrino interactions in
the neutrino transport and the resulting collective neutrino
oscillations~\cite{Duan:2010bg,Duan:2009cd,Balantekin:2004ug,Pehlivan:2011hp,Sasaki:2017jry}. 
It should also be kept in mind that neutrinos can induce fission
during the r-process~\cite{Qian:2002mb,Kolbe:2003jf}. 

 \section{Experimental Nuclear Data}

Because the r-process environment is so closely linked to the properties of the
nuclei involved, an understanding of nuclear properties is necessary. Nuclear physics properties necessary to understanding the production of 
nuclei in an $r$-process site include neutron separation
energies, $S_n$; $\beta$-decay rates $\lambda_\beta$;
neutron capture rates, $\lambda_n$;  and information on the
shell structure of r-process nuclei~\cite{pfeiffer01}. While many
of the
properties of r-process progenitor nuclei have yet to be measured, studying neutron-rich nuclei approaching the r-process path is beneficial as it provides constraints on nuclear structure models used to extrapolate the
properties of r-process nuclei.

From an astronomical standpoint, an understanding of the r-process arises from
observations of visible light absorption spectra from stars
containing r-process elements~\cite{aoki03,aoki10}, 
$\gamma$-ray observations~\cite{Diehl:2014,Diehl:2015}, and
gravitational wave observations~\cite{ligo}.  All of these
observations are used to constrain a different aspect
of the r-process including yields and the nature of the actual r-process site. 
Observational data is also intertwined with nuclear data.  Both 
aspects of r-process research rely heavily on each other.  For this reason,
accurate, thorough, and useful nuclear data are important.
\subsection{Nuclear Masses}

Perhaps most fundamental to understanding the r-process is a knowledge
of nuclear masses.  Knowing nuclear masses of neutron-rich nuclei can allow a determination of neutron separation energies, $S_n$, which will lead to a determination of the classical r-process path as a function of environmental
neutron abundance and temperature.  Further, nuclear mass measurements
can help to constrain models, which are generally used to extrapolate to 
the neutron-rich nuclei~\cite{matos09}.

Detailed mass systematics toward the r-process path is necessary
to constrain the uncertainties with the r-process evolution and
final abundance
distribution~\cite{martin16,mumpower15a,mumpower15b}. As the hot
r-process environment is presumed to start at a temperature
$T_9\approx$10 and proceed to form light nuclei at about
$T_9\approx$2.5 (where $T_9 \equiv 10^9$ K) followed by the formation
of the heavier nuclei before finally freezing out at
$T_9\approx 1$, it is worthwhile to measure nuclear masses to an
uncertainty of $\delta m\lesssim kT\approx 100$ keV.  However, even uncertainties of
several hundred keV are useful as they
can help constrain mass models. Likewise, masses of neutron-rich nuclei that approach -- but are not on -- the r-process path are also useful for constraining mass model trends.

Direct measurements of nuclei include frequency-based and
time-based measurement~\cite{sun15}.  Of the frequency-based
measurements, Penning traps and storage rings are the most
developed and well-known techniques.  Time-of-flight (TOF)
measurements vary in technique, but all rely on measuring the
flight time of nuclei of unknown mass between two points in a
beam path of known rigidity and comparing to that of nuclei for
which the masses are known. 

\begin{figure}[h!]
\caption{Left: Relative uncertainties $\delta$m/m of nuclear measured nuclear masses. Right: Measured decay rates of nuclei.  The approximate location of the r-process path is shown on the left diagram, and the approximate location of the neutron and proton drip lines is shown on the right figure\label{mass_unc}}
\includegraphics[width=0.5\textwidth]{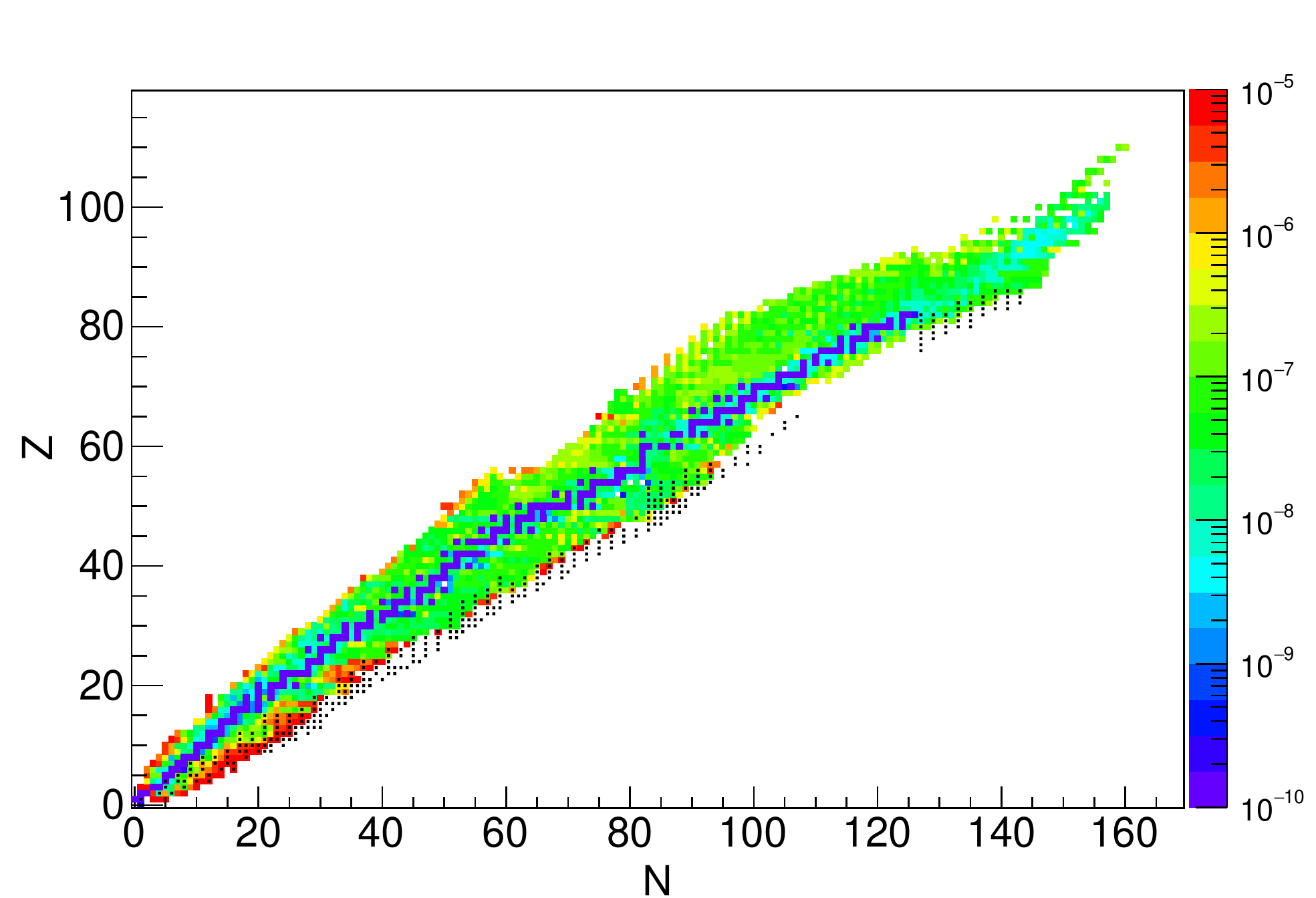}
\includegraphics[width=0.5\textwidth]{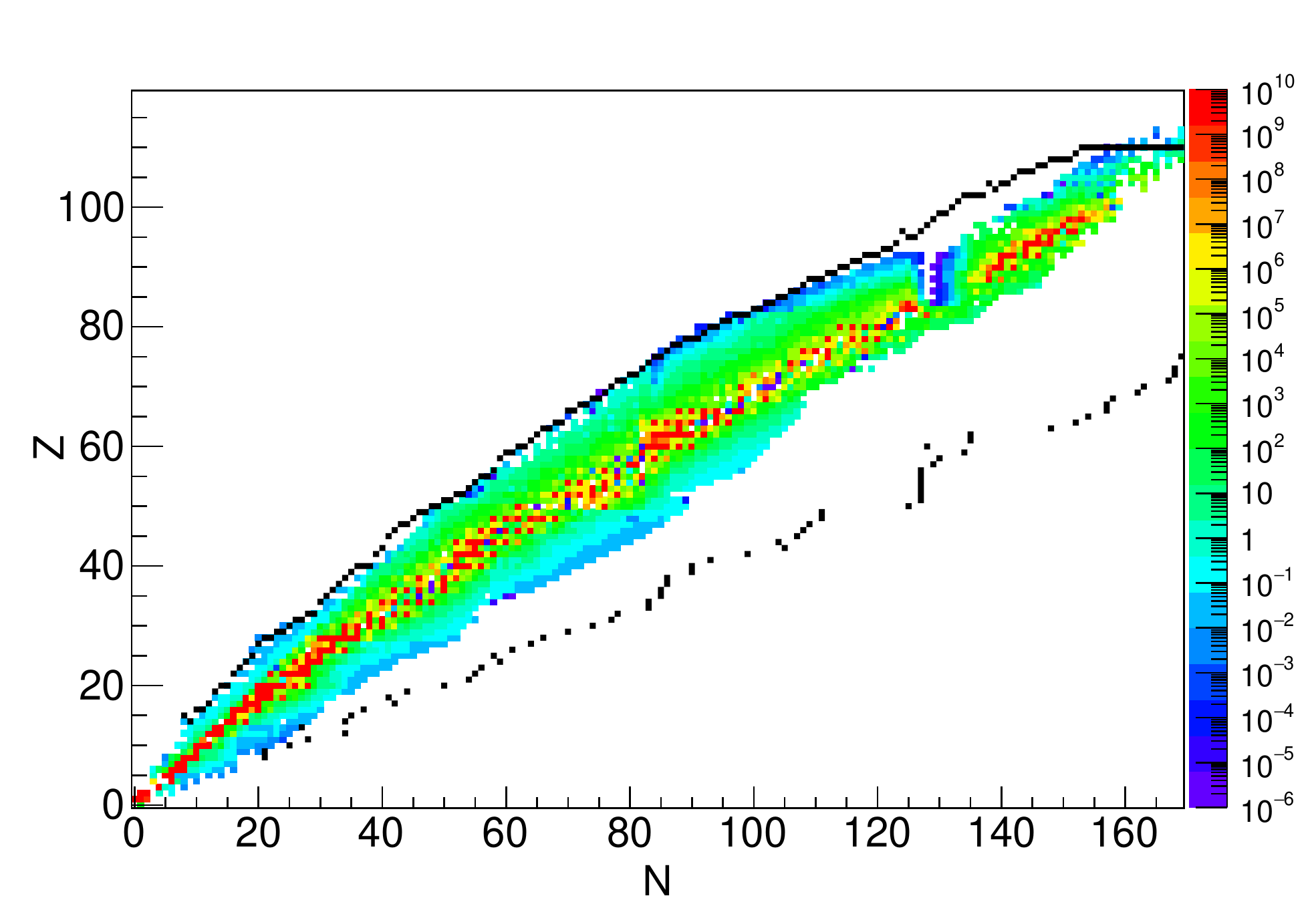}
\end{figure}

Given the number of techniques and efforts to measure nuclear masses 
with greater precision, constraints on nuclear masses and mass models for nuclei approaching the r-process path have seen significant progress in the past decade.  A representation of the relative uncertainty in nuclear mass measurements as of the AME2016 update is shown in \ref{mass_unc}. While the masses of most r-process nuclei have yet to be studied, progress
continues to be made.

\begin{figure}
\caption{Schematic~\cite{lunney03} of approximate resolutions and
lifetimes of various mass measurement
techniques\label{mass_params}}
\includegraphics[width=\textwidth]{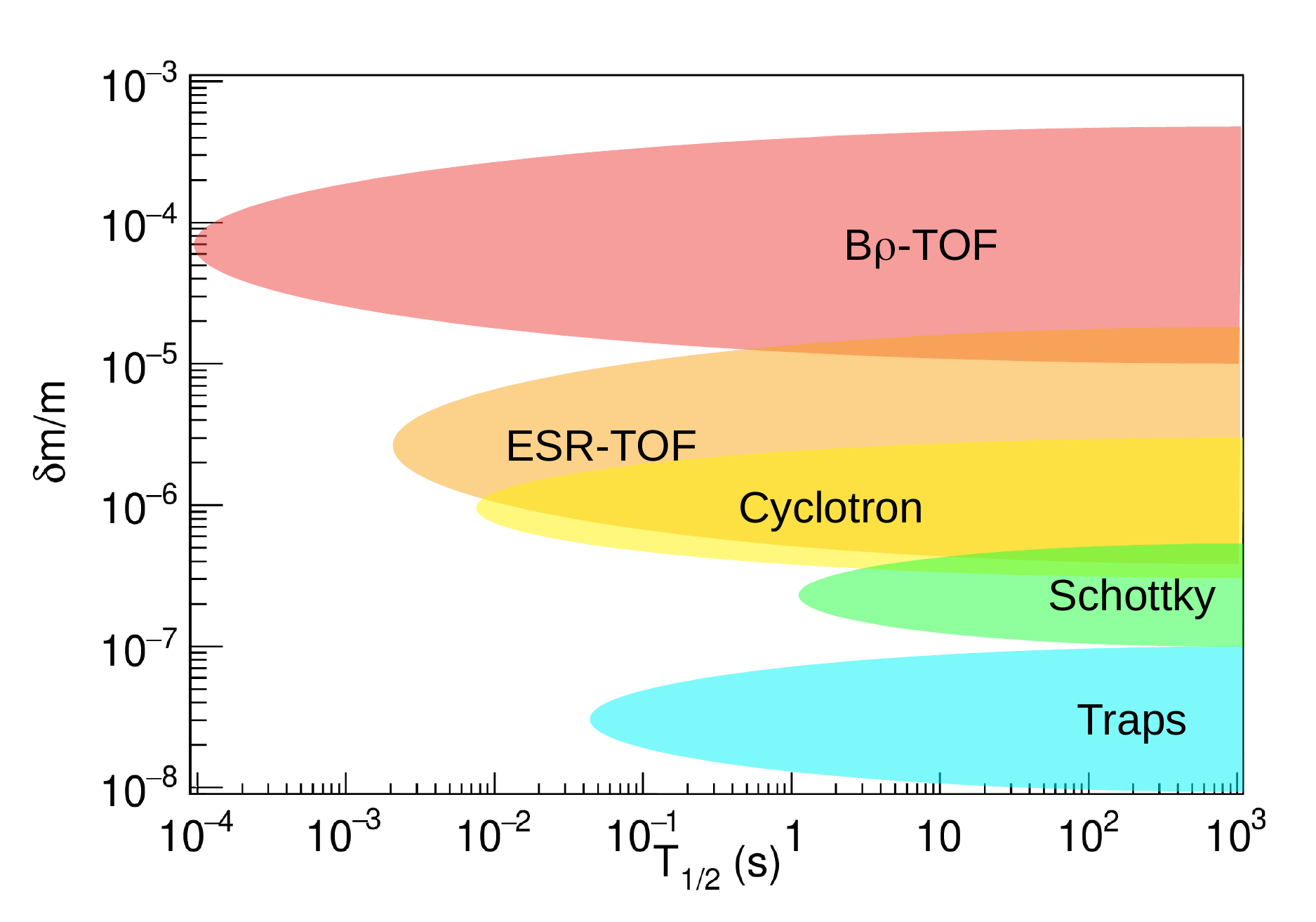}
\end{figure}

Techniques for measuring nuclear masses include the use of Penning traps, beamline
TOF techniques, and storage rings.  While numerous mass measurement techniques
are available~\cite{mittig93}, we concentrate on those currently
used for mass measurements of
exotic nuclei.  Each technique has its own advantages and 
disadvantages.  Mass measurement techniques are limited by the half-lives of nuclei
they are capable of measuring.  Likewise, each technique has a mass resolution limit.
The range of lifetimes and mass resolutions accessible by various techniques is 
 shown 
%\QUERY[6]
in \ref{mass_params}.

Current direct mass measurement techniques have been limited primarily 
by the production rates of nuclei to be measured.  
However, with newer facilities such as the RIBF and FRIB, it may be possible to 
directly measure the masses of r-process nuclei. 
Various mass measurement techniques and recent results from selected facilities
are described below.

\subsubsection{Mass Measurement Compilations and Evaluations}

The field of measuring nuclear masses directly and indirectly is rapidly changing with
multiple measurements made yearly. While mass measurement techniques and programs are
described below with some notable recent measurements, multiple mass evaluations and
compilations are available and described for recent updates.

The Atomic Mass Evaluation was recently updated
(AME2016)~\cite{ame16a,ame16b} to replace the 2003 version 
(AME03)~\cite{ame03,ame03a,ame03b}.  
This volume also contains
the NUBASE2016 tables~\cite{nubase16,ame16a,ame16b} which contain
recommended experimentally known and extrapolated
properties.  This evaluation includes complete experimental information
as well as evaluation procedures and descriptions of calculated data.  Recommended values are reported.  
Downloadable tables for the NUBASE2016 database are available at
multiple online sites~\cite{amdcch,amdcau}.

In addition, the Atomic Mass Compilation 2012
(AMC12)~\cite{amc12} provides an extensive list of experimental
and calculated masses.  This reference could be useful as it also provides in the tabulation the experimental facility that
produced the listed results. Comparisons between AMC12 and the
AME03 have also been noted~\cite{audi15}.

Multiple online resources are also available, including
\texttt{nuclearmasses.org}~\cite{nucmass} and the 
National Nuclear Data Center (NNDC)~\cite{nndc}.
The former provides an interface for examining nuclear properties from multiple datasets.  Another useful
feature of \texttt{nuclearmasses.org} is an up-to-date compilation of new mass measurements.  The NNDC provides
access to multiple reaction and structure databases, as well as interfaces to access nuclear properties in the
Evaluated Nuclear Structure Data File (ENSDF).
\subsubsection{Trap Measurements}

Measurements using Penning traps are capable of very fine mass resolution on the order
of $\Delta M/M\sim 10^{-9}$.  They are, however, limited in the lifetimes of the nuclei that
can be studied with these techniques.  Because the trap must maintain the captured nuclei long
enough to measure it, lifetimes of nuclei measured with
Penning traps are limited to the order of seconds or milliseconds. 

Several trap facilities around the world have measured to extremely good accuracy the masses of
several proton-rich and neutron-rich nuclei.  While the number of nuclei along the r-process
path which have been measured using Penning traps is limited, masses approaching the r-process
path are still useful as they can be used as calibration 
masses for other techniques which are able to reach more 
exotic nuclei.  

A thorough review of mass measurements with Penning traps is
given in reference~\cite{dilling18}.

\paragraph*{JYFLTRAP}
 The JYFLTRAP facility is a double cylindrical
Penning trap installed at the IGISOL facility
at the University of
 Jyv\"{a}skyl\"{a}~\cite{hakala12,aysto13,kankainen11,rahaman07,rahaman08}.
 It is capable of measuring atomic masses up to 
 $A=120$ with accuracies ${\sim} 50$ keV~\cite{jyfltrap}.
 The JYFLTRAP system routinely measures mass uncertainties
 of $\delta m/m\sim 10^{-8}$~\cite{jyfltrap2}.  Currently,
 the JYFLTRAP facility has measured the masses
 of nearly 300 neutron-rich nuclei and those of
 nearly 100 proton-rich
nuclei~\cite{rahaman07a,hakala08,Rahaman2007b,hager06,hager07,hager07b,hakala11,hakala12,wieslander09,rinter07}.
 Of these masses, a portion has been 
 measured near the Z $=$ 50 waiting point of the 
 r-process, $^{131}$Cd, and neutron-rich Sb and Te.
 Nuclear masses have also been measured for
 nuclei near the r-process path for nuclei between the
 first and second r-process peaks.  
\paragraph*{ISOLTRAP}
 The ISOLTRAP facility is a tandem Penning trap mass spectrometer located at the
online isotope separator ISOLDE at CERN.  The masses of over 400 nuclei have been measured with
lifetimes down to ${\sim}$50~ms and mass accuracies typically
of $\delta m/m\sim 5\times 10^{-8}$~\cite{isoltrap,wolf13}.
The trap was recently upgraded by the addition of a
multi-reflection time-of-flight mass
separator/spectrometer~\cite{wolf13}.
This addition significantly improves the trap's purification capabilities while also operating as a mass
spectrometer.  In addition, because of the beam purity achieved by the spectrometer, background-free
decay half-lives can be measured at the ISOTRAP
facility~\cite{wolf16}.  A proof-of-principle measurement of the
half-life of $^{27}$Na was performed using this setup.

The ISOLTRAP facility has measured a wide range of masses of
nuclei between $^{17}$Ne and
$^{232}$Ra~\cite{herfurth05,baruah08}.  Near the r-process
path, the masses of many nuclei in and surrounding the $N=80$ shell closure have been measured.  ISOLTRAP results are 
also pushing the limits of neutron-rich masses past the Z $=$ 82 shell closure. 
Recent measurements of $\mbox{}^{129-131}$Cd have shown a 400 keV
deviation from prior mass determinations using $\beta$-decay
data.~\cite{atanasov15}
However, as with many Penning trap measurements,
the masses of nuclei along the r-process path remain a challenge outside of N $=$ 82 shell closure.

\paragraph*{{LEBIT}}
 The Low Energy Beam and Ion Trap (LEBIT) facility is located at
the NSCL facility in the USA~\cite{ringle13}.
Combined with the fragmentation facility at the NSCL, the LEBIT facility can measure the masses of nuclei produced by
fragmentation of fast beams.  Fragments are then slowed in a gas stopping system. This process avoids any chemistry constraints on
the beam and is able to produce exotic beams from a variety of primary beams.  The LEBIT facility operates a linear gas
stopper and a cyclotron gas stopper~\cite{schwartz16} along with
a laser ablation ion source and a plasma ion source,
both used for stable beams.  An accumulator and buncher downstream of the gas stoppers and ion sources cools the beam
prior to injection into the single Penning trap
system~\cite{schwartz16b}.

The LEBIT facility has currently measured the masses of over 50 isotopes.  Many of these are impossible to produce at
ISOL facilities, a distinct advantage of this facility. Currently, the LEBIT facility has concentrated on precision mass
measurements of proton-rich nuclei, the completion of the FRIB
facility may provide new opportunities for LEBIT~\cite{gade16}.

\paragraph*{{CPT}}
 The Canadian Penning Trap (CPT), located at Argonne National Laboratory near Chicago, IL, is
installed on the ATLAS linac
facility~\cite{savard01,wang04,schelt12}.  While a major
concentration of the CPT
has been on isotopes along the N $=$ Z line, progress has also been
made for nuclei approaching the r-process path~\cite{schelt12}.

The CPT contains a radio-frequency quadrupole (RFQ) trap and a precision Penning trap.  Isotopic species are 
produced in the ATLAS facility and separated in the Enge split-pole spectrograph.  A gas cooler and buncher system
thermalizes the nuclei prior to injection into the trap.  Measurement accuracy has been extended to levels similar to
other facilities, ${\sim} 10^{-8}$.

Recent upgrades of the gas catcher system~\cite{savard03} and the
use of a $^{252}$Cf source to produce fission fragments 
have made possible the measurements of the masses of 40
neutron-rich nuclei with $51\le Z\le 64$~\cite{schelt12}. These
measurements are
useful to constrain mass models for neutron-rich nuclei relevant to the r-process.  Measurements of Te and Sb along
the r-process path were also made.  

\paragraph*{{TITAN}}
 TRIUMF's Ion Trap for Atomic and Nuclear Science (TITAN) is a system of multiple Penning traps 
with a charge breeder
from an electron beam and associate photon counters for simultaneous decay 
spectroscopy~\cite{dilling06,delheij06,froese06,kwiatkowski13,lascar16}.
While the primary purpose of TITAN
is determination of the $V_{ud}$ CKM matrix element, it is also used for nuclear mass measurements and
nuclear mass model constraints~\cite{dilling03}.

TITAN is capable of mass measurements of comparatively short-lived nuclei, having measured the mass of $\mbox{}^{11}$Li,
with $t_{1/2}$  $=$  8.8 ms~\cite{smith08}.  Recently, the TITAN
facility was used to measure the masses of 
the neutron-rich nuclei $\mbox{}^{98,99}$Rb and $\mbox{}^{98-100}$Sr at high
precisions~\cite{klawitter16}.  Such measurements
can constrain mass models for masses approaching the r-process path. Near the N $=$ 82 closed shells, the masses of
$\mbox{}^{125-127}$Cd have been measured~\cite{lascar17}.

The TITAN facility recently underwent an upgrade to install charge breeder capabilities in its electron beam
ion trap~\cite{lascar16}.  This results in higher charge states,
improving the precision of 
mass measurements. The TITAN EBIT also utilizes seven Si(Li) detectors to perform in-trap decay spectroscopy
\cite{brunner11}.  With this
setup 511 keV photons from $\beta$-particle annihilation of decay products can be detected.   

\paragraph*{{SHIPTRAP}}
 Though the primary concentration of the SHIPTRAP facility is on proton-rich nuclei, it is
worth mentioning as this facility has excelled at probing 
massive nuclei~\cite{herfurth05}, displaying the range of masses
accessible by trap facilities. 
The SHIPTRAP facility at GSI is a two-stage Penning trap system set up for precision 
measurements of very heavy ions as well as the search for super-heavy elements.  Singly-charged ions 
produced via fusion-evaporation reactions are delivered to the trap from
the SHIP facility at about
100 keV/u and thermalized in a helium gas cell.  They are then cooled in an RFQ buncher prior to 
delivery to the trap~\cite{dilling00}.  SHIPTRAP was designed and
developed for the measurements of
nuclei with $Z\ge 92$ including the transuranic elements, an uncharted region in the table of the isotopes for mass 
measurements~\cite{schonfelder02}. 

Early measurements of $^{147}$Ho$^+$,  $^{147}$Er$^+$, and $^{148}$Er$^+$ were determined for the first time
using SHIPTRAP~\cite{block05}.  Subsequent measurements have
concentrated on proton-rich nuclei relevant to the
endpoint of the rp-process~\cite{block06}.

\subsubsection{{TOF} Measurements}

Nuclear mass measurement techniques via the time-of-flight (TOF) method -- though less accurate than Penning traps -- can measure the masses of nuclei with shorter
half-lives.  

The TOF method is used to measure nuclear masses through the measurement of the flight-time
of nuclei between two points in a beam facility~\cite{matos09}.
Such a method is generally
employed at radioactive ion beam facilities in which the beamline is used as the primary
detector.  With this infrastructure, the TOF is measured using fast detectors placed at two or more
points in the beamline.  A particular technique, referred to
as the $B\rho$-TOF method, is used to measure masses by measuring the particle rigidity, $B\rho$ along with the flight time of species in the beamline.  Alternatively, in an isochronous mode, 
species
with higher velocities have a longer flight path resulting in all
nuclei with the same mass-to-charge ratio having the same time-of-flight.

The mass is
related to the rigidity and flight time through the relationship:
\begin{equation}
\frac{B\rho}{\gamma} = \frac{mv}{q}
\end{equation}
where $\gamma$ is the usual Lorentz factor for particles in the beamline.  
With an overall TOF uncertainty of ${\sim}$50 ps (after correcting for path length via a measurement of
the particle rigidity), and a TOF ${\sim}0.5~{\mu s}$, the uncertainty in 
the mass from the TOF is ${\sim}$10$^{-4}$.  Greater statistics can reduce the 
statistical uncertainty to values typically ${\sim}$10$^{-6}$ to ${\sim}$10$^{-5}$.
The lifetime limitation
of measured nuclei is due to the TOF between two detectors.  Calibration of 
flight times with this method is accomplished by incorporating known ``reference masses'' in 
the mixed beam.  Because the $B\rho-TOF$ method can measure shorter lifetimes, it can 
extend the range of known masses to more neutron-rich nuclei.  This method is also useful
for measuring multiple masses simultaneously in a single experiment. 

Reference masses for calibration of
the method can be taken from masses measured with Penning traps.  The major limitation on nuclei that can be measured is then the production rates of exotic beams.  
The particle rigidity can be determined by measuring beam position at
dispersive planes in the beamline.   Additional complications of this method may be the 
incorporation of charge states (species that are not fully ionized) in the beam, which must be
separated by measuring the energy deposition in a particle telescope at the end of the beamline.  This method is sometimes referred to as the $B\rho$-TOF-$\Delta E$ method. 

The resolution of the TOF-$B\rho$ method is generally ${\sim}$100 keV/c$^2$, which is roughly the temperature,
$kT$,
of the r-process environment.  This resolution is generally sufficient to constrain
mass models far from stability.

Mass measurements at facilities incorporating the $B\rho$-TOF or $B\rho$-TOF-$\Delta E$ 
method are described below. 

\begin{figure}
\caption{Schematic of the TOF-B$\rho$ experimental setup at the NSCL facility\label{tof_nscl}}
\includegraphics[width=\textwidth]{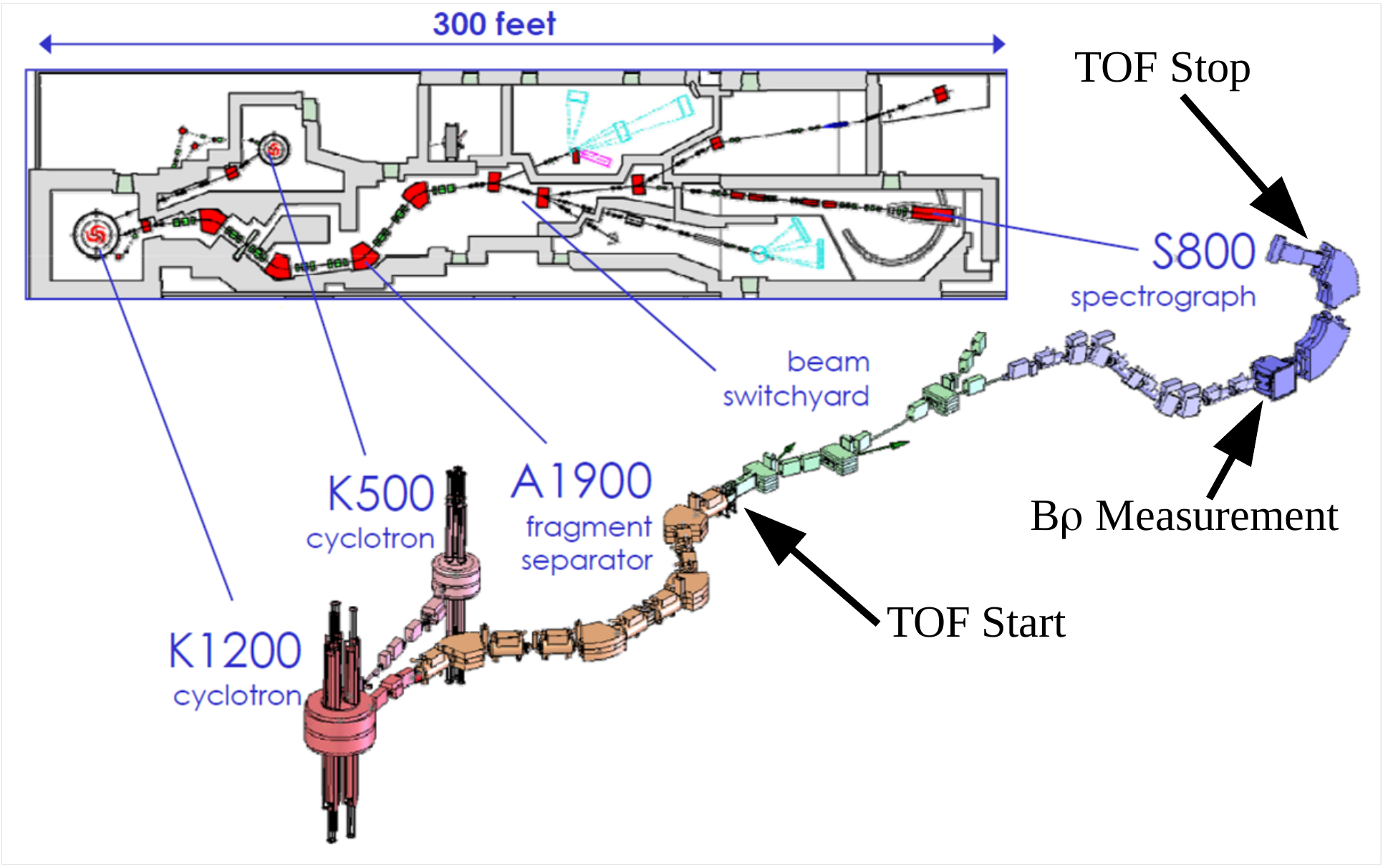}
\end{figure}

\paragraph*{{NSCL}}
 
At the time of this writing, the National Superconducting Cyclotron Laboratory (NSCL) in East Lansing, MI 
utilizes the coupled-cyclotron facility (CCF) and the A1900 spectrometer coupled to the 
S800 spectrometer for creating a particle flight path of about 58 m.  The TOF detectors
are placed at the A1900 extended focal plane and the focal plane of the S800.   Particle 
rigidity is measured at the target position of the
S800~\cite{estrade07,matos08}.  A schematic
of this setup, typical of setups utilizing the beamline for this sort of measurement, is shown in
\ref{tof_nscl}.  For this 
particular setup, the rigidity measurement is done by running
the experiment in dispersive mode, in which the maximum dispersion of the cocktail beam is at the
S800 target position.  A thin, position-sensitive detector is placed at this point.  In order to minimize
the uncertainties due to energy straggling in the detector, the position measurement is done
with a detector that is as thin as possible and as close to the TOF stop as possible.  In this case,
micro-channel plate (MCP) detectors based on secondary electron
emission were used~\cite{shapira00,rogers15}.

As the NSCL upgrades to the Facility for Rare Isotope Beams (FRIB), production rates of more exotic 
nuclei should enable mass measurements of nuclei much farther from stability.  It is predicted that 
nuclear masses for nuclei well past the second peak into the rare
earth region will be possible~\cite{HRSWP}.

The NSCL program has concentrated on a variety of mass
measurements over the past several years~\cite{schatz13}, with
a recent push toward neutron-rich nuclei of increasing
mass~\cite{meisel16}.  Recently, masses of
$\mbox{}^{59-64}$Cr were measured and compared to prior measurements.  In addition, new masses of neutron-rich
Ar, Sc, Mn, and Fe were measured, pushing the limits of mass measurements closer to the drip line for 
potentially early stages in the
r-process~\cite{meisel16,meisel15}.  
The purpose of these measurements was primarily to constrain nuclear reactions in the deep crusts of
neutron stars. Thus, the nuclei measured
are not yet on the r-process path, though they are useful for constraining systematics and mass models
for neutron-rich nuclei.  Prior
measurements~\cite{estrade11,matos12} using the same technique at
the same facility
have determined the masses of $^{61}$V, $^{63}$Cr, $^{66}$Mn, and $^{74}$Ni with resolutions of
about $2\times 10^{-4}$.

\paragraph*{{SPEG}}
 The SPEG (Spectrom\`etre \`a Perte d'Energie du Ganil) is a high-resolution spectrometer
located at GANIL in Caen, France.  The SPEG spectrometer consists of a 82 m flight path following an
alpha spectrometer, so named for its beamline shape.  The SPEG mass measurement program is
established with a long history of mass measurements via the
B$\rho$-TOF method~\cite{savajols01,savajols01b,gomez05}.

The SPEG program uses microchannel plate detectors as the TOF start and stop signals.  These have an 
intrinsic time resolution of ${\sim}$100 ps (FWHM).  The setup also utilized two rigidity measurements.  Like 
the NSCL system, a rigidity measurement is made with the placement of an MCP at a dispersive focal plane.  Another
measurement is made after the TOF stop detector using two drift chambers.  

Initial mass measurements were of light neutron-rich nuclei with
accuracies down to a few 10$^{-5}$~\cite{gillibert86},
and subsequent early measurements of neutron-rich nuclei from boron to phosphorus were conducted
\cite{gillibert87}.  Additionally, the neutron-rich nuclei
$\mbox{}^{29,30}$Ne, $\mbox{}^{34,35}$Mg, $\mbox{}^{36,37}$Al, and 
$\mbox{}^{31-33}$Na were also performed~\cite{orr91}, further
not only extending the region of known masses of neutron-rich
nuclei, but also concluding that a region of known deformation includes $^{30}$Ne.  Subsequent measurements of
masses in this region~\cite{sarazin00} have improved upon or
extended the region of known masses of neutron-rich
nuclei.  In this case, the known mass region was extended to $^{47}$Ar.   

The SPEG collaboration has concentrated heavily on masses of neutron-rich nuclei near N $=$ 16, 20, and 28.
While this mass region is quite low with regard to the r-process, the extension of light mass models 
in the neutron-rich part of the table of isotopes is important.  Such a region may also be 
important for a ``light element primary process'' (LEPP).
%\QUERY[7]
measured neutron-rich masses near the N $=$ 20 shell closure, verifying a region of shape coexistence about
N $=$ 20 and N $=$ 28 nuclei by measuring the masses of nuclei with
$29\le A \le 47$~\cite{sarazin01}. 
These experiments were followed by an additional successful attempt to push the region of 
known nuclear masses of light nuclei 
closer to the neutron drip line~\cite{savajols05,jurado07}.
Preliminary results of this 
model
experiment presented first-time measurements of 7 new 
masses while the masses of 36 nuclei were significantly improved in the mass region A${\sim}$ 10--50. 
Recently, this region was further extended to include mass measurements of $^{19}$B, $^{22}$C,
$^{29}$F, $^{31}$Ne, $^{34}$Na, and other light
nuclei~\cite{gaudefroy12} confirming the halo structure
of several nuclei in this region.  In this measurement, the B and C nuclei measured are presumed to
be at the drip line.   

\paragraph*{Other Facilities}

The Radioactive Ion Beam Factory (RIBF) has recently developed
its SHARAQ spectrometer~\cite{uesaka08,uesaka12,michimasa13}.
Coupling the spectrometer to the BigRIPS beamline, a 105 m flight path can be created.  Time of flight detectors
utilize diamond detectors~\cite{michimasa13b}, which are
radiation hard and can sustain high rates. The high 
resolution of the SHARAQ detector allows for a fine measure of rigidity.  Recent measurements of the masses of
$\mbox{}^{55,56}$Ca have resulted in mass resolutions of
$\sigma\sim$150 keV and 234 keV respectively~\cite{michimasa16}.

The collaboration between Beihang University and~\cite{sun16} the
Heavy Ion Research Facility in Lanzhou (HIRFL)~\cite{xia02} is
another promising activity, particularly as this collaboration
has
concentrated on detector development to improve the time resolution of TOF detectors.  Recently, fast plastic 
scintillators with an intrinsic time resolution $\sigma\approx$ 5.1
ps have been developed~\cite{zhao16}.  Preliminary
tests at HIRFL with multi-wire proportional counters
too measure particle rigidity by position measurements in the dispersive focal plane.  The intrinsic MWPC
resolution is $\sigma\approx$ 1~mm.

The Time-of-Flight Isochronous
(TOFI)~\cite{wouters85,vieira86,wouters88} is an early
device to measure nuclear masses via time-of-flight.  Designed for light-Z, neutron-rich nuclei, this device
operates as a TOF spectrometer in an isochronous mode.  In this mode, slower ions travel a shorter path.  The
TOF is then only dependent on A/Q of the given ion.  This mode is employed in TOF measurements using rings as well
(described in \ref{rings}). As an earlier mass measurement system, the TOFI spectrometer has measured the masses 
of several neutron-rich isotopes between Li and
P~\cite{wouters88,wouters85}. The resolution of these
measurements 
was between about 200 keV and 900 keV.

\subsubsection{Storage Rings}

\label{rings}
Storage ring facilities function on a principle similar to the $B\rho$-TOF method.  In this case, a beam of particles held at a specific rigidity travels in a closed-loop path in a ring.  The frequency of the particle's path about the ring is used to determine the mass.  The 
advantage of the storage ring is that it can operate with fairly low intensity beams to
measure the TOF of the particle multiple times~\cite{ozawa14}. 

Storage rings are capable of measuring multiple masses
in a closed-circuit ring.  Particle frequency, $f$,
about the ring of circumference $C$, is
related to the mass-to-charge ratio~\cite{franzke08}:
\begin{equation}
\label{ring_freq}
\frac{\Delta f}{f} = -\alpha_p\frac{\Delta(m/q)}{(m/q)}
+\left(\frac{1}{\gamma^2} - \alpha_p\right)\gamma^2\frac{\Delta\beta}{\beta}
\end{equation}
where $\alpha_p = 1/\gamma_t^2$ is the so-called momentum
compaction factor. High resolution is achieved via ``cooling''
in which the velocity spread approaches zero, 
$\Delta \beta\rightarrow 0$, or by operating
the storage ring at the ``transition point'' in which
$(\gamma^{-2}-\alpha_r)\rightarrow 0$.  In either
case, narrow variations in the frequency spectrum for a specific
mass are measured.

Storage ring facilities are described below:

\paragraph*{{GSI-ESR}}
  
\begin{figure}
\caption{A schematic setup of the GSI FRS-ESR ring, including downstream
 experimental areas~\cite{geissel08}. (Used with
permission.)\label{gsi_ring}}
\includegraphics[width=\textwidth]{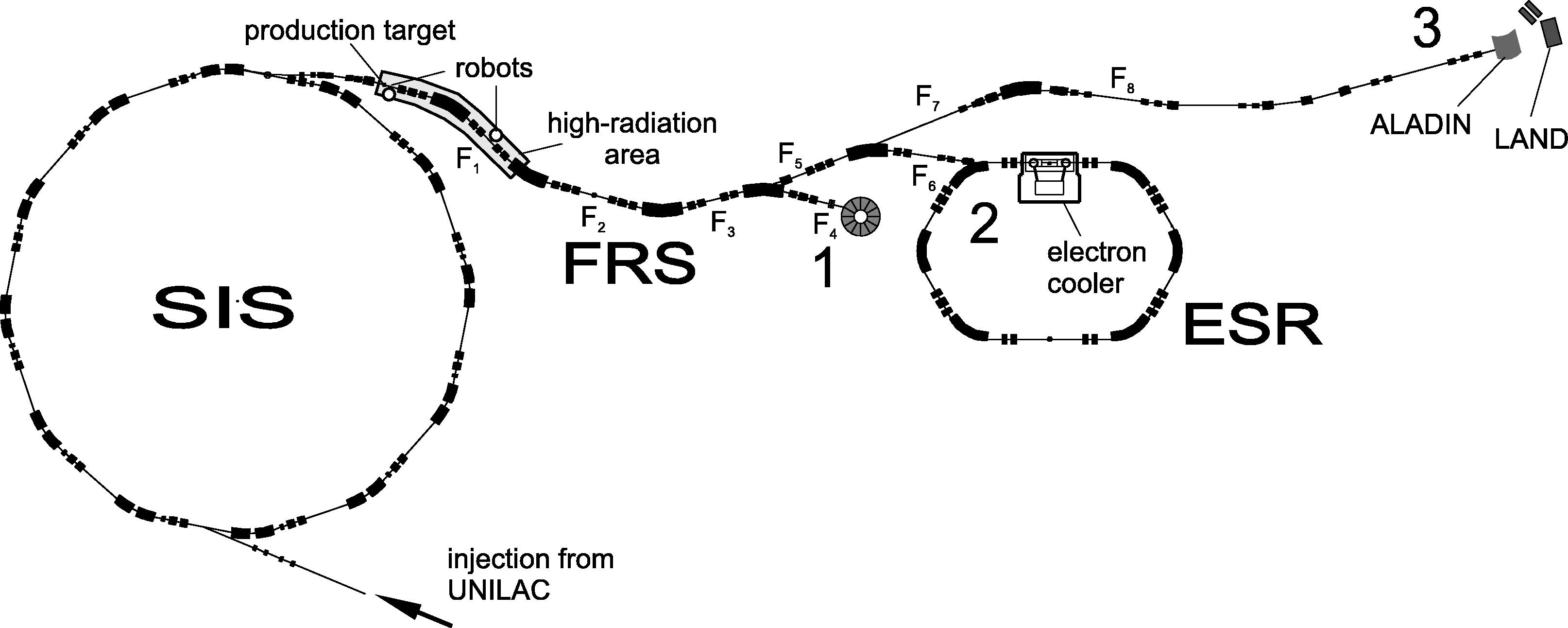}
\end{figure}

The storage ring located at GSI has been in operation since 1990.  
A figure of this experimental system is shown in
\ref{gsi_ring}~\cite{geissel08}.  The heavy
ion synchrotron SIS produces fragments which are analyzed in the fragment separator FRS
and subsequently analyzed in the Experimental Storage Ring (ESR).

The ESR has a circumference of 108 m and a magnetic rigidity of 10 Tm.  It is thus
capable of storing uranium ions at 560 MeV/u.  The rotational frequency of this
ring (used in~\ref{ring_freq}) is upwards of 2~$\times$~10$^6$ s$^{-1}$.

The ESR is capable of running in two modes.  In Schottky mass spectrometry (SMS), measurements are
made of particles with the same velocity.  That is $\Delta v/v \rightarrow 0$.  In 
isochronous mass spectrometry (IMS), measurements are made for particles with the same orbital
period by cutting the rigidity to a very small range.  
In this case, $\gamma\rightarrow \gamma_t$.  In either case, the second
term of~\eqref{ring_freq} approaches zero, and the measurement accuracy can be
improved greatly~\cite{wollnik97,hausmann05}. 

Recently, the FRS-ESR system has measured the masses of nuclei near the N $=$ 82 waiting point
$\mbox{}^{129-131}$Cd with an accuracy of
2~$\times$~10$^{-4}$~\cite{knobel16}.  
This measurement is useful for determining the shell gap at this point.  On the
neutron-rich side of stability, measurements were made using IMS for nuclei
between Se and Ce by fragmenting a $^{238}$U primary beam, 
extending the known mass landscape out by one more
neutron~\cite{knobel16b}.
These measurements could be useful in constraining mass models in this region.
Approaching the r-process path in the rare earth region, masses of neutron-rich
nuclei between Lu and Os were recently measured using SMS to
similar accuracies~\cite{shubina13}

\paragraph*{{HIRFL-CSR}}

The Heavy Ion Cooler-Storage-Ring (HIRFL-CSR) at
Lanzhou~\cite{xia02} is part of the post-accelerator system
at the Heavy Ion Research Facility in Lanzhou (HIRFL).   Beams extracted from the cyclotron system are
stored in the main ring (CSRm) and further extracted for radioactive ion beam experiments or storage
in a separate experimental ring (CSRe).  The main ring has a maximum design rigidity of $B\rho = 10.64$ Tm,
and the experimental ring has a maximum design rigidity of $B\rho = 8.440$ Tm.  These relatively
high rigidities could enable good beam separation and identification for a variety of beam
experiments.  Studies of this device for mass measurement
experiments have been performed~\cite{chen15}.  

This facility was able to measure the masses of the very neutron-rich $\mbox{}^{52-54}$Sc nuclei via the IMS method, finding a strong subshell closure for
the N $=$ 32 Sc isotopes~\cite{xu15}.

The HIRFL-CSR was recently used to produce, identify, and study
neutron-rich nuclei in the $^{18}$O, $^{36}$P, and
$^{29}$Mg region~\cite{sun18} by utilizing the Radioactive Ion
Beam Line in Lanzhou (RIBLL2) to deliver beams to
the CSRe~\cite{zhang16}.  This facility has also been used to
examine the masses of proton-rich nuclei near
$^{58}$Ni~\cite{sun12}.

\paragraph*{{RIKEN rare RI-ring}}

The RIKEN facility was commissioned in 2015~\cite{ozawa12}.  It
consists of a 60 m circumference
ring with 0.5\% momentum acceptance.  At a beam energy of 200 A MeV, the revolution time is 355~ns.
Unlike the GSI and HIRFL facilities, this ring
is supplied by a cyclotron which feeds the BigRIPS.  Particle identification is done
within the BigRIPS rather than the ring.  With this setup, it is possible to 
examine single particles in the ring. It is also worth noting that RIKEN currently has
the highest intensity of $^{238}$U beams in the world, making the rare RI-ring an
attractive site for nuclear mass measurements.  Current plans are to use the Rare-RI ring
for mass measurements in IMS mode to achieve masses with a resolution on the order of 10$^{-6}$.
\subsubsection{Other Similar Systems}

Other systems include variants of TOF systems, including
multi-reflection TOF systems and systems which can
be described as hybrid techniques.  

Two facilities employing similar techniques as TOF and 
cooling ring systems are described below.

\paragraph*{{KEK-MRTOF}}

The KEK Multi-Reflection Time of Flight (MRTOF) system 
is a versatile system for nuclear mass measurements.  It
has a high rate range, a resolution ${\sim}$10$^{-5}$, and
a fast measurement cycle allowing for measurements of
nuclei with lifetimes ${\sim}$30 ms~\cite{schury14}.

The MRTOF functions by injecting trapped ions into 
a drift tube with electrostatic mirrors on either end.  
Reflecting the ion over multiple passages of the
device increases the length of the flight path in
a compact device.  Ejection of the ions out of the end of the
device into a timing detector provides a measurement of
the flight time.  Varying the number of times the
ions can traverse the device will provide multiple
measurements over various time ranges.

Recent measurements with the MRTOF have
yielded first-time mass evaluations of isobaric
chains of fusion-evaporation reaction products 
of actinides in the proton-rich 204$\le A\le$ 206
region~\cite{schury17}.  This provides a proof of the 
capabilities of the MRTOF device for heavy nuclei.

\paragraph*{{MISTRAL}}
 
The ISOLDE MISTRAL spectrometer measures the 
nuclear charge to mass ratio by measuring
the cyclotron frequency of ions in
homogeneous magnetic fields~\cite{lunney01}.
Ion beams are focused onto an entrance slit
in the field and extracted into an electron 
multiplier.  At half-turns in the ion orbit, the ions pass
through slits of an RF modulator, thus determining
their kinetic energy and relating the RF frequency and
the cyclotron frequency~\cite{simon95}.

The MISTRAL experiment concentrates primarily on light,
neutron-rich nuclei.  Recent measurements
have included $^{26}$Ne, $\mbox{}^{26 - 30}$Na, and
$\mbox{}^{29-33}$Mg~\cite{lunney01,gaulard06}.  Uncertainties
off ${\sim}$20 keV are possible with this device.
Other measurements have included halo studies of $^{11}$Li 
and $\mbox{}^{11,12}$Be~\cite{bachelet08,gaulard09}.
\subsection{$\beta$-\lowercase{Decay Rates and Spectroscopy}}

Decay rates of the $\beta$-unstable nuclei along the r-process path are another fundamental
property useful in understanding the r-process.  A knowledge of $\beta$-decay rates
of r-process nuclei will provide a measure of the timescale of the r-process as well as 
a knowledge of the nuclear structure of the involved nuclei.  Additionally, $\beta$ spectroscopy
can provide a measure of the thermodynamics of r-process environments.  

As many $\beta$-decays are to excited states of the daughter nuclei, coincidence measurements of 
$\gamma$-spectra are also useful.  This information can provide a guide for future $\gamma$-ray
astronomy measurements.  Searches for isomers are particularly useful as observational signatures of
the longer-term behavior of explosive events.
\subsubsection{{RIKEN}}

\label{RIBF_beta}
\begin{figure}
\caption{Layout of the RIBF facility showing the coupled
cyclotrons, SAMURAI spectrometer, and RI
ring~\cite{shimizu11}\label{ribf_fig}}
\includegraphics[width=\textwidth]{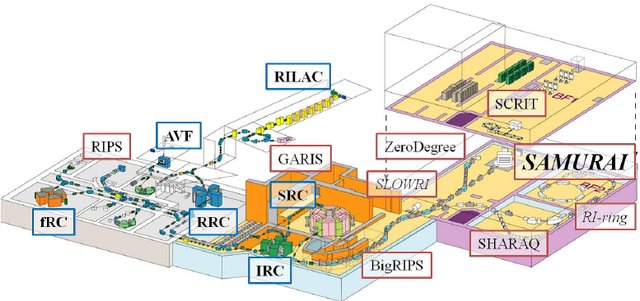}
\end{figure}

Direct measurements of the half-lives of r-process nuclei are difficult as they generally require
production of enough nuclei to obtain an accurate lifetime.  A leader in this field is
the RIBF at RIKEN which has pursued the measurement of $\beta$-decay half-lives
using implantation-type methods~\cite{famiano03,nishimura05}.
The 
layout of the RIBF facility is shown in 
\ref{ribf_fig} \cite{shimizu11}.

\paragraph*{{AIDA}}

The most recent development of the
$\beta$ detection system is the Advanced Implantation
Detector Array (AIDA)~\cite{griffin14}. 
This detector system consists of a stack of double-sided silicon strip detectors (DSSDs) in which
beams from the accelerator facility are implanted.  Decays are measured by timestamping events
from the implantation and subsequent decay.  A system of newly-developed application-specific integrated
circuits (ASICs) reduces the volume of electronics necessary to run this experiment.  Because the
experimental setup can be placed at the end of the beamline, it can operate parasitically while
other experiments are running.  

\paragraph*{{WAS3ABi}}

Similar to the AIDA device is the Wide Range Active Silicon-Strip Stopper Array for Beta and Ion detection
(WAS3ABi)~\cite{wasabi}.  This device is similar in construction
while used in conjunction with the EURICA
spectrometer (described below).

Since their commissioning, the AIDA and WAS3ABi experiments have measured the $\beta$-decay half-lives of over 250 
neutron-rich nuclei, with nearly 150 of these being first-time
measurements~\cite{nishimura11,lorusso14,lorusso15,wu16,wu17}.
These nuclei are close to, or on, the r-process path and cover r-process nuclei near the A$\approx$130 peak, the
A$\approx$165 rare-earth hill, as well as neutron-rich nuclei from Kr to Tc with 90$\le$A$\le$149.
In the case of nuclei near the A$\approx$130 peak, the half-life systematics and shell effects were found to be robust, 
with no substantial changes in the gross nuclear structure as
predicted by theory~\cite{lorusso15}. In the case of
the Kr to Tc region, results indicate an enhancement in the rates in this region, resulting in a potential speed-up of
the r-process in this region. Studies of nuclei in the rare earth region, especially Nd, have been predicted to be important to
the overall r-process~\cite{mumpower15a,mumpower15b}.  The newly
measured half-lives will be useful for future measurements.

\paragraph*{{EURICA}}

In addition to the half-life measurements at RIKEN, $\beta-\gamma$ spectroscopy capabilities have also been added with the
Euroball RIKEN Cluster Array (EURIKA)~\cite{eurica}.  Launched in
2011, EURICA is designed to perform isomer and
$\beta$-delayed
$\gamma$ spectroscopy.  It consists of 12 segmented germanium
detectors extracted from the RISING system at GSI~\cite{rising}.
This creates a nearly 4$\pi$ germanium array which can be mounted around an active beam stopper system (such as WAS3ABi).  The
EURICA/WAS3ABi setup is mounted at the last focal point of the RIBF Zero-Degree Spectrometer. It has an energy resolution of
3 keV at $E_\gamma$  $=$  662 keV and a photo-peak efficiency of 15\% at $E_\gamma$  $=$  662 keV.  With EURICA, $\beta$-decay detection of implanted nuclei 
with coincident or delayed $\gamma$ detection is useful for isomer searches, measurements of decay branching
ratios, shape transitions and coexistence, low-lying yrast and non-yrast states, and single-particle levels.

The initial commissioning of EURICA examined decays of $^{16}$C to $^{16}$N to confirm the 0$^-\rightarrow$2$^-$
transition at 120.42 keV.  The lifetime of this state was measured at $5.25~\rmmu {\rm s}$.  The scientific program
of EURICA began in 2012 to study isomeric states of $^{98}$In, $\mbox{}^{95,96}$Cd, and $^{94}$Ag from the fragmentation of
a $^{124}$Xe primary beam. The EURICA array has been productive in its first year, with neutron-rich nuclei near the r-process
path with 100$\lesssim$A$\lesssim$170~\cite{soderstrom14}.
Much of the EURICA experimentation has concentrated in this
region of the isotopic chart. 

The EURICA experimental configuration makes it quite useful for a variety of studies, particularly those with
cocktail beams.  
Recent results from EURICA include verification and explorations
of the N $=$ 82 shell
closure~\cite{watanabe13,Taprogge14,watanabe14}, multiple isomer
searches of neutron-rich nuclei on or near 
the r-process path around the N $=$ 82
shell~\cite{simpson14,Taprogge14,patel14,patel16,lozeva16,watanabe16,ideguchi16,jungclaus17,patel17},
nuclear structure and deformation studies 
\cite{browne15,moon17,jungclaus16}, studies of collective
motion~\cite{lee15,soderstrom16}, and single-particle and
particle-hole level studies~\cite{jungclaus16b,moon17}.
\subsubsection{{CERN} {ISOLDE}}

Several studies of the $\beta$-decay rates have been
performed with the ISOLDE Decay Station (IDS)~\cite{fynbo17}.
This system is similar
in setup and configuration to the AIDA/EURICA experiment at the RIBF in that it utilized an implantation target surrounded
by HPGE detectors.  LaBr detectors can also be used for fast timing measurements of $\beta$-decay and $\gamma$ coincidences.

Initial experiments of the IDS measured the $ ^{129}$Sn levels
from the decay of $^{129}$In isomer decays~\cite{lica16}.
Using this method, the lowest 1/2$^+$ state of $^{129}$Sn was measured with a lifetime of 19~$\pm$~10 ps.  This was followed
by experiments to measure decay rates of neutron-rich 
$\mbox{}^{148 - 150}$Cs isotopes~\cite{lica17}.  This provided
experimental measurements of the half-life of these nuclei as
well as a measurement of the decay scheme of $^{149}$Ba.  

While the IDS is a more recent implantation system, early experiments at the ISOLDE facility have used a moving tape collector
to examine $\beta$- and $\gamma$-spectroscopid data for
the N $=$ 82 waiting-point nucleus $^{130}$Cd~\cite{dillmann03}.  An
unexpectedly high energy of the 1{\tplus}  level of
$^{130}$In was found.  This level is populated primarily via
the GT transition
of $^{130}$Cd.  Spectroscopy of this nucleus also found a high Q value of 8.34~MeV for this decay, which agrees with
recent mass models that include shell quenching effects.  Additional measurements for the neutron-rich
nuclei near the r-process path with an implantation tape include
half-lives of $^{216}$Bi~\cite{kurpeta00}, nuclear spectroscopy
of $^{133}$Sn~\cite{hoff00}, half-lives and spectroscopy
of $\mbox{}^{130-132}$Cd~\cite{hannawald00,hannawald01},
$\beta$-decay half-lives and $\beta$-delayed neutron
emission, $P_n$, for
the neutron-rich isotopes $\mbox{}^{94-99}$Kr and
$\mbox{}^{142-147}$Xe~\cite{bergmann03}, decay lifetimes and
spectroscopy of
$\mbox{}^{215,217,218}$Bi~\cite{kurpeta03a,kurpeta03b,dewitte04}, decay
half-life and spectroscopy of $^{215}$Pb~\cite{dewitte13}.
\subsubsection{{ANL}/{CARIBU}}

The Argonne National Laboratory CARIBU facility has implemented a decay station to examine decay rates
and spectra of fission fragments extracted from a $^{252}$Cf
source~\cite{mitchell14}.  The two main components of the decay
station
are the X-Array and the SATURN device.  The X-Array is an array of five HPGe clover detectors arranged in a box configuration for
the measurements of primarily low-multiplicity $\gamma$-rays from implanted beams. SATURN is a moving tape system.  
The performance of this system was tested with $\mbox{}^{142,144,146}$Cs $\beta$-decays.  In addition to decay data,
$\gamma$, $\beta\gamma$, and $\beta\gamma\gamma$ coincidences were detected. With this device, the half-life of $^{144}$Cs
was measured to be 1.00(4) s.  Recently, the decay scheme of $^{146}$Ba up to energy levels of ${\sim}$2.5~MeV was also
thoroughly measured with this device~\cite{mitchell16}.  Another
recent measurement with this system has been that 
$^{104}$Nb $\beta$-decay and subsequent $\gamma$
transitions in the $^{104}$Mo daughter~\cite{mitchell16b}.
Measurements of
$^{160}$Eu have recently been reported as well. 

Also, a $\beta$-decay Paul trap, a radiofrequency quadrupole
ion trap, has been developed at Argonne National
laboratory~\cite{scielzo12}.  
This device provides the opportunity to surround a trapped ion cloud with a variety of detectors.  Studies of 
the decay half-lives as well as kinematics of the electron and recoil nucleus provide a means to reconstruct the 
decay spectrum with high precision.  Given the flexibility of this device, it is also possible to measure 
$\beta$-delayed neutron emission
probability~\cite{scielzo14}.  Recent decay measurements with the
ANL Paul
trap include $^{137}$I.  The recoil energy was measured in this proof-of-principle experiment. 
\subsubsection{{NSCL}}

Decay measurements at the NSCL have implemented the NSCL Beta
Counting System (BCS)~\cite{prisciandaro03}.  Combined
with the Neutron Emission Ratio Observer (NERO)~\cite{pereira10}
described in \ref{nero}, decay half-lives as well as 
$\beta$-delayed neutron emission probabilities can be measured.  The BCS is an implantation-type detector for fast
beams.  An upstream PPAC, PIN diodes, and a DSSD provide implantation position and lifetime measurements.  This device 
has been used to measure the half-lives of several nuclei on or near the r-process path 
with 70$\le$A$\le$80 and
90$\le$A$\le$110~\cite{pereira09,pereira11}.  These
include $^{90}$Se, $^{105}$Y, $\mbox{}^{106,107}$Zr, 
$\mbox{}^{107 - 109}$Nb, and $\mbox{}^{108 - 111}$Mo.  These half-lives can be compared to models to gauge the effects of shape
changes, including the onset of triaxiality (in the case of the Nb isotopes). In the case of the neutron-rich Nb isotopes,
the half-lives were similar to the FRLDM model, but lower than the FRDM model by roughly a factor of two.  Results were
similar for the Mo isotopes, with measured rates lower than the FRDM model by nearly a factor of three.  Earlier results
with the BCS include measurements of the half-life of the
doubly-magic $^{78}$Ni nucleus~\cite{hosmer05,pereira08}.
Combining
this system with a $\gamma$-counting system in the same experiment has resulted in spectroscopy measurements of
$^{120}$Pd in an effort to understand proposed neutron shell quenching in the region below the A${\sim}$130 waiting point. 

Further studies with the BCS have included lifetime measurements of the neutron-rich nuclei $\mbox{}^{77-79}$Cu, $\mbox{}^{79,81}$Zn,
and $^{82}$Ga~\cite{hosmer10}.  The rates found have been used
in r-process models to reproduce the 
78$\le$A$\le$80 abundance pattern better than if rates from theoretical models were used.  Half-lives for the
neutron-rich $\mbox{}^{114-115}$Tc, $\mbox{}^{114-118}$Ru, $\mbox{}^{116-121}$Rn, and $\mbox{}^{119-124}$Pd near the proposed r-process path have
also been observed with this method~\cite{montes06}.

The NSCL BCS has also been used to study the decays of
$^{90}$Se~\cite{quinn12}.  The $\beta$-decay half-lives of
this nucleus
have been studied to search for evidence of a subshell at N $=$ 56, which would result in changes in the predicted r-process
abundances, particularly as regards a possible weak r-process.  While this method found no subshell at N $=$ 56, the method
is readily applicable to other similar explorations.  Comparisons of an r-process network calculation using these
experimental results and half-lives determined using a QRPA
calculation~\cite{moller90} have shown little difference
in the final abundance distribution between the two.
\subsubsection{{GSI}}

The GSI silicon implantation beta absorber 
SIMBA detector array~\cite{caballero17} is an implantation decay
station similar to those mentioned above.  It 
consists of two single-sided silicon strip detectors with strips oriented orthogonally and is used for x-y tracking. 
Two SSSDs were placed after these as front absorbers, followed by two double-sided silicon strip detectors used as implantation
layers, and two additional rear absorbers.  This device is used in conjunction with the GSI FRS and two MUSICs and TPC
detectors which provide particle identification.  This can also be used with the beta-delayed neutron (BELEN) system
\cite{gomez14}, which is a $^3$He counter used for
neutron detection to determine $P_n$.

Recently, the decay half-lives and $P_n$ for several neutron-rich nuclei with A$>150$ have been measured at the GSI-SIMBA
facility~\cite{caballero16,caballero17}.  This area is more
difficult to reach experimentally because the production rates of 
the nuclei of interest are lower.  In this work, the half-lives of $\mbox{}^{204-206}$Au, $\mbox{}^{208-211}$Hg, $\mbox{}^{211-216}$Tl,
 $\mbox{}^{215-218}$Pb, and $\mbox{}^{218-220}$Bi were measured, nine of which were measured for the first time.  The values
 of $P_n$ were also measured for $\mbox{}^{210,211}$Hg and $\mbox{}^{211-216}$Tl.  It was found that the values of $P_n$ were comparable to
 or smaller than those predicted by global models.
 
 The GSI experimental program has also incorporated an active stopper with the $\gamma$-ray detection array RISING
~\cite{kumar09,pietri07}.  With this setup, half lives
 for a broad range of nuclei near the doubly-magic
 $^{208}$Pb nucleus have been measured in an effort
 to push decay measurements closer to the third r-process
 peak.  These have been
 either first-time measurements or improvements on previous
 measurements.  Nuclei that have been
 measured include $\mbox{}^{194-196}$Re, $\mbox{}^{199,200}$Os, 
  $\mbox{}^{199-202}$Ir, $\mbox{}^{203,204}$Pt, $\mbox{}^{211-213}$Tl, and 
  $^{219}$Bi 
~\cite{kurtukian09,podolyak09,aldahan12,benzoni12,benzoni14,kurtukian14,morales14}.
  Further experiments with the same setup have examined
  the decay systematics of
   $\mbox{}^{211-213}$Tl, $^{215}$Pb, and $\mbox{}^{215-219}$Bi including
   the $\gamma$-decay states of the daughter nuclei 
~\cite{morales14b}.
   These experiments have been used to examine decay 
   systematics in comparison to those extracted from the
   FRDm+QRPA mass model, which is commonly used
   r-process nucleosynthesis simulations.  Also, a comparison with theoretical models can
  be used to gauge the importance of first-forbidden transitions
in this region~\cite{kurtukian14}.  Near the
   N $=$ 126 shell, shorter-than-expected half-lives can account for shifts in the third r-process peak as well as
   evidence of the significance of first-forbidden transitions in this region.
\subsubsection{{Jyv\"{a}skyl\"{a}}}
The JYFLTRAP system can also be used for the study of $\beta$-decay spectroscopy using the total absorption
technique~\cite{rice17}.  With this technique, direct
measurements of the $\gamma$ and $\beta$ energies of the
daughter products are made or deduced. Measurements of this type
at the IGISOL facility~\cite{aysto01} of
the University of Jyv\"{a}skyl\"{a} were made by first purifying nuclei in the JYFLTRAP Penning trap 
\cite{kolhinen04,eronen12}.  After purification, nuclei are then
implanted into a moving tape, similar to 
techniques mentioned previously.  The tape then transports implanted nuclei to a Si detector system surrounded
by a Ba${\rm F}_2$ segmented $\gamma$ spectrometer.  On the neutron-rich side of stability, the decay energies, 
$\beta$ spectra, subsequent $\gamma$ energies, and branching ratios of $^{86}$Br and $^{91}$Rb have been determined
\cite{rice17}.  While these nuclei are more relevant to reactor
physics, than the r-process as they are not neutron-rich
enough for the r-process, the principle can be applied to r-process nuclei as well.  Also, decay of r-process
progenitors can proceed through these nuclei. 
\subsubsection{Other Facilities}

Similar to the program at Jyv\"{a}skyl\"{a}, the Oak Ridge Holifield facility has performed studies of neutron-rich 
nuclei relevant to reactor physics.  The Low Energy Radioactive
Ion Beam Spectroscopy Station (LeRIBSS)~\cite{beene11} was
equipped with a moving tape system along with two $\beta$ detectors and four HPGE clover detectors for the study of nuclear decay spectroscopy.  Early measurements of the $r$-process nuclei $\mbox{}^{82,83}$Zn and $^{85}$Ga were determined
for the first time~\cite{madurga12}.  These results, combined
with new models, have produced a significant change in the
predicted abundances of the third r-process peak in a network calculation.  

Although much of the work at the LeRIBSS facility has concentrated on neutron-rich nuclei near -- but not on -- the $r$-process path, other results are mentioned here as many of the nuclei measured may provide structure information for
$r$-process post-processing.   Using this facility,
spectroscopy measurements have been made of $^{93}$Br and
$\mbox{}^{93,94}$Kr~\cite{miernik13}.  The literature values of the
decay rates and
branching ratios were in good agreement with the measurements. This system was also used to examine the lighter, but 
very neutron-rich nucleus $^{85}$Ge and its decay
systematics~\cite{korgul17}, providing a partial level scheme of
the 
daughter nucleus with several new transitions.
For nuclei approaching the N $=$ 82 closed shell, this system was used to measure the decay spectroscopy from decays of
$^{124}$Cd and $\mbox{}^{124,126}$Ag~\cite{batchelder14,batchelder16}.
The transport tape collector has also been used to study 
the decay spectroscopy of $^{137}$I and
$^{137}$Xe~\cite{rasco17}.
\subsubsection{Charge Exchange}

Weak interaction rates affect the r-process through
neutrino captures during deleptonization of the 
core~\cite{balasi15} as well as $\beta$ decays during
the post-burst processing~\cite{martin16}.  The former
directly affects the electron fraction $Y_e$ of the
environment while the latter limits how quickly
the r-process moves to the high-mass region of the
isotopic chart.  For much of the $r$-process, weak
interaction rates are dominated by Gamow--Teller 
transitions.  For this reason, significant understanding
of the r-process can be gained by understanding the 
GT-strength functions of r-process progenitor nuclei.

Charge exchange reactions, such as (p,n), ($^{3}$He,t) or ($^7$Be, $^7$Li) reactions and their inverses
have become very useful for mapping the GT-strength functions
of many nuclei, providing a measure of the weak-interaction rates
of many nuclei.  The transitions brought about by these reactions
mimic those produced in $\beta$-decays, including transitions
to excited states in daughter nuclei.  In these reactions, the
reaction cross section can be related to the strength function
for $\beta$ transitions~\cite{bertsch87,osterfeld92}.

An advantage of these types of reactions is that
the heavy beams can be exotic beams at intermediate energies.  Difficulties include the fact that these reactions are generally run in inverse kinematics to accommodate
radioactive ion beams.  As a result obtaining good resolution is difficult.  Many charge
exchange studies on neutron-rich nuclei have been limited to the
lighter nuclei~\cite{shimoura98,takeuchi01,teranishi97}.
However,
this technique has been developed and pursued at multiple facilities,
and it remains an active field of study~\cite{tanihata16}.
\subsection{$\beta$-\lowercase{Delayed Neutron Emission}}

\label{nero}
Closely related to the studies of $\beta$-decay rates are measurement of $\beta$-delayed neutron emission
probabilities.  This is important, as it adds additional neutrons to the r-process site.  While it is
important to measure the $\beta$-delayed neutron emission probability, $P_n$, for nuclei along the r-process path,
post-processing may also be influenced by additional neutrons emitted in the decay process.  The results of these 
measurements are able to provide constraints on fundamental properties of nuclear mass models.

Early measurements of $P_n$ concentrated on neutron-rich nuclei before the N $=$ 82 shell closure and approaching the r-process
path. This region is important not only because of prior studies of ``shell quenching'' in this region, but also
because this is a region of production via fission of heavy r-process progenitors.  The $P_n$ values of 
$\mbox{}^{99,100g,101,102g,103}$Y, $\mbox{}^{104g,104m,104m,105-110}$Nb, and $\mbox{}^{109-112}$Tc were measured and compared to theoretical results.
These measurements were implemented by a simultaneous $\beta$-decay rate measurement and a neutron detection
measurement using $^3$He counters embedded in a polyethylene moderator.  In general, the $P_n$ values measured were
higher than those predicted with QRPA calculations using the
FRDM~\cite{moller95} and ETFSI~\cite{aboussir95} mass models.
The single exception was
$^{110}$Tc.

A significant amount of work to measure $P_n$ has been conducted at the NSCL.  Recent measurements of 
$P_n$ have been performed on $^{104}$Y and
$\mbox{}^{109,110}$Mo~\cite{pereira09,sarriguren10} with upper limits
established for 
$^{105}$Y, $\mbox{}^{103-107}$Zr, and $\mbox{}^{108,111}$Mo.  Half-lives have also been measured in these experiments simultaneously
using the implantation technique described in the previous section.  The values of $P_n$ were measured with the 
NSCL Neutron Emission Ratio Observer (NERO)~\cite{NERO}, an array
of $^3$He and B${\rm F}_3$ gas tubes arranged in a
polyethylene moderator matrix.  The combination of the NSCL BCS and NERO provides the capability to measure half-lives
concurrently with neutron-emission probabilities.  With this measurement, quadrupole deformation parameters could
be determined for nuclei in this region.  Triaxiality of nuclei in this region was also shown to contribute to the
$\beta$-decay properties. 

At GSI, the BELEN detection system~\cite{gomez14} is used.  This
system is similar to NERO in that a polyethylene moderator is
used to contain a detector matrix.  The BELEN system differs in
that it does not utilize B${\rm F}_3$ tubes.  Recent
measurements
with BELEN have measured the $P_n$ values of $\mbox{}^{210,211}$Hg and $\mbox{}^{211-213}$Tl, targeting the region with $N\ge 126$.  This
measurement provided evidence that there is 
no particular model which can be globally applied to the entire r-process path that
satisfactorily predicts $\beta$-decay half-lives and neutron branchings on both sides of the N $=$ 126 shell closure.  
\subsection{Fission Barriers and Distributions}

Fission cycling in the r-process, particularly in sites with very large neutron-to-seed ratios (such as neutron star mergers),
can be responsible for injecting additional neutrons into the r-process, causing
additional neutron captures.  Currently, measurements of fission fragment distributions and fission barriers are sparse for nuclei with
masses above A${\sim}$260.  This creates uncertainties in r-process calculations.

Earlier measurements of fission yields of $^{235}$U have
investigated the yields of $\mbox{}^{124 - 132}$In~\cite{shmid83}
as well as $\mbox{}^{92-99}$Rb, $\mbox{}^{94-100}$Sr, $\mbox{}^{142-148}$Cs, and
$\mbox{}^{143-149}$Ba~\cite{shmid81}.  These types of experiments are
useful for establishing the systematics of fission fragment distributions.  For example, studies of the In isotopic line
have shown that the yields follow a Gaussian distribution, while those of the Rb, Sr, Cs, and Ba isotopes have shown that the 
distributions have a ``wing effect'' in which the decline of fragment yield with mass is less steep than that predicted by
theory.

The IGISOL-JYFLTRAP facility has progressed in the study of the proton-induced and deuteron-induced 
fission of Th and
$^{238}$U~\cite{penttila10,penttila12,lantz16}.  In these
studies, fission yields have been deduced by stopping fission products in the JYFLTRAP device.  Fission products formed in a reaction chamber are separated in a dipole magnet before being cooled in an RFQ buncher and passed into the Penning trap. 
Isobaric chains of more than 20 different elements have been observed with this technique.   The IGISOL facility is following up
these measurements to develop the capability for measuring
neutron-induced fission~\cite{gorelov16}.

Recent techniques have also been employed at
GANIL~\cite{caamano13} and GSI~\cite{geissel92,pellereau17} to
study the fission of $^{238}$U
via transfer reactions with a $^{12}$C target (at GANIL) or via
Coulomb excitation (at GSI~\cite{schmidt00}).  These experimental
setups differ in the energy of the incident U beam, which is
fully relativistic at GSI.   Electromagnetically induced
fission on $^{238}$U is the equivalent of studying neutron-induced fission on $^{237}$U with this technique.
Fission fragments at GSI are identified using multiple sampling ionization chambers (MUSIC) with multi-wire proportional counters.

Elemental and mass distributions have been determined with a high degree of resolution and over
a complete range of products.  A good comparison to the
\texttt{GEF}~\cite{schmidt16} code was found with some
discrepancies related to the odd--even
mass staggering as well as discrepancies in the symmetric valley.

Fission fragment mass distributions have been measured for nuclei
as heavy as $\mbox{}^{256,258}$Fm~\cite{schmidt12}.  A shift in the
systematics is noted for these two nuclei.  While symmetric fission was observed for $^{256}$Fm, a shift to symmetric
fission was observed for $^{258}$Fm.  These systematics are useful for constraining fission models in this region.
\subsection{$\beta$-\lowercase{Delayed Fission}}

The study of $\beta$-delayed fission is difficult experimentally as the branching ratios for this type of reaction are 
relatively rare.  The most recent studies have established
experimental techniques using
$\mbox{}^{178,180}$Tl~\cite{liberati13,elseviers13},
$\mbox{}^{192,194,196}$At~\cite{andreyev13,truesdale16}, $\mbox{}^{200,202}$Fr, and
$\mbox{}^{186,188}$Bi~\cite{andreyev13b}.  
As fission experiments push to heavier, more exotic nuclei, the systematics of
the fragment distribution as well as the fission rates can be better constrained. Recent progress in the development of higher-purity beams and detection techniques synchronized with beam characteristics is enabling fission studies of a larger variety of nuclei.

The $\beta$-delayed fissions of $^{180}$Tl were studied at
ISOLDE with a high-purity beam~\cite{andreyev10}.  Asymmetric
fission fragments from the $^{180}$Hg daughter were observed.  In this experiment, a rotating implantation target 
system was used.  The Tl fragments were implanted into a target surrounded by Si detectors capable of 
detecting electrons, $\alpha$-particles, and fission fragments.  Germanium detectors were also used to detect photons in
coincidence with the fission fragments.   In addition to the fission fragment distribution, the most probable neutrons emitted in 
$^{180}$Hg was found to be one.   The ISOLDE group has also studied the $\beta$-delayed fission of $\mbox{}^{200,202}$Fr, finding 
112 fission fragments from $^{200}$Fr and 6 from
$^{202}$Fr~\cite{andreyev13b}.

At GSI, the $\beta$-delayed fission probabilities and energies for $\mbox{}^{192,194}$At were studied using the recoil-fission
correlation technique~\cite{andreyev13}.  Isotopes produced via
fusion reactions were separated in the SHIP velocity 
filter~\cite{munzenberg79} prior to implantation in a
position-sensitive silicon detector array.  The ensuing
$\alpha$ and fission decays were then measured in this array.
As with the previous experiment, germanium detectors were used to
detect subsequent $\gamma$-decays.  With this technique, the
$\beta$-delayed fission of $\mbox{}^{192,194}$At was identified,
and an upper limit was placed
on the total kinetic energy of the fission fragments. The $\beta$-delayed fission probabilities are then estimated based upon systematic
arguments. 

The same experimental technique at the SHIP facility at GSI has also been used to study $\beta$-delayed fission of $\mbox{}^{186,188}$Bi,
verifying fission with four events in $^{188}$Bi and three events from $^{186}$Bi.  The probabilities of $\beta$-delayed fission
in both cases were determined qualitatively to within a factor of five.

For the neutron-rich nuclei, recent measurements include
$\mbox{}^{228,230}$Ac~\cite{shaunggui01,yangbing06}.  After chemically
separating $^{228}$Ac from the decays of $^{228}$Ra, and using mica fission track
detectors and HPGe detector arrays, the $\beta$-delayed fission probability for $^{228}$Ac was estimated at
(5~$\pm$~2)$\times$10$^{-12}$.  A similar technique was used to extract $^{230}$Ac from $^{230}$Ra decays.  The
$\beta$-delayed fission probability of $^{230}$Ac was tentatively found to be (1.19~$\pm$~0.40)$\times$10$^{-8}$.
Much earlier measurements of the half-lives of $\beta$-delayed fission of neutron-rich nuclei were 
made for $\mbox{}^{256m}$Es~\cite{hall89}, and
$\mbox{}^{234gs,234m,236,238}$Pa~\cite{gangrsky80,baasmay85}.

An excellent, recent compilation of known $\beta$-delayed
fission precursors is found in reference~\cite{andreyev13b}.
\subsection{Neutron {Ca}pture Rates}

Neutron capture cross sections and rates are necessary for 
constraining the r-process path in both speed and location. 
Evaluated rates -- both theoretical and experimental -- are
available on a number of databases, including the
\texttt{REACLIB}~\cite{cyburt10} database and the 
\texttt{KADoNiS}~\cite{dillmann06,rauscher12,szucs12} database.
Known rates can also be useful in evaluating the efficacy of
sites based on their neutron richness.  
\subsubsection{{n\textbackslash TOF}}
Direct measurements of neutron capture cross sections
on r-process progenitors are currently difficult as the
required target is so short-lived.  However,
direct measurements of neutron capture cross-sections
on stable targets or radioactive targets close to 
stability are possible.  This is done
at the CERN n{\textbackslash}TOF facility.  A recent measurement
of the neutron capture cross section on $^{206}$Pb was
performed using the n{\textbackslash}TOF neutron
source~\cite{domingo08}.  While $^{206}$Pb is predominantly
an s-process isotope, a precise measurement of
the neutron capture cross section can provide
a means to subtract the s-process contribution
from solar abundances, thus constraining the r-process
contributions and calculations in the actinide region.

The n{\textbackslash}TOF facility has also been used  
in a similar fashion to determine the r-process
contribution of $^{209}$Bi~\cite{domingo06}. With
a Maxwellian averaged cross section, the r-process contribution to $^{209}$Bi was estimated to be 81~$\pm$~3\%.
\subsubsection{{GSI}-{LAND}}

Coulomb dissociation can also be used to determine neutron capture cross sections.  With this technique, a radioactive ion beam incident on a target is excited above the neutron emission threshold.  This technique requires a neutron detection system.  The LAND setup
at GSI employs this method.  Recently, Coulomb
dissociation of $^{18}$C has been used to determine
the thermally averaged neutron capture cross section
$\sigma_{n\gamma}$ on $^{17}$C~\cite{heine17}. These results were
used to determine that the $^{17}$C(n, $\gamma$) reaction
has no pronounced influence on the second and third peaks of the
r-process in neutrino-driven wind models and in neutron star
mergers.
\subsubsection{Other Facilities and Techniques}

Using the photoactivation technique, a beam of photons is used to excited a target above the neutron threshold.  
As an example, branch point nuclei in the s-process path nuclei can be studied. These are then used to accurately separate contributions to  nuclear abundance distributions from neutron-capture processes.  One particularly interesting experiment was the study of
the $^{187}$Re($\gamma$,n) reaction to  
determine the $^{186}$Re neutron capture cross 
section~\cite{muller06}.
While this nucleus is not an r-process progenitor, 
its capture cross section is important in understanding
the s-process contribution to the r-process
cosmochronometer
$^{187}$Re.

The activation technique has been used to study
the (n, $\gamma$) cross section of the long-lived
isotope $^{182}$Hf~\cite{vockenhuber07}.  Because of
its long half-life, this nucleus can be used as
a target which was illuminated with neutrons at
$kT=25$ keV.  With this technique, the Maxwellian averaged cross sections have been found.

A particularly interesting proposal for the measurement of
neutron capture reactions in inverse kinematics involves the use
of a storage ring with a neutron target~\cite{reifarth14}.  The
storage
ring, such as the GSI ESR~\cite{geissel08} would be capable of
storing an intense beam of short-lived nuclei.  The neutron
target could be implemented by passing the beamline of the
storage ring through a reactor
core.  With this method, assuming a neutron flux, $\phi = 10^{14}$ cm$^{-2}$s$^{-2}$ and a stored particle intensity of 10$^{13}$ s$^{-1}$, a daily reaction rate of 20$\times\sigma$ per day
could be measured, where $\sigma$ is the estimated neutron capture cross section in mb.  Thus, in a single day, cross sections on the order of
mb could feasibly be measured.  The neutron-induced reaction would
be detected via the change in the $B\rho$/q of the beam particles.
\subsection{Nuclear Structure Studies}

Far from stability, in addition to nuclear masses and decay
half-lives, much can be known about the r-process path through
measurements of neutron separation energies and shell structure.
Shifting and appearance of neutron major and minor shells near
the r-process path can result in a shift in predicted neutron
separation energies, and hence, the location of the path itself.
Shape transitions are closely related to this, as deformations
may result in a modification of the presumed shell structure.
\subsubsection{Neutron Separation Energies and Shell Closures}

As the effects of neutron capture rates are most affected at the closed shells, this part of the nuclear chart has seen a significant amount of study in the
past several years.   Direct, accurate measurements
of (n, $\gamma$) cross-sections on r-process nuclei are
currently challenging.  However,
these cross-sections can be determined
indirectly via neutron pickup reactions such as (d,p) reactions
in inverse kinematics~\cite{chiba08,jones11,kozub12}.  In this
case, the shell
structure and the robustness of the N $=$ 82 shell closure for the
Sn nuclei have been determined. 

In-flight fission has made it possible to push the limits of neutron-rich nuclei for the purposes of exploring structure far from stability.  For example, the WASABi detector (mentioned in \ref{RIBF_beta}) was used to determine the $\gamma$ energy spectrum
of $^{126}$Pd and $^{128}$Pd produced via in-flight fission of 
$^{283}$U~\cite{watanabe13}. Photons were detected from nuclei
implanted in the WASABi endstation.  Intensities 
of the Pd $\gamma$-decays were then used to assign spins to nuclear levels.  As with the (d,p) reactions mentioned above, the robustness of
the N $=$ 82 shell closure was confirmed in the Pd isotopic chain.
\subsubsection{Isomers}

Isomers could be a possible observational signature of light
curves resulting from an astrophysical
r-process~\cite{jungclaus07}.  While atomic absorption lines are
generally able to distinguish individual elements, isomeric
transitions can be used to identify isotopes in a potential
r-process site.  Astronomical $\gamma$-ray observations
(described in \ref{gamma_obs}) present the possibility for
isotopic identification of the 
r process occurring in real time~\cite{diehl05}.

In many cases, isomers can be found in conjunction with other measurements.  For example, the TITAN Penning-trap mass spectrometer has found a low-lying 80 keV isomer of the neutron-rich nucleus $^{98}$Rb
while measuring the masses of the Rb and Sr isotopes in the
$A\approx$100 mass region~\cite{klawitter16}.  The EURICA
experiment has also been equipped with  former EUROBALL HPGe
detectors and has been used to find isomers in the neutron-rich
nuclei $\mbox{}^{126,128}$Pd nuclei, providing further evidence of a
robust 
N $=$ 82 shell closure~\cite{watanabe13}.

Former EUROBALL detectors have also been used to find high-spin states of the neutron-rich $^{204}$Pt nucleus at the N $=$ 126 closed shell.  This implantation-type experiment has identified a $55~\rmmu {\rm s}$ high-spin isomer with 
$J^\pi$ $=$ 7$^-$~\cite{steer08}.
While this may be short-lived for observational purposes, the versatility of implantation experiments and simultaneous isomeric searches is evident with this setup.
\subsection{Current and Future Facilities}

With the advent of new radioactive ion (RI) beam facilities
worldwide, the possibility of directly probing the r-process path
is becoming a reality.  Some facilities are already operating,
while others are under construction~\cite{blumenfeld13}.  While
the number of RI beam facilities is numerous, 
we will mention only a representative sample here of recent upgrades and planned construction of
RI beam facilities.
\subsubsection{{RIBF}}

The Radioactive Ion Beam Factory (RIBF) in Saitama, Japan is an 
in-flight RI beam separator~\cite{yano07}.  It is a coupled
cyclotron
facility capable of producing heavy primary beams with 
energies up to 350 MeV/A.  Exotic beams are produced via
fragmentation and separated in the BigRIPS  fragment separator.  The
facility also has the capability of producing fission products
via in-flight fission.
\subsubsection{{FRIB}}

The Facility for Rare Isotope Beams (FRIB), located at
the site of the NSCL in Michigan, USA, is currently undergoing
construction with a target completion date of 2020. It consists of a superconducting
linear accelerator capable of primary beam energies of up to
200 MeV/A for uranium. This facility will be coupled to a 
reaccelator for studies with low-energy ($<$12 MeV/nucleon) 
secondary beams.  The low-energy facility currently exists and
is being used coupled to the existing facility.  
\subsubsection{{RAON}}

The RAON facility in Korea is currently under construction.  It will contain a variety
of instruments, including KOBRA (KOrea Broad acceptance Recoil spectrometer and Apparatus), LAMPS (Large Acceptance Multi-Purpose
Spectrometer), and ZDS (Zero Degree Spectrometer)~\cite{RAON}.
Each of these devices 
can be used for a variety of experiments with ranges of acceptance, resolution, and 
energy.  This facility is predicted to be capable of primary beams of 
600~MeV protons at 600 $\rmmu$A and 200 MeV/A uranium beams at 8.3 $\rmmu$A.   
\subsubsection{{GSI}-{FAIR}}

When complete, the Facility for Antiproton and Ion Research
(FAIR)~\cite{FAIR} at GSI will be one of the largest research
projects worldwide.  Full operations are planned for 2025
following
a planned commission in 2022.  This facility will accommodate several experimental programs
including QCD studies, decay spectroscopy, production of exotic nuclei, isomeric studies, trap experiments, and a host of other experimental programs relevant to nuclear astrophysics.  

The FAIR facility is expected to provide uranium beams with an 
intensity of up to 10$^{12}$ ions/s with a high energy of 1.5 GeV/A.  Intense secondary beams can be produced from this.

 \section{{Ca}ndidate Astrophysical Environments}

 The unambiguous identification of a single dominant site for  $r$-process  nucleosynthesis (if there is one) has remained elusive. Often the field has  been convinced of its certain identification only to be later disappointed.  
The widespread hopes, after the observation of a universal characteristic  pattern of elemental abundances for elements heavier than iron have not yet been fulfilled.
There is currently a wellspring of support for binary neutron
star mergers  as the primary site for the $r$-process
based upon a number the observations such as  the GRB170817
kilonova~\cite{Cowperth17,Chornock17,Nicholl17,Smartt17,Pian17,Troja17,abbott17,Villar17}
associated with  gravitational waves from GW170817, and the
apparent rare event deduced to have occurred in the
Reticulum-\textit{II} dwarf
galaxy~\cite{roederer16,ji18,Frebel18}.  

The $r$ process is best characterized as a sequence of rapid neutron captures. 
It requires both an explosive environment and neutron-rich matter at high densities, appear plausible ingredients to guarantee such $r$ process conditions. 

There are a number of viable astrophysical environments that can achieve this.
These include: shock ejected material from core collapse supernovae, neutrino-driven winds in supernovae, shock-induced explosive helium burning in supernovae, magnetic-turbulence driven ejecta in magneto-hydrodynamic jets from supernovae and collapsars, accretion disks of neutron stars or black holes, tidal or neutrino driven ejection from the mergers of two neutron stars or a neutron-star plus black-hole binary, or even neutron-rich regions in inhomogeneous big-bang models.  

Over the past half century some of these models have been ruled out as the physics of the environments have been better elucidated.  
On the other hand, core-collapse supernovae and neutron star mergers (NSMS) have remained, the two most  favored  candidates.  
Many researchers are now even convinced that NSMs are the
dominant site for the $r$-process~\cite{Thielemann17}. 
As noted previously, observations~\cite{sn08} showing the arrival
of heavy-element $r$-process abundances in very
low-metallicity stars would favor the short stellar lifetime of
core-collapse supernovae (CCSNe) as the $r$-process
site.  However, identifying the occurrence of an
$r$-process environment and its location within models
of CCSNe has been difficult~\citep{arn07,Thielemann11}.  

In order to prepare a proper ground for a review and discussion this search for an explanation of the observed $r$ process characteristics through such astrophysical models, we now summarize the basic astrophysical considerations common to all such models. 
Thereafter we devote a section to the observational constraints in more detail. 
Thus, the section that follows will take up the above model list, and deepen the discussion of each of the candidate models.

\subsection{The Basics}

\label{basics}

The $r$ process involves  a sequence of rapid neutron
captures in an explosive
environment~\citep{Burbidge:1957,Mathews85}.   Although many
sites have been proposed for the $r$-process, whatever
the environment, it can be shown that  the Solar-System
$r$-process abundances are for the most part well
reproduced by simple beta-decay flow in a system that is in
approximate $(n,\gamma) \leftrightarrows (\gamma,n)$ equilibrium.

Under these conditions, the relative abundances of isotopes of a given element are simply determined by the equations of  nuclear statistical equilibrium (NSE).
\begin{equation}
\frac{n(Z,A) }{n(Z,A+1)} =  \frac{1}{n_n}  \biggl(\frac{2 \pi \mu kT}{ h^2}\biggr)^{3/2} 
\times \frac{G_{A}G_{n}}{G_{A+1} } e^{-Q_n/kT}~~,
\label{rproc}
\end{equation}
where $\mu$ is the reduced mass of the neutron plus
isotope $^{A}Z$, $h$ is Planck's constant,
$k$ is the Boltzmann constant, and $T$ is the
temperature.  The quantity $G_A$ is the partition function
for nucleus, $^{A}Z$,  $Q_n$ is the neutron capture
$Q$ value for isotope $^{A}Z$ (or equivalently the
neutron separation energy for the nucleus $^{A+1}Z$),  and
$n(Z,A)$ represents the number density of an isotope
$^{A}Z$.   Note, however, that this formula neglects a small
correction~\cite{Mathews11} for the difference between Maxwellian
and Planckian distribution functions for the photons.

\eqref{rproc} defines a sharp peak in abundances for one (or a few) isotopes within an isotopic chain.   The flow of beta decays along these peak isotopes is then known as the $r$-process path.  

The location of the $r$-process path peak is roughly
identified~\cite{Burbidge:1957} by the condition that neutron
capture ceases to be efficient once  $n(Z,A+1)/n(ZA)^<_\sim 1$.  
Taking the logarithm of Eq.~\ref{rproc} and inserting the numerical terms, the $r$-process path can be identified by the following relation
\begin{equation}
\biggl(\frac{Q_n}{kT}\biggr)_{\mbox{\footnotesize path}} = 
  2.30 \biggl(35.68 + \frac{3}{2}\log{(\frac{kT}{\mbox{\footnotesize MeV}})} 
  -  \log{(\frac{n_n}{\mbox{\footnotesize cm}^{-3}})} \biggr)~~.
 \label{path} 
\end{equation}
The  elemental abundances $n(Z,A)$  along this path are then determined by the flow of beta decays,
\begin{equation}
\frac{dn(Z,A)}{dt} = \lambda_{Z-1} n(Z,A-1) - \lambda_Z n(Z,A)~~,
\end{equation}
where the total beta decay rate of each element along the path is given by the weighted sum of  beta decay rates for each isotope $\lambda_\beta(Z,A) = 1/\tau_\beta(Z,A)$:
\begin{equation}
\lambda_Z = \sum_A n(Z,A) \lambda_\beta(Z,A)~~.
\label{betaflow}
\end{equation}
For a typical $r$-process temperature of $T_9\sim 1 $, the requirement that the $r$-process path reproduces the observed abundance peaks at $A = 80, ~130$ and $195$, implies that the $r$-process path halts at waiting points in the beta flow near the neutron closed-shell nuclei $^{80}$Zn, $^{130}$Cd and  $^{195}$Tm.  For a neutron density sufficiently high ($n_n \ ^>_\sim 10^{20}$ cm$^{-3}$) so that the neutron capture rates exceed the beta-decay rates for these isotopes, the peak abundances along the $r$-process path must be for isotopes with $Q_n \sim 1\mbox{--}3~\mathrm{MeV}$, and thus $(Q_n/kT)_{path} \sim 10\mbox{--}30$.  
 
 This constraint on $Q_n$, however, concerns the conditions near \emph{freeze-out}, when the final neutrons are exhausted at the end of the $r$-process.   At this point, the system falls out of NSE and nuclei along the $r$-process path decay back to the line of stable isotopes.  

 Earlier in the $r$ process the neutron densities can be
quite high and the $r$-process path shifted to more
neutron-rich nuclei.    For example, in the neutrino driven wind
(NDW) models of~\cite{woosley94}, the $r$-process
conditions begin with a neutron density  of $n_n \approx 10^{27}$
cm$^{-3}$ and a temperature of $T_9 \sim 2$.  The density
is also much higher ($> 10^{32}$ cm$^{-3}$) when the
material is first ejected from the proto-neutron star.   Such
conditions can also be achieved for an $r$ process which
occurs during
NSMs~\cite{Freib99a,Freib99,Rosswog99,Rosswog00,Korobkin12,Piran13,Rosswog13}.  
 
 Of course, as the $r$ process freezes out, one must make a detailed accounting of the full $r$-process reaction network, i.e.
\begin{eqnarray}
\frac{dn(Z,A)}{dt} &=& n(Z,A-1)\phi_n \sigma_{n,\gamma}(Z,A-1) \nonumber \\
 &&+   n(Z,A+1) \phi_\gamma \sigma_{\gamma,n}(Z,A+1) \nonumber \\
 &&+   n(Z-1,A) \lambda_{\beta}(Z-1,A) \nonumber \\
 &&+  {\rm terms}~{\rm with}~ (n,p), (n,\alpha), (p,\gamma), (\alpha,\gamma), \nonumber \\
 &&+  (n,{\rm fission}), (\beta,n), (\beta, {\rm fission}),~~{\rm etc.} \nonumber \\
&&- n(Z,A)[\phi_n \sigma_{n,\gamma}(Z,A) + \lambda_\beta(Z,A) \nonumber \\
&&+ \phi_\gamma \sigma_{\gamma,n}(Z,A) + \cdots ]
\end{eqnarray}
 where $ \phi_n$ and $\phi_\gamma$ are the time-dependent neutron and photon fluxes, respectively.  
 
 A number of network codes exist in the literature.  For more details about the implementation
 of such codes see, for example, reference~\cite{meyer12}.

 \section{Astronomical Observations}

\label{observations}
\subsection{Messengers of the $r$~Process}
   Cosmic nucleosynthesis processes including the $r$~process are recognized from their characteristic abundance patterns of atomic nuclei:  Nuclear reactions rearrange the nucleons of the original composition of nuclei in ways that characterize the nuclear burning conditions, i.e.~the \emph{process} and \emph{site} of nucleosynthesis. Therefore, the ideal messenger of the $r$-process would be the observation of the abundance pattern of isotopes from an $r$-process event.

However,  such nucleosynthesis events are typically launched from a hot and dense environment, which is optically thick to all types of radiation, including characteristic nuclear or atomic lines. Only the explosive dilution makes us register messengers from the origin of the nucleosynthesis process.
Neutrinos may escape from dense nucleosynthesis sites, and have
been direct messengers for nucleosynthesis in our
Sun~\citep{Haxton13} and for SN1987A~\citep{Arnett89}. 
Next to this, $\gamma$~rays from the radioactive decay of
short-lives isotopes created in the nucleosynthesis event may
provide a rather direct, isotopic, and characteristic
messenger~\citep{Qian:1999}. 
These have only been measured for few supernova explosions and
their remnants, for SN1987A~\citep{Matz:1988}, Cas
A~\citep{Iyudin:1994,Grefenstette:2014,Siegert:2015}, and
recently for the first time for a supernova of type Ia with
SN2014J~\citep{Diehl:2014,Diehl:2015}, for isotopes of iron-peak
nuclei.

Being a neutron capture process, the characteristic signatures for the $r$-process are expected to show up in the nuclei well beyond the iron peak. 
The abundance of such heavy isotopes is generally low compared to lighter elements of the iron peak or below.
Additionally, the short radioactive decay times of neutron-rich heavy isotopes disfavor the direct observation of their decay, due to a substantial envelope absorbing radiation from the inner nucleosynthesis products of the core-collapse supernova of a massive star.  For supernovae of type Ia with less of an overlying envelope, conditions for an efficient $r$~process are not generated, and the predominant ejecta composition peaks at around iron nuclei based upon the conditions of nuclear statistical equilibrium (NSE) realized in the SN~Ia environment.

Merging neutron stars, however,  could provide more favorable conditions for the direct observation of nuclear decays, as neutron-capture nucleosynthesis may occur near the surface of the event.  This material is only occulted by matter liberated from the merging compact stars in the event itself.
Nevertheless, limitations of instrumental sensitivities require nearby $r$-process events for such direct messengers ($\leq$3~Mpc), and have not been successfully recorded.

Next to this, the \emph{afterglow} of a nucleosynthesis event provides information about the event and its nucleosynthesis processes, as the radioactive energy of newly created isotopes is absorbed in an envelope, and converted through scattering and absorption processes into radiation at lower energy, where observing capabilities are well developed.
This is abundantly exploited in the case of supernovae, and has
recently also been realized  for the case of an
$r$-process event through observation of GW170817 in a
broad range of electromagnetic radiation, following the trigger
from a characteristic combination of gravitational-wave and
$\gamma$-ray burst signatures~\citep{abbott17}.

For supernovae, the afterglow can be traced from the photospheric until the nebular phases.   That is, optically-thin phases of the supernova spectra can be observed for months after the event to obtain a kind of tomography.
For $r$-process events, both the amount of the envelope and radioactive material may be smaller, and typical radioactivity may be more short-lived.  Hence, the observing window is much shorter, 
consisting of measurements for a few days only. 
Nevertheless, the GW170817 kilonova has provided us with a unique opportunity, which is reminiscent the way in which  observations of  SN1987A  brought insight into the astrophysics of core-collapse supernovae.

The main body of astronomical observations that have established the study of $r$-process nucleosynthesis is based upon atomic-line spectroscopy of starlight from metal-poor stars which are enhanced in their heavy-element abundances. 
Since Fraunhofer's discovery of characteristic absorption lines in the solar spectrum, the astronomical window between about 350 and 700~nm has served as the main tool for cosmic elemental abundances. 
With a spectral resolution of $\lambda/\delta\lambda \gtrsim$50,000, atomic lines can be discriminated for their detailed analysis. 
We note that these are elemental rather than isotopic abundances. 
Isotopic line offsets are mostly smaller than the line widths that result from thermal and turbulent motions (few km~s$^{-1}$) in these stellar atmospheres.
From optical to infrared spectroscopy, rarely is spectral resolution sufficient to disentangle isotopic signatures of absorption lines from heavy element species.
However, Isotopic ratios of an element (e.g.~Eu)
can be estimated from the profile details of photospheric spectral
lines.  This is because as the hyperfine splitting from the total
nuclear mass and nuclear spin is sufficiently large so as to
shift the line position enough  that the line shape distortion
can be measured \citep[e.g.~][]{sneden02,aoki03}.

The integrated area of an absorption line then provides a measurement of the abundance of the absorbing atomic (or molecular) species in the photosphere of the star.
Such absorption lines from heavy elements have been measured in metal-poor stars (i.e.~where the abundance of iron was much lower than in typically-observed stars).  In particular, there is a small subset of these stars (${\sim}$few percent) which show substantial enhancements in heavy-element signatures.
Stars of low mass (one M$_\odot$ and below) evolve rather slowly.  Their evolution time is comparable to or exceeds the age of the universe. Their structure changes very slowly if at all, and their present day photospheric composition for the most part still reflects the gas from which they were formed. 
If the photosphere then shows a low content in heavy elements in general, as most-easily traced through absorption lines of neutral and ionized iron, then we call such stars \emph{metal-poor}. 
The iron abundance in stellar-photosphere data has been found to
have a dynamic range of eight orders of magnitude, and the
currently-observed record lowest iron abundance is 10$^{-7.8}$
times that of our Sun~\citep{Keller14}. 

We call it \emph{Galactic archeology}, when stars across a range of metallicities are analyzed for their signatures from nucleosynthesis before their original formation: The material from which such a star has been formed is the object that is observed.
A difficulty in the spectroscopic measurement of neutron-capture elements, however,  is the weakness of the  spectral lines of these elements and the severe contamination due to  features of other elements (e.g.,~Fe, Ti, molecules). 
This difficulty is reduced in the analysis of metal-poor objects having an excess of  elements attributed to the $r$-process such as Eu. 
As a result, many spectral lines of neutron-capture elements that are not measured even for the solar spectrum are used in the analysis of Eu-enhanced metal-poor stars.

These observations have then shown that the relative abundances among heavy elements, that is, the abundance pattern, appears to be very similar to the $r$-process abundance pattern observed for our Sun, after the $s$-process contributions to solar abundances have been subtracted.
This pattern remains the same, even though metallicity (as traced through the iron abundance from its well-measured absorption lines) changes over orders of magnitude.  
This  is a strong indication of universality of the underlying
nucleosynthesis~\citep{Cowan95}, and has been the main stimulus
for searching the origins of this universal $r$-process
nucleosynthesis. 
We discuss these results in more detail in Sec. \ref{metal_poor}.

The elemental abundance pattern of the $r$~process characteristically shows several peaks, related to the neutron magic numbers piling up abundances along the reaction path of neutron captures and $\beta$ decays that characterize the $r$-process:
 $N=50$ leads to a peak in the element range $Z\sim$35 with elemental masses around 80, i.e.~$^{80}$Se, characteristic elements being Sr, Y, and Zr that can be measured in stellar photospheres.
 $N=82$ leads to a peak in the element range $Z\sim$54 with elemental masses around 130, i.e.,~$^{130}$Te, characteristic elements being Xe, Te, and I; as these are difficult to observe in stellar spectra, Ba and La lines and their abundances are often taken to characterize this second $r$-process peak.

There is an intermediate, minor, abundance peak in the elemental mass range of mass numbers 150--170, the rare elements, characterized by La and the lanthanides.
The magic number at $N=128$ leads to a peak in the element range $Z\sim$78 with elemental masses around 195.  This is the third $r$-process peak at $^{195}$Pt. Characteristic elements near this peak
that can be observed through spectral lines are Os and Ir. Beyond the third peak, the actinides U and Th have  also been measured in stellar photospheres. The entire range of elemental abundances between barium and europium ($Z=56\mbox{--}68$) was considered to be characteristically different from the abundance pattern of elements as expected from the $s$-process.  Hence this range is also often used as indicative of the $r$-process.
 In terms of easy observational access, Eu has a prominent role, with its two main stable isotopes $^{151}$Eu and $^{153}$Eu.   Both are unshielded by other stable nuclei against $\beta$~decay from the $r$-process path.
 
The difference with respect to nucleosynthesis from the $s$-process is an observational necessity to analyze nucleosynthesis contributions from the $r$-process.
Therefore, it may sometimes be difficult to draw significant conclusions on the $r$-process.  For example, Pb abundances are considered to be 80\% due to the $s$-process.  
In general, however, roughly equal portions of the elements heavier than iron are considered to originate from either the $s$- or $r$-process.  Hence, for many elements the abundance pattern of the $r$-process is a significant constraint. 

Photospheric abundances inferred from absorption line spectroscopy are subject to some systematic uncertainty.  This is  the necessity to model the thermodynamic conditions under which the atomic lines are excited. 
In order to reduce such systematics, one often resorts to similar types of stars. That is, one considers similar spectroscopic types of stars, among which such systematics should be small. 
Also, selecting field stars is helpful.  The hope is  to find a
typical gas composition to the star that is being observed.  In
the extreme case, \emph{solar twins} have been used for precision
spectroscopic abundance determinations, reaching a level of
0.01~dex~\citep{Jofre17}.  However this is done  at the cost of a
small sample size.

The formation environment of stars might vary, as they could either be forming from well-mixed, or from less representative interstellar gas.  The latter case would then not trace the earlier nucleosynthesis history and be biased, either from a nearby group of young stars and their ejecta, or from a nearby supernova event, or from a companion star transferring mass.  This is in particular evident in stars in the Galactic halo, in which the impact of an individual event is more significant due to the overall-low metallicity, and presumably slower compositional evolution of the halo gas.

In the disk and bulge, stellar activity and feedback is large.  Thus, a mixture of sources and processes may contribute to a diverse imprint on the gas composition out of which the next-generation stars have been formed. These are still alive today so that their photospheres provide a messenger of the $r$-process. 
Recently, observations from $^{26}$Al radioactivity (decay time
$\tau\sim$~My) in the Galaxy have indicated that massive-star
and supernova debris preferentially are ejected into large
superbubbles, as these stars are formed in
clusters~\citep{Kretschmer:2013}. It may thus take tens of My or
more for such ejecta to cool and be recycled into the
next-generation stars. 
It is unclear how efficient the ejecta from rare and non-clustered nucleosynthesis sites may mix and merge into star-forming material.

\subsection{The Solar Abundance Reference}
\begin{figure}
\caption{Solar system isotopic abundance pattern. From~\protect\cite{arlandini99}.
\label{r-pro-abun}}
\includegraphics[width=\textwidth]{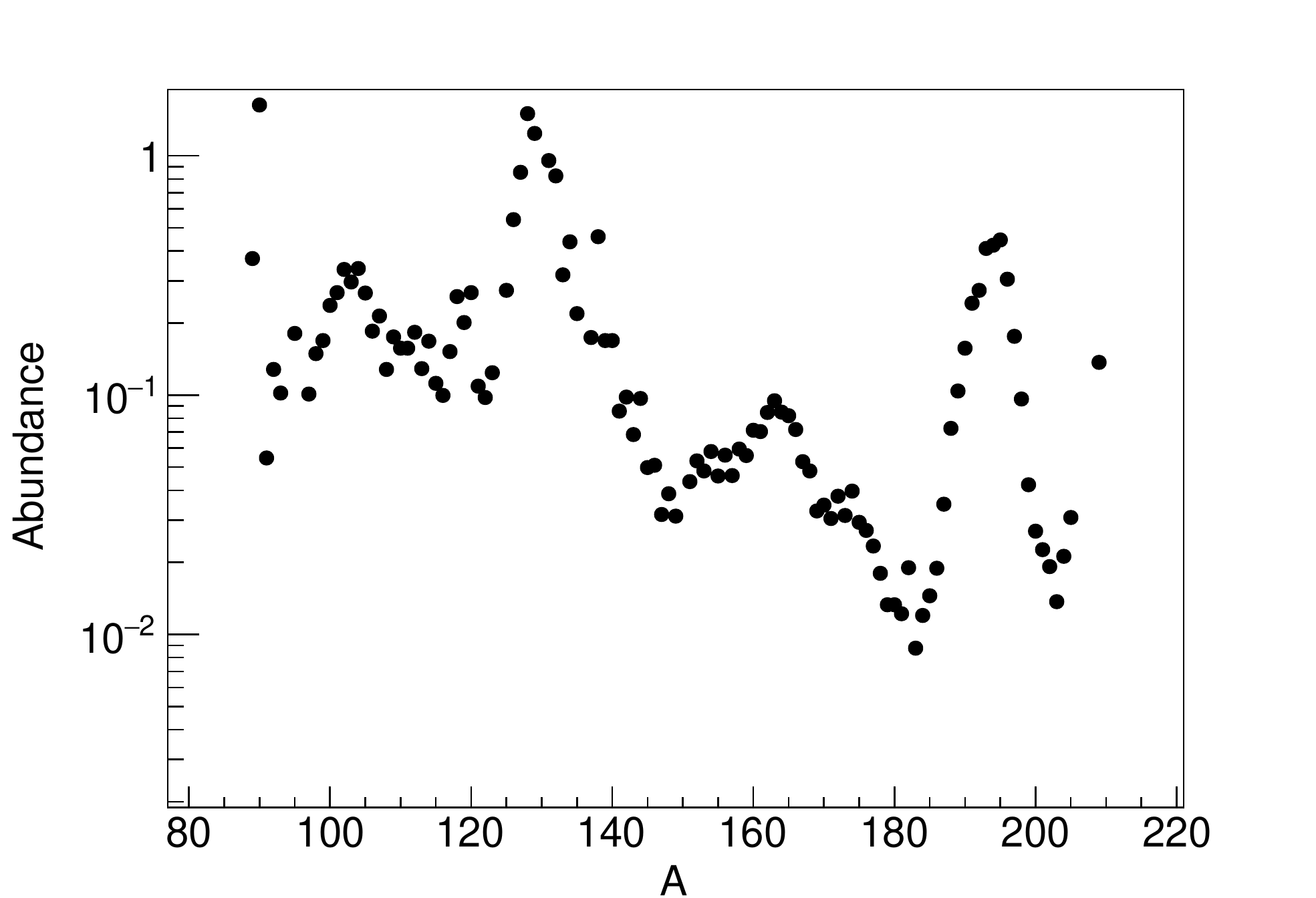}
\end{figure}

  \ref{r-pro-abun} shows an example of the isotopic abundance
distribution of solar-system $r$-process material
obtained~\cite{arlandini99} after subtraction of the
$s$-process contribution.  Newer versions of this
decomposition can be found in  the literature (e.g.~\cite{sn08}).
This distribution is, perhaps, 
the most important reference with which to compare the predictions of $r$-process
models (e.g.,~\cite{wanajo06};~\cite{goriely11}). 

Heavy elements with mass number ($A$) larger than
${\sim} 70$ originate from neutron-capture reaction sequences from both the $s$- and $r$-processes. 
The contributions of the two processes are apparent
from the double abundance peaks corresponding to the
neutron magic numbers (50, 82, and 126). These appear at slightly different mass numbers due to the different reaction paths: The $r$-process reaches magic number nuclei during a neutron irradiation burst.  Thereafter,  $\beta$~decays  shift the abundance peaks toward lower mass numbers, while the $s$-process path directly produces abundance peaks at magic-number nuclei.

Since the $s$-process yields are basically determined by
reactions along the region of stable nuclei, modeling of the $s$ process is less uncertain and better established than that of the $r$-process. 
Because the nuclear properties of neutron capture and binding are the drivers of the $s$-process, and irradiating nuclei with neutrons at such a low fluence leaves sufficient time for intermediate $\beta$~decays,  little detail needs to be assumed about the astrophysical reaction environment. 
These model calculations then can be calibrated, or normalized, to observational data with great reliability.  This is particularly true because  a number of stable isotopes away from the valley of stability and toward the neutron rich side provide shielding against the $r$-process path and its contributions.  Thus one can fit  the parameters of the model to
best represent this set of about 35 isotopic abundances of
$s$-only nuclei. \citep[see discussion of different
methods by~][]{arn07}.

This enables an estimation of the
fraction of abundances produced by the $s$-process for other nuclei
\cite{kaeppeler89}. 
The $r$-process component of each isotope is then determined
by subtracting the $s$-process component from the abundance of each
isotope for solar-system material.
The contribution of the $s$-process to solar abundances
of elements heavier than iron has been determined to be about two
third overall, with specific element contributions of 85\% for
barium (a typical $s$~element) and 6\% for Europium (a
typical $r$~element)~\citep{arlandini99,Bisterzo14}.
Nevertheless, this derivation of the $r$-process abundances is not without problems: Strictly speaking, those abundances are just the ``not from the $s$-process'' abundances, and could be a combination of nucleosynthesis processes (and sites).

The uncertainty of the $s$-process models affects the accuracy of
the estimates of the $r$-process abundance patterns of solar-system
material. This is particularly significant, or even dominant, in the case of the nuclei in which $s$-process
contributions dominate.  This is because a small error in the estimate of the
$s$-process component can result in a large error in the obtained
$r$-process fraction. One example is the element Pb ($s$-process third peak).   For this element the $s$-process
component should be dominant in solar-system
material.  On the other hand, a pure origin from the $r$-process cannot be excluded. 
The production of Pb by the $s$-process is dependent on models and is not well estimated from other elements. 
This makes it difficult to constrain the models for
production of heaviest stable nuclei by the
$r$-process~\citep{arlandini99,arn07},
due to the absence of sufficient $s$-only nuclei for model calibration in this region.
Also, tracing the uncertainty of the ``$r$-process abundances'', one must properly add the uncertainty of the primary isotopic abundance to the uncertainty of the model fit, plus the uncertainty of the neutron capture cross sections that are inherent in the model.

Solar abundances are determined from photospheric absorption line
spectroscopy~\citep{Asplund05,Asplund09} and from meteoritic
sample analyses~\citep{Lodders03}.
An essential contribution from the analysis of meteorites is that it can determine isotopic abundances and ratios. 
This is the basis of the above constraints on the models
for the $s$-process. 
Recent assessments of the two methods exhibit fairly good agreement
for heavy neutron-capture elements in general~\cite{Asplund09}.
In
particular, the updates of atomic-line data, including the
effects of hyperfine splitting and isotope shift, contributed significantly to the improvements of the estimates of the solar 
$r$~abundances~\citep[e.g.~][]{Lawler:2008}.

\subsection{Observing an $r$-Process Event}
  \subsubsection{Gamma Rays from Radioactivities}

\label{gamma_obs}
Characteristic $\gamma$ rays from radioactive decays could be the most-direct measurement of decay, and hence of freshly-injected nuclei, from a source.
The line energy characterizes the isotope, and  the intensity of
the decay marks and supports the assignment to a particular
isotope~\citep{Qian:1999}. 

Most likely, $\gamma$ ray emission will be dominated from the decay of isotopes that are close to the valley of stability, rather than the ones close to the $r$-process path. 
This is because isotopes along the $r$-process path  are expected to be very short-lived.
The nucleosynthesis during the $r$-process is derived
from nuclear-reaction network calculations. Measurements in
laboratories can be made of the level schemes that might be
populated along the $\beta$~decay sequence as it approaches
its end near or at stability of the daughter isotope. This
provides estimates of the characteristic $\gamma$-ray
spectra~\citep{barnes16}. 

Knowing the line energy, spectrometers with adequate resolution are able to set much more direct constraints on the composition of material created in the event.   This is better than 
 absorption line spectroscopy from atomic lines resulting from the transport of $\gamma$-rays of the radioactivity from thermalized photospheric gas.
Sensitivities of current $\gamma$ ray spectrometer instruments are about 10$^{-6}$~ph~cm$^{-2}$~s$^{-1}$ for observing times of 10$^6$s. This provides a window of opportunity only for radioactive isotopes with decay times below several weeks, and for sources ejecting 10$^{-3}$M$_\odot$ of such material at about 5~Mpc distance. 
The $r$-process events are plausibly rare.   Therefore, no such gamma-ray signal has been seen.  The chance to detect  a nearby event would be low at the required at rates below a few Mpc$^{-3}$ yr$^{-1}$.

\subsubsection{Kilonova Emission and {GW}170817}

\label{GW170817_kilonova}
\begin{figure}%f7

\caption{The afterglow of the neutron star merger event GW170817 shows a characteristic decline from radioactive energy deposition, re-radiated by envelope material in optical and infrared emission. The spectrum peaks at around $1~\rmmu {\rm m}$ wavelength, which is thought to reflect a composition rich in heavy elements. Suggestive features may indicate the presence of Cs and Te, although the line signatures of heavy elements are quite uncertain. From~\protect\citep{Smartt17}.\label{fig_kilonova}}
\includegraphics[width=\textwidth]{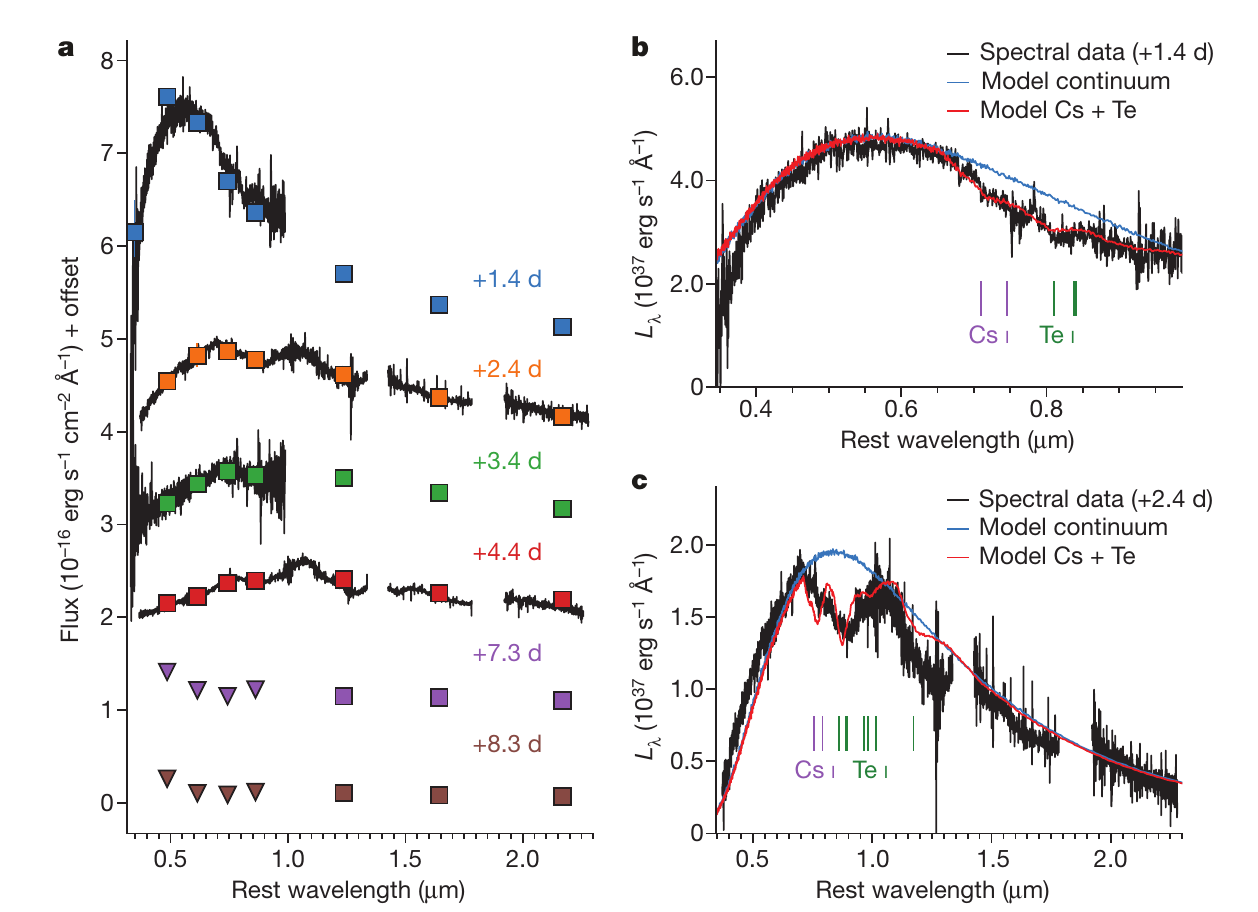}
\end{figure}

  The key to a successful broad observational dataset for a single source event was the GRB trigger detected by Fermi and the followup search for the characteristic and unique chirp signal of gravitational wave emission, as it emerged from the collision of two neutron stars. 
On August 2017, the LIGO and VIRGO gravitational-wave facilities
reported such a signal~\citep{abbott17}, which led to a wide
range of follow-up observations. 
The coincident detection of a $\gamma$-ray burst signal of
the `short' type by the Fermi-GBM and INTEGRAL/SPI gamma-ray
instruments~\citep{Goldstein:2017,Savchenko:2017a} provided more
clues as  to the nature of the event.  These  supported the
earlier hypothesis that sGRBs  originated in neutron star
mergers. 
Several hours were necessary to digest the significance of those two event triggers and their apparent correlated nature.  Then a fleet of optical telescopes and other facilities were ready to follow up on the event (as it had been daytime in Chile when the event occurred).
An electromagnetic signal attributed to the event was detected
with GROND and other telescopes and attributed to the same
event~\citep{smartt:2017}; this signal is called a
\emph{kilonova}, from its brightness ranging between that of
novae and supernovae. The distance to the event then was
estimated to be 40~Mpc, from the association of the kilonova with
the elliptical galaxy NGC4993. 
The light curve and spectra obtained from these observations provided a hallmark for the study of neutron star mergers, as well as for the study of the origins of the $r$-process. 
They showed a characteristic decay of the kilonova within a few days, consistent with radioactivity as the source of energy heating the kilonova (see \ref{fig_kilonova}). 

The spectrum peaked in the infrared regime around 1--$2~\rmmu {\rm m}$, which hints at envelope material rich in heavy elements.
With the spectrum, broad features can also be recognized.  These are reminiscent of the lines of heavy elements Doppler-broadened by ejection velocities of order 30\% of the speed of light.
Note, however, that the heavy isotopes presumably created in the
neutron star merger event have atomic line signatures which are
poorly known, with a large number of lines from the typically
90--110 atomic-shell electrons and their orbital
variety~\citep[e.g.~][]{kasen17}.
 Therefore, improvements of radiation transport models as well as atomic physics knowledge of heavy elements are required, before such associations can be claimed.

Nevertheless, the observations of the overall spectral energy
distribution appear plausibly consistent with expectations from
the kilonova model~\citep{kasen17}. 
Upon more detailed comparison to models, it turns out that the afterglow is more persistent and bright than predicted in models, particularly in the red part of the spectrum.
This might be due to a relatively-large aspect angle, compared to
the jet axis.   Such a jet  is also expected to represent the
main direction of bright gamma-ray emission from the associated
short~GRB~\citep[see~][for a detailed discussion on the GRB
emission and its puzzles]{Burgess:2017}.
At large aspects, the ejecta from a neutron star merger are a
mixture of dynamical ejecta as the neutron stars collide, disk
wind ejecta as the black hole forms, and the unbound torus with
nucleosynthesis enrichments~\citep[see~][for phases of neutron
star merging and material ejection]{Rosswog:2018,Metzger:2017}.

Clearly, more such events would need to be studied from observations and models, to disentangle the aspects of different ejecta and nucleosynthesis, as they appear from different viewing angles.
Nevertheless, as in the case of supernovae, much can be learned
from afterglow spectroscopy.   Supernova studies of radiation
deposits and re-radiation have paved the path toward these
studies~\citep[see~][for a recent model]{Kerzendorf14}.

\subsection{Metal-poor Stars in our Galaxy}
\label{metal_poor}
\begin{figure}
\caption{Abundances of Ca and Eu as a function of Fe abundance in Milky Way field stars. Abundance
 data are taken from the SAGA database~\cite{suda17}. Dashed and
dotted lines indicate [X/Fe] $=$ 0 and 1, respectively, for X $=$ Ca and
Eu. Dozens of stars with [Fe/H]$\sim -3$ show enhancement of
Eu ([Eu/Fe]$>+1$)\label{fig_Mg-Eu-abundances}}
\includegraphics[width=\textwidth]{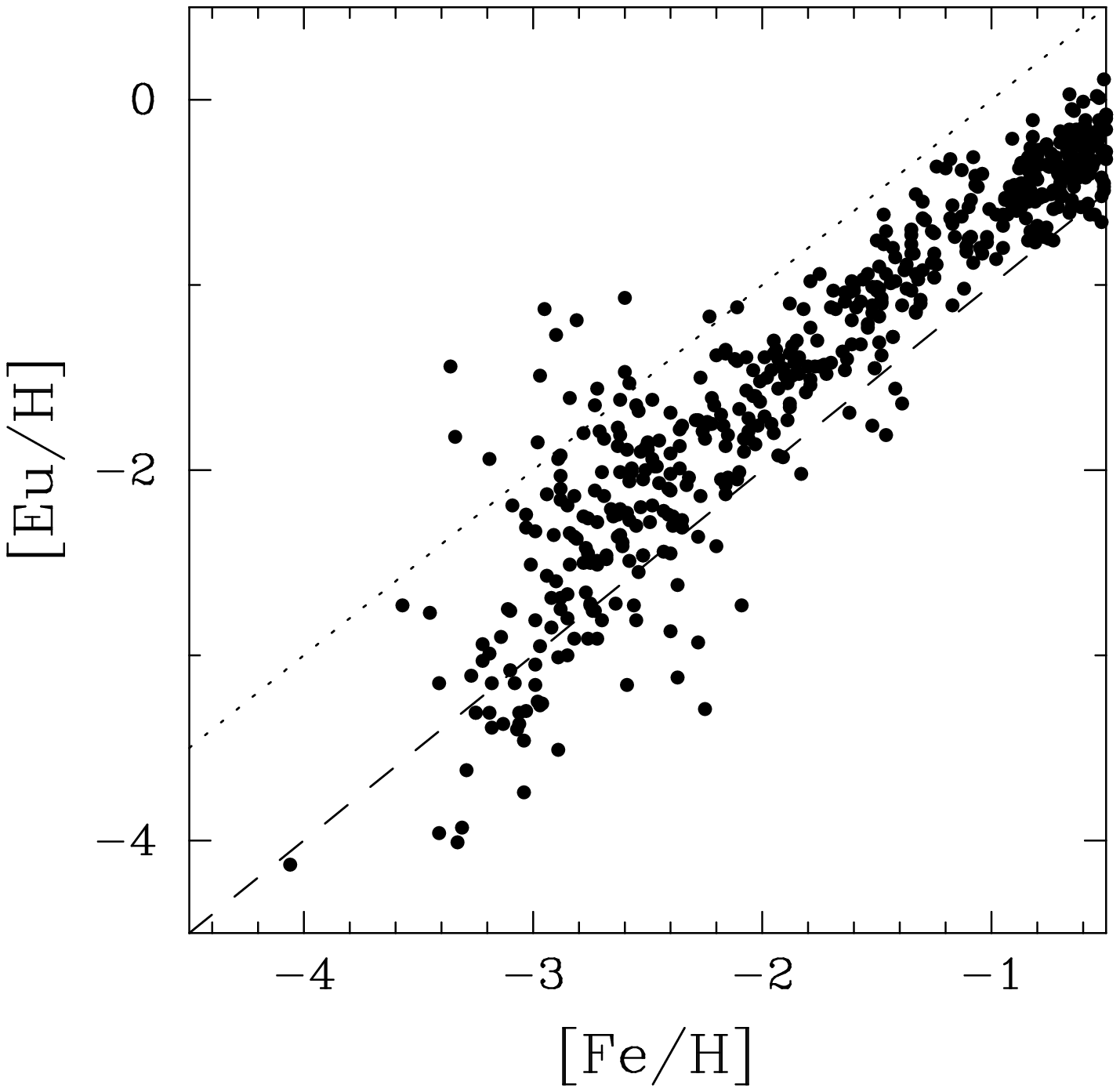}
\end{figure}

  Currently, there are about  1000 metal-poor stars in our
Galaxy for which a measurement of most of the elemental
abundances has been obtained~\citep{Frebel18}.
\ref{fig_Mg-Eu-abundances} shows the abundance distribution for
Fe and Ca for example. 
 The abundance patterns of such metal-poor stars are very useful
references to study the yields of the sources of the
$r$-process and their possible variation~\citep{sn08}.
Details about the abundance distributions are discussed in
\ref{sec:abundance_distribution}. 

\begin{figure}

\caption{Elemental abundances pattern characteristic of the $r$~process.
Open circles show data from halo stars, red boxes are data from stars in dwarf galaxy Ret~II. The black line shows the scaled solar abundance pattern after subtraction of the s~process abundances. From~\protect\citep{Frebel18}.\label{fig_rProcAbundPattern}}
\includegraphics[width=\textwidth]{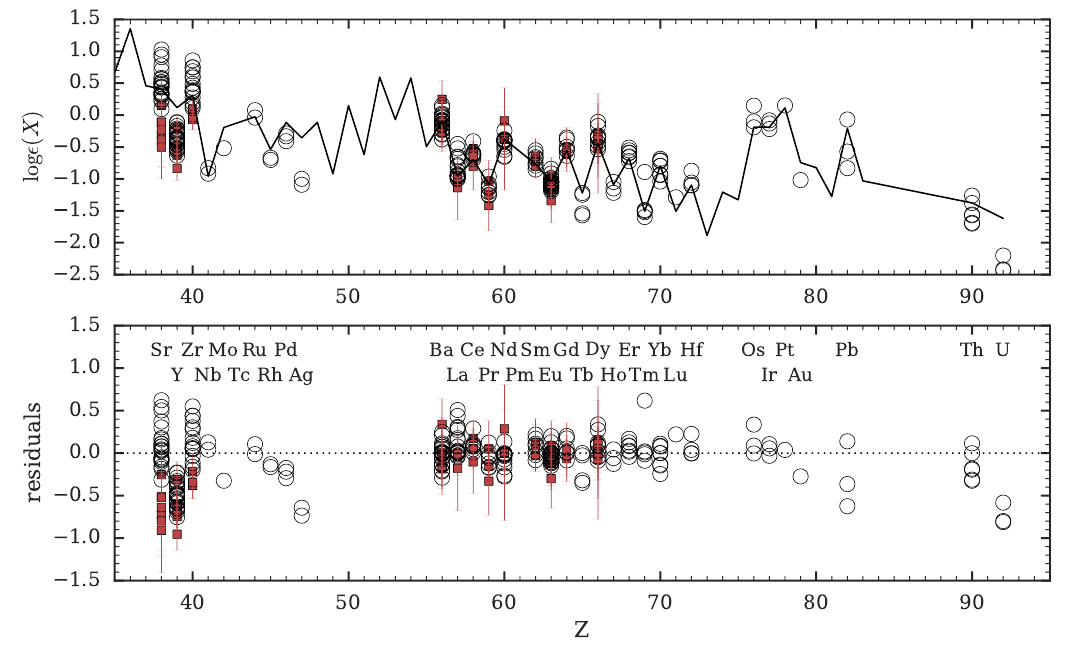}
\end{figure}

\subsubsection{The Abundance Pattern in the Stellar Archeological Record}

Photospheric spectroscopy has established that in a subset of metal-poor stars, heavy element abundances appear to be enhanced relative to general metallicity. (We take metallicity as indicated through the iron elemental abundance). \ref{fig_Mg-Eu-abundances} depicts Eu abundance of Galactic stars versus the Fe abundance. 
Since Eu is most-easily measurable among elements whose origin is predominantly $r$-process, 
Eu-enhanced metal-poor stars found in [Fe/H]$<-2.5$\footnote{Abundance ratios of two elements
  are given as [A/B]  $=$  $\log(N_{\rm A}/N_{\rm B}) -\log(N_{\rm
    A}/N_{\rm B})_{\odot}$. [Fe/H] is referred to ``Fe abundance''.} in the figure are ideal objects to study the abundance
patterns produced by the $r$-process.  
\ref{fig_rProcAbundPattern} shows a typical example of such an
elemental abundance pattern~\citep{Frebel18}.

A surprising result obtained for Eu-rich stars from the early studies of such objects is that the abundance patterns of neutron-capture elements of these stars are very similar
to that of the $r$-process component in solar-system material
\cite{Sneden98,sn08}. 
The agreement is in particular evident in elements from the second  and third peaks corresponding to the neutron magic numbers 82 and 126 (Ba-Pt). 
We note that the measurements of the elements at the second
$r$-process peak (Te-Xe) are only available by UV
spectroscopy with the Hubble space telescope HST for a few of the
brightest stars~\cite{roederer12}.

\ref{fig_rProcAbundPattern} also shows the `solar $r$-abundance pattern' for comparison (black solid line). The residuals (lower panel) indicate that this pattern is well reproduced in many metal-poor stars, although for the light $r$-process elements, the larger residuals indicate more variety.
Compared to the elements from the second and third peaks, abundances of lighter elements show star-to-star scatter. This could be due to the variation of the abundance pattern produced by the $r$-process event. There could also be contributions of another process that provides light neutron-capture elements in the early Galaxy (see \ref{sec:light_n_capture_elements}). 

Another interesting result is the abundance variation of actinides, Th and U. The abundance ratios of these elements with respect to the second-peak elements such as Eu are generally lower in very metal-poor stars, as expected from their decay since  these stars are older than the Sun (\ref{fig_rProcAbundPattern}). 
However, some objects have excess of Th compared to the expected
value from the solar-system $r$-process pattern
(including the decay effect), and are called
\emph{actinide-boost} stars~\citep{schatz02}.  

The variation found in the current sample of Eu-enhanced stars is, however, at most a factor three, whereas the production of actinides is predicted to be severely dependent upon parameters in the $r$-process models. 
This result suggests that the abundance ratios of actinides produced by the $r$~process in the early Galaxy are \emph{regulated} by some mechanism. 
Another observational part of the $r$-process puzzle is the relatively low abundances of Pb compared to actinides (\ref{fig_rProcAbundPattern}). 
Since Pb is the product of the decay of actinides and other
heaviest nuclei, it is difficult  in the models to reduce the Pb
production while keeping the abundance of actinides. Hence,
possible problems in abundance measurements for Pb need to be
considered~\citep{mashonkina12}.

\subsubsection{Abundance Pattern Variabilities}
 \label{sec:abundance_distribution}

The trends and scatter in the Eu abundance distribution shown in \ref{fig_Mg-Eu-abundances} also provide a useful constraint on the origins of the $r$-process. The enrichment of an element depends on the time scale for the corresponding nucleosynthesis event. 

A recent, very precise ($\sigma\sim$0.01~dex), determination of
elemental abundances for 12 elements has been obtained for 79
solar-twin stars at different metallicities~\citep{spina18}. This
not only sets the abundances, but it also allows for tracing the
galactic evolution over 10~Gyr, even separating thick and thin
disk aspects.
From their analysis, the authors conclude that in the thin disk $s$~process contributions prevail during the entire history of the Galaxy, while in the thick disk, enhancements from the $r$-process are clearly a characteristic common feature.
This suggests that in the early galactic evolution, the
$r$-process was highly active~\citep{spina18}.
In the more metal-poor stars of the Galactic halo, the main source of neutron-capture elements was from the $r$-process. 

A remarkable result obtained by the intensive spectroscopic studies
for very metal-poor stars in the past two decades is the discovery of large star-to-star scatter in abundance ratios of neutron-capture
elements. 
\ref{fig_Mg-Eu-abundances} shows the Ca and the Eu abundances as a function of metallicity (Fe abundance). 
Here we adopt Eu as an indicator of the $r$-process yields.
We find a clear correlation, with little scatter, in the abundances between Ca and Fe. Both of these originate from massive stars and their supernova explosions at low metallicity. 
Contributions of type Ia supernovae are significant at high metallicity, and result in a change of the slope for [Fe/H]$>-1$ (see next section). 
By contrast to the Mg abundances, the Eu abundances
show a large star-to-star scatter at low metallicity. 
The object having the highest Eu enhancement with respect to Fe
has [Eu/Fe]$\sim+2$~\citep{aoki10}.  
Metal-poor stars with high Eu abundances are interpreted as objects that were born from gas clouds significantly polluted by an $r$-process event in the early Galaxy. 
This indicates that the gas clouds were not chemically
homogeneous, and insufficient mixing appears to be a characteristic of the early Galaxy.
The $r$-process events should then be independent of those providing Fe, which in the early Galaxy would be the common core-collapse supernovae.

Recent studies suggest that $r$-process-enhanced 
stars have been formed in small stellar systems such as the currently observed ultra-faint dwarf galaxies. 
These probably were enriched in heavy elements only by a small number of $r$-process events (see next subsection).

The detection probability of Eu is lower in metal-poor stars
having a low Eu abundance, because Eu spectral lines are quite
weak in such objects.  
The fraction of $r$-process-enhanced stars is estimated
to be about 3\% including this bias~\citep{abohalima17}. 
This indicates that the Eu-producing $r$-process events are
rare, but each event could provide the Galaxy with a large
amount of Eu. 

\subsubsection{Light Neutron-Capture Elements}
 \label{sec:light_n_capture_elements}

The universality of $r$-process abundances does not seem to extend over the entire range of elements heavier than Fe-group elements. In general, the lighter of the heavy elements, in the Sr--Y group (see \ref{fig_rProcAbundPattern}) conform less to the solar $r$-process pattern. The abundance of the even lighter element Ge does not correlate well with Eu, and may indicate a special, i.e.~non-$r$-process, origin for Ge, although measurements of this element are limited to a small number of metal-poor stars.

On the other hand, there are many metal-poor stars that have low
abundances of heavy neutron-capture elements like Eu, but have a
large excess of light elements~\cite{honda06,aoki13}. The origin
of the excess of these lighter elements is still a controversy,
and is sometimes attributed to a  \emph{Light Element Primary
Process}~\citep{Travaglio:2004}
or \emph{weak r-process}~\citep{wanajo06}. Continuous
observational studies 
have been made to reveal the characteristics of this process. A
recent study~\cite{aoki17} shows that metal-poor stars have wide
variations in the abundance patterns from light to heavy
neutron-capture
elements including intermediate ones (e.g.,~Mo, Pd). A correlation of the Mo ($Z=42$) abundance to those of Sr, Y and Zr ($Z=38$--40) was reported
along with a gradual decrease of abundances with increasing atomic number. 

\subsection{Stellar Abundances in other Galaxies}
   Small stellar systems, dwarf spheroidal galaxies, are found near the
Milky Way Galaxy, sufficiently close so that interactions with the Galaxy are plausible. 
These have a wide
range of stellar mass ($10^{3}$--$10^{7}$ solar masses), and are characterized by an absence of interstellar gas. Their stellar population thus has formed long ago, and probably in the early evolution of that system. 
They could be survivors of small stellar systems born in the very early phase of
the Milky Way formation.  
About 30--40 such galaxies are known at present.
Such small stellar systems could be significantly affected by a small number of nucleosynthesis events.
Therefore, even a single $r$-process enrichment might be revealed in observations of metal-poor stars from nearby dwarf galaxies.
Accretion of such stellar systems could be the origin of the $r$-process enhanced very metal-poor stars in the Milky Way halo.
Therefore, they are a useful complement to $r$-process studies from stars in the Milky Way galaxy.

\begin{figure}
\caption{Eu abundance ratios of stars in dwarf galaxies as a function
  of metallicity ([Fe/H]). Abundance data are taken from the SAGA
 database~\cite{suda17}.  Different symbols mean stars in
different
  galaxies (open triangle: Fornax; open circle: Carina; filled circle:
  Sculptor; open square: Draco; filled triangle: Leo I; stars:
  Ursa Minor; red filled diamonds: Reticulum II). Upper limits of Eu abundances of the two most metal-poor stars in Reticulum 2 are also shown.  The other seven objects in Reticulum II are very metal-poor but show large excess of Eu. A typical error of [Eu/H] is 0.2~dex, although it depends on data quality and strength of
  spectral lines. The dashed line indicates [Eu/Fe]$=0$. Many metal-poor stars in faint dwarf galaxies have similar [Eu/H] values around $-1.3$  (box shown by dotted line)\label{fig_Eu-dSph}}
\includegraphics[width=\textwidth]{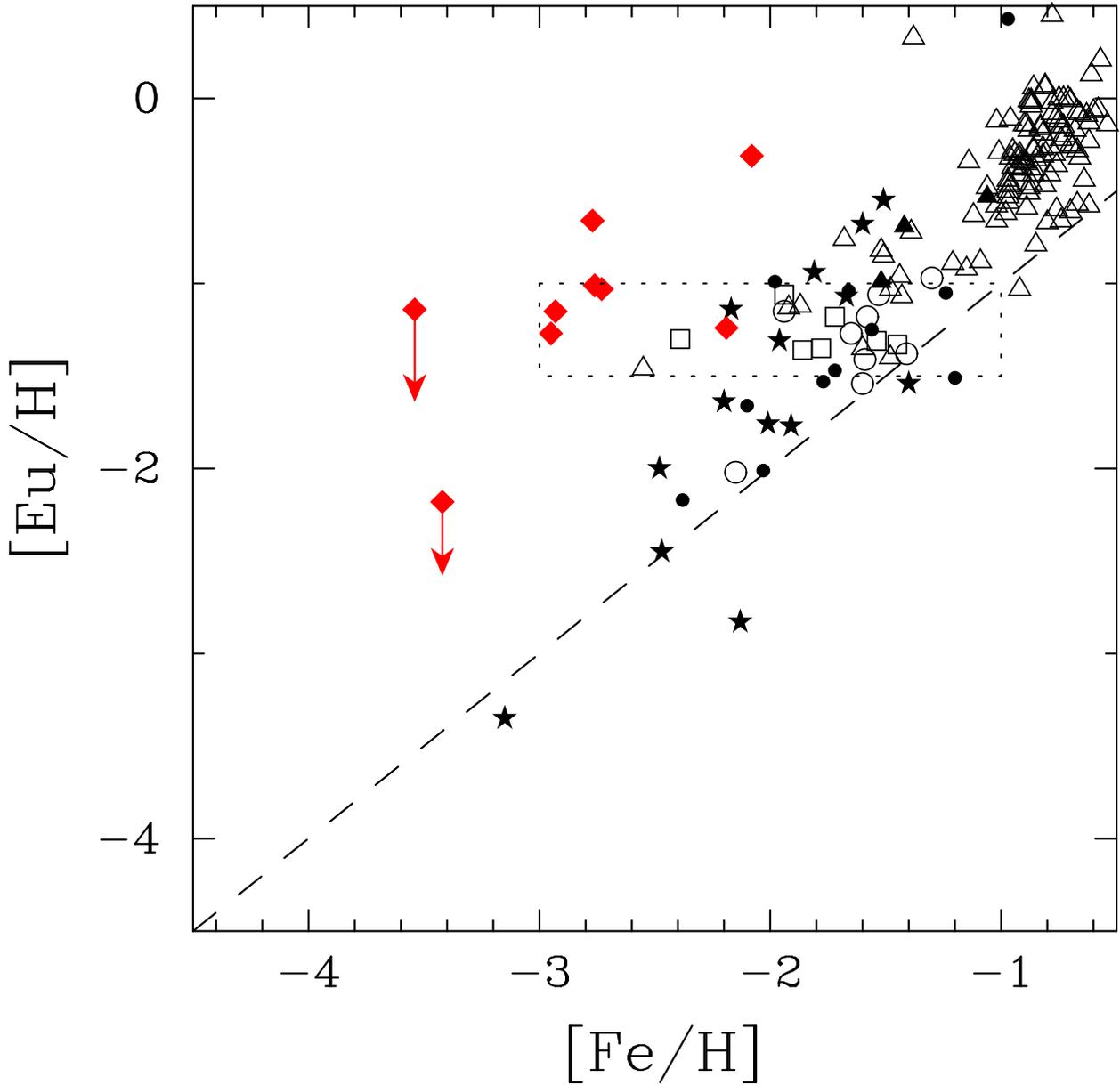}
\end{figure}

\ref{fig_Eu-dSph} compiles the Eu abundances measured for dwarf galaxy stars. 
As in the case of galactic field and halo stars, generally, the Eu abundances increase with an increase of Fe abundances. 
The number of stars with low Eu abundances at low metallicity ([Fe/H]$<-2$) is small, compared to Galactic field and halo stars shown in \ref{fig_Mg-Eu-abundances}.
This, however, reflects a bias in the sampling due to the less-sensitive detection limit of Eu lines for dwarf-galaxy stars, which are apparently much fainter than field halo stars studied with high-resolution spectroscopy. 

It is remarkable that dwarf galaxy stars with $-2\lesssim$ [Fe/H] $\lesssim -1$ have almost
constant [Eu/H] values ($\sim -1.3$), independently of their Fe
content~\citep{tsujimoto14}. 
Subsequent observational studies for a larger sample of dwarf
galaxies further support this
feature~\citep{tsujimoto14,Tsujimoto17}. 
It appears that particular and large enrichment from the $r$-process has occurred early, while later evolution increases both Fe and Eu as found in Fornax stars.

The dwarf galaxy Reticulum II appears to be a unique object.
Here, an entire group of very metal-poor stars
($-3\leq$[Fe/H]$<-2.5$) show a surprisingly large excess
of Eu that is about 1--2 orders of magnitude higher than that
seen in other dwarf spheroidals~\citep{ji18,roederer16}. 
At first glance, this appears to reflect a massive and single $r$-process enrichment event, providing the seed for the material out of which all these stars formed in an early phase of the evolution of this galaxy when [Fe/H]$\sim -3$ or $-2$.
Such Eu-enhanced stars are also known in the
Milky Way halo, as mentioned above, and here their fraction is estimated to
be 3\%--5\%~\citep[e.g.,~][]{barklem05,abohalima17}.
In Reticulum II, abundances were measured for nine stars that must have formed early at low metallicity ([Fe/H]$<-2$), and seven of them
have high Eu abundances. The remaining two stars have the lowest
metallicity ([Fe/H]$<-3$) of the observed sample.  Hence, these  may have formed earlier.
A large set of 41 elemental abundances in the brightest star of
Reticulum~II at a visual magnitude of 16 were measured, improving
precision in this case by a factor~2~\citet{ji18}. This study
confirms a general $r$-process enrichment with the known
abundance pattern, including abundances up to the third
$r$-process peak.
Eu abundances in $r$-process-enhanced stars in faint dwarf galaxies including Reticulum II are found at values [Eu/H]${\sim} -1.3$. This corresponds to a Eu/H mass ratio of $10^{-10.6}$. 
Assuming the original mass of hydrogen gas 
in the galaxy to be $10^{6}$ solar masses, this yields an Eu mass estimate to be provided by
the $r$-process of $10^{-4.6}$ solar masses. This
almost constant value of Eu abundance could be explained if the
supply of $r$-process elements  is due to  a single
event in an early evolutionary stage of these galaxies. Moreover,
the amount of Eu is comparable with the expected
$r$-process products of neutron star
mergers~\citep{goriely11}.

\subsection{Abundances of Interstellar Gas in the Galaxy}
   Interstellar gas assembles the ejected material from individual nucleosynthesis events, including that of the $r$-process. 
The stellar abundance measurements discussed above assume that the stars being observed formed out of material that either was sufficiently mixed to represent the chemical composition of the star forming region at that time, or else had a special enrichment from a single, additional, $r$-process injection event (for the `$r$-enriched' abundance results).

Mixing of interstellar gas is believed to be driven by turbulence created by winds and explosions, mostly, with some low-frequency contributions from large scale kinematics and spiral wave transits.
But since injection of fresh nucleosynthesis material probably occurs where the interstellar gas is being energized and heated, the process of star formation generally requires interstellar gas to be cold so that gravitational attraction can form dense (protostellar) clumps within molecular clouds. In between, therefore, interstellar gas is required to both mix and cool. Typical times discussed for this are on the order of 10$^5$ to 10$^8$ years. 
With a supernova rate on the order of
10$^{-2}$~y$^{-1}$~\citep{Diehl:2006} within such
settling time of freshly-enriched interstellar gas, there is a
significant probability of other supernovae occurring. 
Specifically, these would be from massive stars, which generally are produced as a group and hence in the same region. Therefore, multi-event enrichments appear plausible before stars are being formed, with a corresponding bias.
It is therefore (i) not surprising to find differences in compositional signatures between stars and gas, and (ii) rare nucleosynthesis events and their enrichments may be more easily recognized in stellar material, as anomalies from the well-mixed and multiple-source averaged typical stellar abundance signature.

On the other hand, the available abundance data from interstellar
gas includes its own bias and systematics. In particular there
are biases from pre-solar grains, terrestrial deposits, and
emission line physics. Absorption line spectroscopy against
background sources such as stars, AGNs, and gamma-ray bursts are
scarce, but will provide a useful complement to such data.   ALMA
is going to play an important role here~\cite{kaminski18}.

\begin{figure}
\caption{The rate of events injecting $^{244}$Pu into the terrestrial atmosphere.From \protect\citep{Wallner:2015}. \label{fig_rProc244Pu}}
\includegraphics[width=\textwidth]{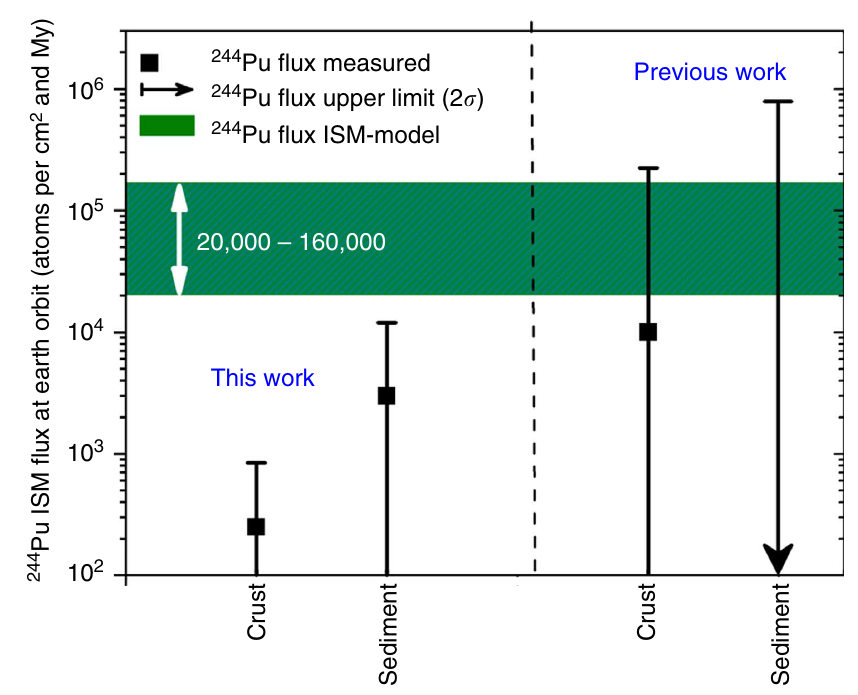}
\end{figure}

\subsubsection{$r$-Process Material on Earth}\label{oceancrust_obs}

Material on Earth also can provide astronomical, or cosmic, information, if anthropogenic distortions can be separated or eliminated.
The field of nucleo-cosmochronology has been established 
from the study of long-lived terrestrial radioactivity,
t~\citep{Clayton:1985}, as was originally suggested by Rutherford
in 1929.
This made use of the U and Th radioactive decays to estimate, on one hand, the epoch of their nucleosynthesis long before the solar system formed (if isotopes of the same element are used), and, on the other hand, the production ratio of U/Th, can constrain the $r$-process itself.

An interesting current-day \emph{astronomical} measurement is added from material that falls to Earth, if as is the case for pre-solar grains (see below), and for material that can be excluded to arise from cosmic ray spallation reactions. 
The latter is the case for all nuclei well above the iron group and specifically abundant $^{54}$Fe.
Even terrestrial  $^{60}$Fe thus provides information on supernova nucleosynthesis, and has been measured in ocean-crust deposits and lunar samples.  However,  we do not discuss this here, as $^{60}$Fe plausibly is attributed to the $s$-process.
Ocean crust material has been shown to slowly accumulate material from the Earth's atmosphere, cosmic-ray age dating of depth layers thus establishes a record of infalling interplanetary gas over millions of years. 
The probes taken from locations distant from coastlines and at great depth (thousands of km, and $>$5~km, respectively) have been shown to be free from contamination, and thus suitable for study.
Most interesting for the $r$~process is $^{244}$Pu with a radioactive half-life of 81 million years.
After several searches in 2015 the method of accelerator mass
spectrometry obtained a signal that could be combined with other
deposits to yield an incident flux of $^{244}$Pu at Earth from
currently-active $r$-process
nucleosynthesis~\citep{Wallner:2015}. 
\ref{fig_rProc244Pu} shows this result, as compared the core-collapse supernova injection rates derived from earlier measurements, and the predictions are shown as the horizontal (green) band. 
The tacit assumptions are that interplanetary dust particles reaching Earth are representative of the current interstellar medium composition in the solar vicinity, and that this is sufficiently well mixed to represent nearby nucleosynthesis over the past 10$^8$ years.
Expectations can then be set from the Galactic rate of core
collapse supernovae of about 1--2 per century~\citep{Diehl:2006},
if one (the most-frequent) $r$-process scenario is used
as a reference. 
\citet{Wallner:2015} show that the measured rate falls 2--3 orders
of magnitude below such expectations. This clearly implies
(within the above assumptions) that the rate of nearby
$r$-process events must be much smaller than the rate of
supernovae.

As discussed by~\citet{Hotokezaka:2017a}, the abundance of
$^{244}$Pu (T$_{1/2}\sim$81~My) also available at the time of
solar system formation, provides information.
Its relatively large abundance relative to $^{238}$U (T$_{1/2}\sim$4.5~Gy) suggests that at the time of solar system formation, an $r$-process injection into the proto-solar molecular cloud occurred within ${\sim}$100~My, while currently the above mentioned ocean crust value suggests a lower rate. 

This early solar system presence indicates that the
$r$~process did indeed occur in the solar vicinity at
the time of solar system formation, The universality of the
$r$-process abundance pattern  suggests it was ongoing
from the epochs of the young Galaxy until solar system formation
(see however late injection scenarios discussed for the early
solar system,~\citet[e.g.~][]{Huss:2009}.) 
\citet{Tsujimoto17} provide an analysis of all available
radioactive clocks which may be used to constrain the
heavy-element injections into the proto-solar nebula, and hence
in meteoritic material. They then 
separate an `average' from a `latest-event' injection, comparing the different heavy-element abundances. Thus, they derive a constraint on the rate of $r$-process injections relative to the rate of supernova material injections. Their value of 1/1400 requires that $r$-process injections are rare, compared to supernovae.  Rare variants (at a level of 10$^{-3}$ or below) of supernovae, and also neutron star mergers, can fulfill both these constraints.

We note, however, that the conversion of the ocean crust measured count of $^{244}$Pu atoms to an ejected amount of $^{244}$Pu from a source involves several steps. Each of these has considerable uncertainty:
(i) The conversion to $^{244}$Pu flux incident on the top of the Earth's atmosphere involves ocean and atmospheric transport; this is probably well calibrated through $^{10}$B and $^{53}$Mn, which are produced by cosmic rays in the upper atmosphere, and the uptake factor for the deep ocean probe can be measured.
(ii) The transport of interplanetary dust particles, the presumed carriers of $^{244}$Pu, in the heliosphere and a Local Bubble is uncertain. Magnetic fields and the net electric grain charge will be key.
Additional steps not only involve the ocean crust constraint, but, in the same way, the constraints that can be derived from meteoritic analyses and of early solar system material:
(iii) The transport of nucleosynthesis ejecta into the interstellar medium, and into the Local Bubble, therefore, involves (a) the physics of the turbulent and dynamic interstellar medium, and (b) the initial slowing down and cooling of hot matter by the ISM. 

The first aspect (a) is being addressed by
magneto-hydro-dynamical
simulations~\citep[e.g.~][]{deAvillez:2005}, and turns out to be a
major astrophysical challenge, across scales beyond a few hundred
parsecs. Then, one may adopt different locations of the solar
system within the ISM structures, leading to
a variety of conditions as shown by~\citet{Kuffmeier:2016} for
the case of the $^{60}$Fe/$^{26}$Al ratio from supernova
nucleosynthesis.
The second aspect (b) may be explored from supernova remnant
observations. Tracing ejecta flows into the ISM can be done
through radioactivity studies. Typically, supernova remnants show
observable characteristics in radio and X-ray emission from young
ages until several tens of millenia. But this does not imply
ejecta mixing within 10$^5$ y or thereabouts.  Only the
particle energies fall below the radio and X-ray emission
thresholds. Tracing $\gamma$-ray emission from radioactive
$^{26}$Al (T$_{1/2}\sim$700,000~years), the large Doppler shift
velocities exceeding typical galactic
rotation~\citep{Kretschmer:2013} have been interpreted as an
indication that nucleosynthesis from massive stars and their
supernovae are typically injected into kpc-sized superbubble
cavities~\citep{Krause:2015}. Multiple such injections of
nucleosynthesis ejecta would occur into the same superbubble.
Therefore, mixing could be delayed by the fate of the
superbubble's evolution.
Then, cooling of fresh ejecta may not occur on time scales below 10$^8$~y.
Thus the composition of material would be better characterized by the mixing properties of the ISM, than by the rate of injection events. 

Clearly observations of terrestrial material are proof that $^{244}$Pu abundances are not a sample of what is called the generic $r$-process abundance pattern. This could be related to nucleosynthesis transport in the interstellar medium, or to scarcity of nucleosynthesis injections.

 \subsubsection{Constraints from Pre-solar Grains}

While the analysis of the isotopic composition of material samples in the laboratory clearly obtains the most precise values for abundances among different elements and isotopes, the path from cosmic interstellar gas to the terrestrial laboratory includes numerous possibilities to affect the composition. 
The first step is the formation of dust grains.  This  requires  appropriate thermodynamic conditions, as chemical reactions 
to operate for sufficient time to form molecules that are refractory, and  nucleation to form a dust grain.

Grain surface chemistry is then  important to the composition of interstellar gas in its cold phases, and ice mantles of specific species (water, ammonia) may grow to make larger grains. 
Interstellar grains, unlike gas, then propagate ballistically, although the grains are not  electrically neutral due to their higher mass. 
In the interstellar medium, shocks and radiation from nearby stars heavily processes dust.  This may fragment grains, remove ice mantles, and generally change the grain size distributions. For most of these processes, the isotopic composition of an element is rarely affected, with exceptions from isotope-sensitive chemical reactions of light species.
Therefore, nucleosynthetic studies on material samples of pre-solar origins require normalization. Deviations of isotopic ratios within an element from the solar reference are the key information.

In early analyses of meteoritic material, such deviations had been interpreted as a general indication that the meteoritic material included an unknown and variable contribution of pre-solar grain material. The deviation from solar isotopic ratios could be attributed to isotopic under- or over-abundances in pre-solar grains, as the contamination fraction and origin varied.
Only with the development of nanometer scale spatial resolution of laboratory mass spectrometry could determine multiple isotopic ratios for individual pre-solar grains became possible. 
The harsh chemical sample preparation for such analyses includes
chemically dissolving much of the bulk meteoritic material,
leaving behind the most refractory parts. This led to the
isolation of stardust grains in the forms of diamonds, silicon
carbide, and graphite grains~\citep[e.g.~][]{Zinner:2014}. 

Generally, presolar grain composition analyses have failed to
show clear evidence of any $r$-process
contributions~\citep{Zinner:2008}.
While some anomalies exist for Xe, Mo, and Ba isotopic ratios, the case for a clear signature of a rapid neutron capture origin is too uncertain at least for Ba and Mo: 
Barium in general is considered a typical s~process product, as
all stable isotopes lie along the region of stable isotopes with
no unstable isotope in between, and only shielding from the
$r$-process path by $^{134}$Xe and $^{136}$Xe
plausibly can imprint $r$-process information for these
s-only isotopes.  These are  superimposed on a mix of r~and
s~process origin for the other isotopes. In the case of
$^{100}$Mo, only one unstable isotope separates this isotope
from the valley of stability, and episodic strong neutron
irradiations in shell burning regions of massive stars have been
invoked to explain the abundance deviation of
$^{100}$Mo~\citep{Meyer:2000}. 
The Xe isotope abundance pattern, finally, appears to be the only
clear $r$-process signature from pre-solar grains.
However, the absence of a similarly-high excess in the
$^{136}$Xe abundance is puzzling, and requires a specific
explanation for attribution of the Xe abundance signature to be
related to $r$-process~\citep{Gilmour:2007}.

\subsection{Astronomical Summary on the $r$~Process}

The elemental-abundance pattern seen in the currently-available sample of heavy-element enriched metal-poor stars of our Galaxy is remarkably reminiscent of the abundance pattern seen in solar material after subtraction of the (modeled) contribution from the $s$~process.
This suggests that some \emph{r-process} nucleosynthesis has been operating in the Galaxy since its earliest evolution and until today.

The observed scatter in abundances of $r$-process elements appears significantly larger than the observed abundance scatter for elements attributed to massive stars and their supernovae, and to supernovae of type~Ia. Thus, it appears as if normal supernovae of any of the two major types are not the sources of the $r$-process.
Rather, events that occur at substantially lower frequency, and/or mix significantly less-well with star-forming interstellar gas in the Galaxy, must be responsible for providing material from the $r$-process.
The observed low content of $^{244}$Pu, as compared to $^{60}$Fe, in ocean crust material on Earth points in the same direction, i.e.,~$^{244}$Pu-contributing event and  less frequent or less efficient deposits of interstellar material within the current solar system.
Also, an apparent enrichment in heavy elements by a rare, probably single, event in one of the small satellite galaxies of the Milky Way galaxy point to a source of the $r$~process that occurs at a rate well below 1/10$^4$~y and ejects on the order of 10$^{-4}$--$10^{-2}$~M$_{\odot}$ of material enriched in $r$-process nuclei.

The neutron star merger event GW170817 and its kilonova have contributed valuable and detailed data from a currently most-plausible source of $r$-process material. The spectrum of the kilonova light and its rapid fading in brightness are consistent with models which have an $r$-process which has synthesized radioactive material and ejected heavy elements as a key ingredient.

No further $r$-process relevant observations could be contributed to date from pre-solar grain or gamma-ray line studies.  Nevertheless,  in principle, these are promising candidates for measuring the signature of an $r$-process rather directly, i.e.~as isotopic information. 

 \section{Sites for the $r$~Process}

\label{models}
There are a number of viable sites that satisfy the basic
requirements to enable an $r$~process (as presented in
\ref{metal_poor} above). They each are likely to contribute at least
some to the total abundance of $r$-process elements in
the Galaxy. (cf.~\cite{Shibagaki16}).  
 In the following sub-sections we give the essential ingredients in a 
 number of candidate models, and we summarize their respective status on the basis of the current literature, then discuss their weaknesses and strengths in the light of the observational constraints.

\subsection{Neutrino Driven Winds}

A favored model for  $r$-process nucleosynthesis for a
number of years was the neutrino driven wind (NDW)  above the
newly formed proto-neutron star in CCSNe~\citep{woosley94}.  A
neutron star is formed by the collapse of the iron core of a
massive star.   The  proto-neutron star then cools by the release
of ${\sim} 10^{53}$ ergs in neutrinos on a timescale of ${\sim} 10$
s.  These neutrinos interact with material behind the outgoing
supernova shock.  This  generates a hot bubble that helps to
drive the explosion~\citep{Bethe85,Wilson03}.  It also leads to
the ablation of material from the proto-neutron star into the hot
bubble.  This  is the so-called  neutrino-driven wind (NDW).  The
high entropy and large number of neutrons per seed in this wind
seemed at one time to be an ideal $r$-process
site~\cite{woosley94}.  Neutrino nucleus interactions, however,
are key~\citep{Balsi15} to the success or demise of this model.

Although the original formulation was quite successful,
subsequent
calculations~\citep{Fischer10,Hudepohl10,Fischer12,Hempel12,Wanajo13}
have found the  NDW  to be inadequate as an $r$-process
site.   The downfall of the NDW has been
attributed~\citep{Fischer10,Fischer12} to the implementation of
higher-order  modern neutrino transport methods, and effects of
the neutron--proton mass difference.  These cause the neutrino
luminosities and energies to diminish at late times.  The
resulting electron fraction and  neutrons per seed ratio in the
neutrino driven wind become too small for a robust
$r$-process, although this can be mediated somewhat by
nuclear medium effects on the neutrino
opacities~\cite{Roberts12}. Another difficulty is the alpha
effect: 
In a core-collapse supernova during the epoch of alpha particle
formation as the temperature drops protons produced by
$\nu_e$ capture on neutrons can in turn capture more
neutrons to bind into alpha particles, increasing the electron
fraction~\cite{Fuller:1995ih}. This ``alpha effect'' could be a
significant impediment to achieving the required r-process
yield~\cite{Meyer:1998sn}. Density fluctuations may amplify this
effect~\cite{Loreti:1995ae}. However, it is possible to get
around the alpha effect by assuming presence of sterile neutrinos
which reduce the $\nu_e$ flux 
\cite{McLaughlin:1999pd,Fetter:2002xx,Caldwell:1999zk}. 

There is also some indication~\cite{Olson17} that the diminished
neutrino luminosities  depend upon  nuclear equation of state  as
well.  Neutrino luminosities and energies were studied for
variety of Skyrme density functional equations of state
in~\cite{Olson17}.  These were chosen as they are  consistent
with constraints on the symmetry energy from the combination of
isobaric analog states, pygmy resonances, and heavy ion
collisions  and also satisfy neutron stars with masses as large
as $2.01 \pm 0.04$ M$_\odot$ as required by
observations~\citep{Demorest10,Antoniadis13}. For all of these
models a drop in the late time  neutrino energies and luminosity
results even in a low order multi-group flux-limited diffusion
scheme.  Hence, the likelihood of a NDW $r$-process
seems less likely in the current models for core-collapse
supernovae.

The desired conditions of high entropy and  neutron-rich
composition~\citep{Otsuki03,Wanajo13} do not occur in standard
neutrino energized wind. Nevertheless, it is quite likely that a
\emph{weak $r$-process} occurs in the NDW producing
neutron rich nuclei up to about
$A \sim 125$~\citep{Wanajo13,Shibagaki16}.

\subsection{Magnetic Neutrino-Driven Wind}

Although standard models of neutrino-heated winds from proto-neutron stars  do not reach the requisite neutron-to-seed ratio for the production of r-process nuclides beyond the $A \approx 130$ peak, the abundance distribution created by the $r$- $rp$-, or $\nu p$-processes in  proto-neutron star winds depends sensitively on the entropy and dynamical expansion timescale of the flow, which may be strongly affected by high magnetic fields. 

In~\cite{Thompson17} magnetohydrodynamic simulations were made of
nucleosynthesis for non-rotating neutrino-heated winds from
proto-neutron stars with strong dipole magnetic fields
($10^{14}\mbox{--}10^{16}$ G). It was found that the strong field forms a
closed zone and helmet streamer configuration at the equator,
with episodic dynamical mass ejections in toroidal plasmoids.
This dramatically enhanced the entropy in these regions and
conditions favorable for the production of the third-peak
r-process could be obtained if the wind was neutron-rich.  For
$B\gtrsim10^{15}$ G, it was found  that $\gtrsim10^{-6}$ M$_{\odot}$ and
up to ${\sim}10^{-5}$ M$_{\odot}$ of high entropy material was
ejected per highly-magnetized neutron star birth in the wind
phase, providing a possible mechanism for prompt rare heavy
element nucleosynthesis even in the neutrino-driven wind
scenario. 

\subsection{$r$-\lowercase{Process in Low-Mass Prompt Supernova Explosions}}

One of the earliest numerical supernova models for the
$r$-process~\cite{Hillebrandt84} was based upon the
shock-induced ejection of neutron-rich material in low-mass
(${\sim} 10$ M$_\odot$) core collapse supernovae. Although
this model is not often referenced in the literature of the past
decade, it remains as a viable model for the
$r$-process~\cite{Sumiyoshi01,Wanajo05}. 

Among core-collapse models,  low-mass O-Ne-Mg cores are the
easiest to explode~\cite{Kitaura06,Dessart06,Burrows07}.  They
have  smaller cores and  weaker gravitational potentials, and are
in nuclear statistical equilibrium at the time of core bounce. As
such, these  stars have the possibility  to explode via
hydrodynamics rather than by delayed neutrino
heating~\cite{Bethe85} as in the NDW $r$-process.  This
avoids some of the issues with neutrino interactions in the NDW
models.  In this scenario the prompt shock itself is sufficiently
robust as it bursts through the outer layers of the proto-neutron
star to directly eject neutronized matter which subsequently
experiences an $r$-process.   

\subsection{Neutron Star Mergers}
  
In recent years by far the most popular model
(cf.~\cite{Thielemann17}) for  $r$-process
nucleosynthesis is based upon binary neutron-star mergers.  This
scenario can involve either two neutron-stars   ($NS+NS$) or
a neutron-star plus black-hole ($NS + BH$)
binary~\citep[e.g.~][]{goriely11,Korobkin12,Piran13,Rosswog13,Rosswog14,Goriely13,Wanajo14,Nishimura16bh,Kawaguchi16,Thielemann17,Lippuner17,Kyutoku18}.  

Indeed, since the
observation~\cite{Cowperth17,Chornock17,Nicholl17,Smartt17,Pian17,Troja17,abbott17,Villar17}
of the gravitational waves emanating from  GW170817 and the
associated kilonova and GRB from GW170817 there has been a flurry
of models~\cite{Cote17,Metzger17,kasen17,Bovard17} describing the
$r$-process in NSMs and the associated hydrodynamic
ejection of $r$-process material into the interstellar
medium~\citep{Radice16,Sekiguchi16,Murguia17,Chang18}.

 Ejected matter from  NSMs can be  very neutron-rich
($\langle Z/A \rangle \equiv Y_e \sim 0.1$).  Hence, the $r$-process path can proceed
along the neutron drip line all the way to the  region of fissile
nuclei ($A \approx 300$).  Indeed, in such models, fission recycling
can occur.  That is, after the $r$-process terminates by
beta-induced or neutron-induced fission, the fission fragments
each continue to experience neutron captures until  fission again
terminates the $r$-process path and   the process
repeats.  In this process the abundances can become dominated by
the fission fragment distributions after a few cycles.  Hence,
the beta-decay flow near the neutron closed shells is not as
important.  In that sense, the nuclear physics for NSMs is quite
robust~\cite{Mendoza15}.  However, this paradigm is  very
sensitive to theoretical models of extrapolated nuclear
properties near the termination of the $r$ process,
particularly  fission barriers and fission mass
distributions~\cite{Shibagaki16}.  

A number of
studies~\citep{Freib99a,Freib99,goriely11,Korobkin12,Goriely13,Wanajo14,Nishimura16bh}
have indicated  that the $r$ process in NSMs can produce
a final abundance pattern that is similar to the solar 
$r$-process abundances, but only for heavier A$>130$ nuclei.  
However, the distribution of  nuclear fission products strongly affects the final abundance pattern
\cite{Eichler16}. 
Extrapolation of fission fragment distributions (FFDs) to the
vicinity of the $r$-process
path~\citep[cf.~][]{Martinez07,Erler12,Eichler16}, however, is
fraught with uncertainty.  It has been
suggested~\citep{Shibagaki16}, for example, that incorporating a
model that leads to a  broad distribution of fission fragments,
smooths out the effect of the neutron closed shells, thereby
providing a means to fill in the isotopes bypassed in the main
$r$ process.

Current microscopic calculations (e.g.~\cite{Martinez07} tend to
produce lower fission barriers and narrower fragment
distributions as discussed below.  However, the
$r$-process study of~\citet{Shibagaki16} made use of
$\beta$-decay rates,  $\beta$-delayed neutron emission
probabilities, and  $\beta$-delayed fission probabilities
taken from 
~\cite{chiba08} based upon fits to known fragment distributions.
The spontaneous fission rates and the $\alpha$-decay rates
were  taken from~\citet{Koura04}. For these rates
$\beta$-delayed fission  is the dominant nuclear fission
mode near the termination of the $r$
process~\citep{chiba08b}. However, this is a phenomenological
model.  
 
 From the perspective of galactic chemical evolution, it may be
difficult for NSMs to reproduce the observed r-process abundance
distribution of metal-poor stars at [{Fe}/{{H}}]$<$~-3.
A possible explanation  is that the time scale of metal
enrichment in the most metal-poor stars does not follow a simple
age-metallicity relation as a simple one-zone model for the halo.
That is, some of the enrichment of $r$ elements may
occur in dwarf galaxies or the IGM that later are incorporated
into the Galactic halo. This explanation has been
corroborated by kinematic studies~\cite{Roederer18} of the
Galactic halo indicating that the $r$-enhanced stars are
the result of late mergers.

For example,~\citet{Hirai15} considered the  chemical evolution
in dwarf spheroidal galaxies (dSphs).  Such dSphs  are the
building blocks of the Galactic halo and have a much lower star
formation efficiency than that of the Milky way halo. That paper
showed  that  when the effect of metal mixing was taken into
account,  the enrichment of $r$-process elements in
dSphs by NSMs could reproduce the observed [Eu/Fe]
vs.~metallicity distribution with a merger delay time of as
much as 300 Myr. This is because metallicity is not really
correlated with the time ${\sim} 300$ Myr from the start of the
simulation
due to the low star formation efficiency in dSphs. They  also confirmed that this model is consistent with observed properties of dSphs such as the radial profiles and metallicity distribution. A merger time  of ${\sim} 300$ Myr and a Galactic NSM rate of ${\sim} 10^{-4}$ yr$^{-1}$ could reproduce the abundances of metal poor $r$-process enhanced stars and is consistent with the values suggested by population synthesis and other nucleosynthesis studies.

 Similarly, Komiya and Shugiyama~\cite{Komiya16} re-examined the
enrichment of r-process elements by NSMs considering this
difference in propagation using a chemical evolution model based
upon  hierarchical galaxy formation. They found that  observed
r-process enhanced stars around [{Fe}/{{H}}]-3 are
reproduced if the star formation efficiency is lower for low-mass
galaxies using a realistic delay-time distribution for NSMs. They
showed that a significant fraction of NSM ejecta escape from its
host proto-galaxy and  pollute intergalactic matter as well as
other proto-galaxies. The propagation of r-process elements over
proto-galaxies was shown to reduce  the abundance distribution at
[{Fe}/{{H}}]$<$ -3.  They could obtain a  distribution
compatible with observations of the Milky Way halo stars. In
particular, the pre-enrichment of the intergalactic medium
explains the observed scarcity of extremely metal-poor stars
without Ba and the abundance distribution of r-process elements
at [{Fe}/{{H}}]$\le$ -3.5. 
 
\subsection{Black hole-Neutron Star Mergers}

During the merger of a black hole and a neutron star, baryonic
mass can become unbound from the system. Because the ejected
material is extremely neutron-rich, the r-process rapidly
synthesizes heavy nuclides as the material expands and
cools~\cite{Kyutoku18,Nishimura16bh,Nishimura17,Fernandez17,Roberts17}. 

In~\citet{Roberts17}, models of black hole-neutron star mergers
were mapped into a Newtonian smoothed particle hydrodynamics
(SPH) code and the evolution outflows were followed. It was found
that the ejected material produces r-process nucleosynthesis even
for unrealistically high neutrino luminosities, due to the rapid
velocities of the outflow. Neutrinos could, however, have an
impact on the detailed pattern of the r-process nucleosynthesis. 

In~\citet{Fernandez17} an investigation was made into  the ejecta
from black hole--neutron star mergers by modeling the formation
and interaction of mass ejected in both the  tidal tail and a
disk wind. The nucleosynthetic yields in the outflows were
obtained using a post-processing nuclear reaction network.  The
kilonova emission was also computed with a radiative transfer
code.  A large initial tail mass resulted in the fall-back of
matter into the disk that was   then ejected in a disk wind.
Relative to the case of a disk without dynamical ejecta, the
combined outflow had lower mean electron fraction, faster speed,
larger total mass, and larger absolute mass free of high-opacity
lanthanides or actinides. They found that in most cases, the
nucleosynthetic yield was dominated by the heavy r-process
contribution from the unbound part of the dynamical ejecta. A
solar-like abundance distribution could be obtained when the
total mass of the dynamical ejecta was comparable to the mass of
the disk outflows. 

In~\citet{Kyutoku18} a fully general-relativistic
neutrino-radiation-hydrodynamics simulation was made of the
merger of black hole-neutron star binaries   throughout the
coalescence.  This work particularly focused on the role of
neutrino irradiation in dynamical mass ejection. This study
confirms  that the ejecta from black hole-neutron star binaries
are indeed very  neutron rich and should accommodate a strong
r-process unless magnetic or viscous processes contribute
substantially to the mass ejection from the disk. 

In~\citet{Wu16bh} $r$-process nucleosynthesis in
outflows from black hole accretion disks formed in double neutron
star and neutron star-black hole mergers was studied. They found
that these outflows, powered by angular momentum transport
processes and nuclear recombination, can be a dominant source for
the total mass ejected by the merger. They calculated  the
nucleosynthesis yields from the disk outflow using thermodynamic
trajectories from hydrodynamic simulations.  It was found that
outflows can produce a robust abundance pattern around the second
r-process peak (mass number A ${\sim}$ 130), independent of
model parameters, with significant production of A $<$
130 nuclei. This suggests that dynamical tidal ejecta with a high
electron fraction may not be required. Disk outflows reached the
third $r$-process peak (A ${\sim}$ 195) in most
simulations, although the amounts depended strongly on the model
parameters.

\subsection{Magneto-hydrodynamic Jet Models}

The magneto-hydrodynamic jet (MHDJ) supernova
model~\citep{Nishimura06,Winteler12,Nishimura15} remains as a
viable alternative.  In this scenario magnetic turbulence leads
the launch of neutron rich material into a jet.  As this material
in the jet is transported  away from the neutron star it can
undergo $r$-process  nucleosynthesis.  Since material is
transported away quickly it avoids the problems associated with
neutrino interactions near the proto-neutron star as in the  NDW
model.  Moreover, the required 
timescale, neutron density, temperature, entropy, electron
fraction  are easily  achieved in this model. Such jet models
also can naturally provide a site for a strong but rare
$r$-process early in the  early history of the Galaxy as
required from astronomical observations~\citep{Frebel18}.

The  MHDJ supernova  model 
is  described in detail in~\cite{Nishimura06}
and~\cite{Winteler12}.  In~\cite{Nishimura06},  two-dimensional
MHD simulations were  carried out from the onset of the core
collapse through the shock propagation into the silicon-rich
outer layers (${\sim} 500~{\rm ms}$ after bounce). The
$r$-process nucleosynthesis was calculated in the later
phase by employing  two different paradigms for the
extrapolations of  temperature and density in time starting from
the composition produced  the explosion.   

It was found~\citep{Nishimura06} that a jet-like explosion could
be formed from 
 the combined effects of rapid rotation  and a  strong initial magnetic field.  A ratio of rotational energy to gravitational energy $T/W \approx 0.5$\% was adapted along with a magnetic field of ${\sim} 10^{13}$ G. 
 As the ejected low $Y_e$ material  in the jet emerged  from the silicon layers, an $r$ process occurred that reasonably reproduced the solar  $r$-process abundance distribution up to  the third ($A \approx 195$) peak. 

In~\citet{Winteler12} the MHDJ simulations were elaborated by
utilizing a three-dimensional magneto-hydrodynamic core-collapse
supernova model and also including an approximate treatment of
the neutrino transport.    Similar to the two-dimensional
calculations,   bipolar jet formation required  a  rare
progenitor configuration involving  a high rotation rate
($T/W \approx 0.8$\%) and a large magnetic field ($5 \times 10^{12}$ G).
The conservation of magnetic flux amplified the initial magnetic
field to ${\sim} 10^{15}$ G during collapse.   Similar to the models
of~\cite{Nishimura06},  the low $Y_e$ material ejected
with the jet experienced  $r$-process nucleosynthesis
that reproduces  the solar $r$-process element
distribution.  
However, many of the jet simulations tend to under-produce nuclides just below and above the $r$-process abundance peaks.  This tendency seems to be a generic weakness of the MHDJ model.

\subsection{$I$-\lowercase{Process}}

In~\citet{Hempel13} a study was made of an intermediate
$r$-process ($I$-process) motivated by
observed carbon-enhanced metal-poor (CEMP) stars in the Galactic
Halo.  These stars display enrichments in heavy elements
associated with either the s or the r neutron-capture process
(e.g.,~barium and europium, respectively), and in some cases they
display evidence of both. The abundance patterns of these
CEMP-s/r stars, are particularly puzzling, since the $s$
and the $r$ processes are thought to occur in very
different astrophysical conditions. They investigated whether the
abundance patterns of CEMP-s/r stars can arise from the
nucleosynthesis of the $I$ process, which is
characterized by neutron densities intermediate between those of
the $s$ and the $r$ processes. They considered
different  constant neutron densities  ranging from
$10^7$ to 10$^{15}$ cm$^{-3}$. They showed that the
$I$-process models successfully reproduced the observed
abundance patterns  of 20 CEMP-s/r stars. 

\citet{Nishimura17} investigated r-process nucleosynthesis in
magneto-rotational supernovae, based on an explosion
induced by the magneto-rotational instability (MRI). A series of axisymmetric magnetohydrodynamical
simulations were performed that numerically
resolved the MRI. Neutrino-heating dominated the explosions and enhanced by magnetic fields, showed mildly neutron rich
ejecta producing nuclei up to A ${\sim}$ 130 (i.e.,~the weak r-process), while explosion models with stronger
magnetic fields reproduced a solar-like r-process. However, the most common abundance patterns in their  models
were  between the weak and regular r-process.  These models  produced light and intermediate mass nuclei. A variety of abundance distributions were identified.  Some of these  were consistent with $I$-process abundance patterns in r-process enhanced
metal-poor stars. These
models indicated that magneto-rotational supernovae may provide the desired environment for $I$-process nucleosynthesis.

\subsection{Collapsar $r$ Process}

There has been some
interest~\citep{Fujimoto06,Fujimoto07,Fujimoto08,Surman08,Ono12,Nakamura13,Nakamura15}
in the possibility of $r$-process nucleosynthesis  in
the relativistic jets associated with the collapsar (failed
supernova) model for gamma-ray bursts. \citep[see~][for a
review]{Nakamura13,Nakamura15}. 

Collapsars~\citep{Woosley93,Paczynski98,MacFadyen99,MacFadyen01,Popham99,Aloy00,Zhang03}
are  a favored model for the formation   of observed
long-duration gamma-ray bursts (GRBs). In the collapsar paradigm
the central core of a massive star collapses to a black hole.
However, the angular momentum of the progenitor star leads to the
formation of a heated accretion disk around the nascent black
hole.  Material is material from a polar funnel region by
magnetic field amplification and heating from the pair
annihilation of thermally generated neutrinos  emanating from
this accretion disk.  This   then leads  to an outflow of
neutron-rich
matter from the accretion disk into a relativistic jet along the polar axis.  

A large volume of work has explored  the formation of such collapsars 
[e.g.~\cite{Takiwaki04,Sawai05,Obergaulinger06,Burrows07c,Takiwaki09}; see also reference in \citet{kotake06}], 
and the development of the associated relativistic jets [e.g.~ \cite{MacFadyen99,MacFadyen01,Popham99,Aloy00,Zhang03,Proga03,Hawley06,Mizuno07,Fujimoto06,McKinney07,Komissarov07,Nagataki07,Barkov08,Nagakura11}].

The axi-symmetric special relativistic magneto-hydrodynamics
(MHD) $r$-process study  of~\citet{Nakamura15} made use
of  a model from~\citet{Harikae09}  for slowly rotating collapsar
models.   A method developed in~\citep{Harikae10}  was applied
to compute the  detailed neutrino-pair heating by ray-tracing
neutrino transport.   Hydrodynamic studies  of material in the
heated  jet along with  the associated nucleosynthesis were
evolved out to the much later times and lower temperatures
associated with the $r$ process.  In particular, that
paper explored whether the highly relativistic jet heated via
neutrino-pair annihilation of the collapsar model is capable of
generating  the high entropy per baryon and neutron-rich material
required for an $r$-process.

It was found~\citep{Nakamura15} that  this environment could
indeed produce an $r$-process-like abundance
distribution.  However, the very rapid time scale and high
entropy caused the abundances to differ from the solar abundance
distribution.  This is an extreme example of a model with a very
rapid freezeout that underproduces isotopic abundances above and
below the main $r$-process peaks.  

There is a persistent problem in the MHDJ  model, or any
model~\citep[e.g.~][]{Otsuki03} in which the $r$-process
elements are produced on a short time scale via the rapid
expansion of material away from the neutron star.  Most such
models under-produce isotopic abundances just below and above the
$r$-process abundance peaks.  

This can be related to the conditions of freezeout when the
neutrons are exhausted and the synthesized nuclides  begin to
beta decay back to the region of stable isotopes.   Neutron
captures and photo-neutron emission proceed in equilibrium for
nuclei with a neutron binding energy of about 1--2~MeV.  For
$r$-process models with a rapid transport time, the
density diminishes rapidly so that a sudden freezeout occurs
close to the $r$-process path.  As this occurs there is
a region along the $r$-process path above and below 
closed neutron shells where  the $r$-process path shifts
abruptly toward the closed shell from below (or  away from the
closed shell for higher nuclear masses).  This shifting of the
$r$-process path causes isotopes with $N = 70\mbox{--}80$
($A \sim$110--120) or $N=$90--100 ($A \sim$140--150)
to be bypassed in the beta-decay flow.   Indeed, this was a
consistent feature even in the original realistic NDW models
of~\cite{woosley94}.  Indeed, this effect has been apparent in
most $r$-process calculation since the
1970s~\citep[cf.~review in~][]{Mathews85}.

\subsection{$r$-\lowercase{process from Dark Matter induced Black Hole Collapse}}

In~\cite{Bramante16} it was suggested that a new paradigm may
occur as an explanation of the high $r$-process
abundances in the Reticulum-\textit{II} dwarf galaxy.  They
argue that the binary NSM scenario is extremely unlikely, because
binary stellar evolution models require a significant supernova
kick to produce NS--NS or NS-black hole (BH)
merger~\cite{Benjamini16,Safarzadeh17}.  However such kicks would
remove the compact binary system from the weak gravitational
potentials of the ultra-faint dwarf spheroidal.  They considered
a novel mechanism whereby neutron stars in regions of high
dark-matter density implode after accumulating a enough mass of
DM. They found  that r-process proto-material is ejected by the
tidal forces when a single NS implodes into a BH. The rate also
matches that required for the r-process abundance of both
Reticulum II and the Milky Way. They also noted that such DM
models which collapse a single NS may also solve the missing
pulsar problem in the Milky Way Galactic Center. 

\subsection{The $tr$-Process}

The abundance patterns for some observed stars do not fit the
standard $r$-process. In~\citep{Boyd12,Famiano16} it was
hypothesized that such abundance patterns  might  be produced by
failed supernovae.  In particular they considered  stars for
which their core at first collapses to form a neutron star.
However,  subsequent   infall onto the proto-neutron star causes
it to  collapse to a black hole.  This is the so-called
\emph{fall-back supernova}. 

Stars in this class span a mass range from roughly 25 to 40 solar
masses~\citep{Heger02} for low-metallicity stars. When the
neutron star collapses to a black hole the ongoing
$r$-process ceases, terminating either when the
$r$-processed regions are swallowed by the black hole or
when the electron anti-neutrinos fall below the event horizon.
Thus, this truncated $r$-process, or
$tr$-process, nucleosynthesis could terminate a
neutrino-driven wind $r$-process.  This scenario,
however  depends upon the precise time at which the black hole
prevents further $r$-process production or the emission
of nuclides into the interstellar medium. The delayed collapse to
a black hole, combined with  the difficulties in observing the
higher mass rare-earth elements, could suggest  a cutoff in the
$r$-process distributions at around $A\approx 165$.
To implement this scenario, the authors  simply assumed that mass layers  that produce the lighter $r$-process nuclei are ejected prior to those  that produce the heavier
$r$-process nuclides. 

 As a possible additional benefit of the $tr$-process, it
was noted in~\citet{Boyd12}  that in some cases the stars
produced no nuclides in the $A =130$ peak or beyond. This
could have the effect of enhancing  the yields of the lightest
$r$-process nuclides relative to the main $r$
process.  Hence, this could provide an alternative means to fill
in nuclei in the $A=  110\mbox{--}120$ region. Indeed, evidence in metal
poor stars of  the production  of nuclides in the $ A= 110\mbox{--}120 $
mass region and lighter may indicate
 a $tr$-process origin.  This conclusion, however, is
very dependent upon nuclear properties of the light neutron-rich
nuclei along the $r$-process path and requires a much
more detailed simulation of the relevant astrophysics.
In~\citet{Famiano16} it was shown that the scatter of [Sr/Ba] in
metal-poor stars is very sensitive to the adopted equation of
state, and hence,   might be used as a diagnostic.

 \section{Summary and Prospects}

 In summary, we are at a very unique time in history with regard to unraveling the mystery of $r$-process nucleosynthesis.  
For the first time Nature has revealed $r$-process nucleosynthesis in real time in the form of 
a binary neutron star merger and the associated kilonova.  
 For the first time we have opportunities to directly and indirectly probe properties of nuclei 
far from stability with the advent of radioactive ion beam accelerators already in existence and soon to come online.  
This forefront is accompanied at the same time with a leap in astronomical observations of metal poor stars in the halo of our Galaxy and in dwarf galaxies in the Local Group.  These have revealed a complex dynamics of mergers and nucleosynthesis and the existence or rare $r$-process events that eject large amounts of $r$-process material in a poorly mixed interstellar medium.

Nevertheless, uncertainties and mysteries remain.  Just to name
a few: Is the $r$-process abundance distribution in the solar system the result of one or multiple environments?  What are the roles of fission fragment yields and neutron-induced 
or $\beta$-induced fission on the termination of the $r$-process, and what are the associated fission fragment mass distributions?  Is there fission recycling in the $r$-process?  What are the actinide-boost stars?  Do they indicate the existence of fission recycling?  What are the roles of individual nuclei and their neutron-capture and $\beta$-decay rates in determining the final $r$-process abundances? Does a more accurate description of the neutrino transport, including oscillations and other quantum effects significantly change the predicted r-process yields?

  In conclusion, although great progress has been made and is being made,  there is still much work to be done to unravel the mystery of $r$-process nucleosynthesis in the Galaxy.  Even so, with the current advance 
  of astronomical instrumentation,  nuclear measurements, and nuclear theory, a definitive understanding of the $r$-process is within reach.

\section*{Acknowledgments}
This work was supported by Grants-in-Aid for Scientific Research of {JSPS {Japan}} ({15H03665}, {17K05459}).
The work of GJM was supported in part by {DOE {USA}}  nuclear theory grant
{DE-FG02-95-ER40934}. The research of ABB was supported in part by the
{U.S. National Science Foundation {USA}} Grants No. {PHY-1514695} and
{PHY-1806368}.  MAF is supported in part by the {U.S. National Science
Foundation {USA}}  Grant  No. {PHY-1712832} and by the {Core Fulbright U.S.
Scholar Program  {USA}}. All authors are supported in part by the visiting
professor program at the {NAOJ {Japan}}.
%\QUERY[8]

%\back{}

\nocite{*}

\bibliography{review_bib}
\end{document}